\documentclass[11pt,a4paper,final]{iopart}

\usepackage{iopams}  
\usepackage{xspace}
\usepackage{graphicx}
\usepackage{enumitem}
\usepackage{amsthm}

\newtheorem{theorem}{Theorem}[section] 
\newtheorem{lemma}[theorem]{Lemma}

\newtheorem{proposition}[theorem]{Proposition}

\usepackage[bookmarks=true,
            bookmarksopen=true,colorlinks=true,breaklinks=true,linkcolor=blue,%
            citecolor=blue]{hyperref}

\hypersetup{%
  pdftitle = {Smooth Gowdy symmetric generalized 
  Taub-NUT solutions}, 
  pdfsubject = {},
  pdfauthor = {F. Beyer and J. Hennig}, 
  pdfkeywords = {}%
}

\newcommand{\R}{\ensuremath{\mathbb R}\xspace}
\newcommand{\C}{\ensuremath{\mathbb C}\xspace}
\newcommand{\So}{\ensuremath{\mathbb S^1}\xspace}
\newcommand{\St}{\ensuremath{\mathbb S^2}\xspace}
\newcommand{\Sth}{\ensuremath{\mathbb S^3}\xspace}
\newcommand{\SU}{\ensuremath{\mathrm{SU}(2)}\xspace}
\newcommand{\SoXSt}{\ensuremath{\mathbb S^1\times \mathbb S^2}\xspace}
\newcommand{\Tt}{\ensuremath{\mathbb T^2}\xspace}
\newcommand{\scalarpr}[2]{\left<#1,#2\right>}
\newcommand{\diag}{\ensuremath{\mathrm{diag}\xspace}}
\newcommand{\RR}[1]{\ensuremath{\mathcal{R}[#1]}\xspace}
\newcommand{\Fe}{\ensuremath{F_0}\xspace}

\newcommand{\Fo}{\ensuremath{F_1}\xspace}
\newcommand{\Eqref}[1]{Eq.~\eref{#1}}
\newcommand{\Eqsref}[1]{Eqs.~\eref{#1}}
\newcommand{\Sectionref}[1]{Section~\ref{#1}}
\newcommand{\Propref}[1]{Proposition~\ref{#1}}
\newcommand{\Theoremref}[1]{Theorem~\ref{#1}}
\newcommand{\Lemref}[1]{Lemma~\ref{#1}}
\newcommand{\Conditionref}[1]{Condition~\ref{#1}}

\newcommand{\keyword}[1]{\textit{#1}\xspace}

\newcommand{\ee}{\mathrm e}
\newcommand{\ii}{\mathrm i}
\newcommand{\dd}{\mathrm d}
\newcommand{\E}{\mathcal E}
\newcommand{\Al}{\mathcal A_1}
\newcommand{\Ar}{\mathcal A_2}
\newcommand{\Hp}{{\mathcal H_\mathrm{p}}}
\newcommand{\Hf}{{\mathcal H_\mathrm{f}}}
\newcommand{\p}{_\mathrm{p}}
\newcommand{\f}{_\mathrm{f}}

\begin{document}

\title{Smooth Gowdy symmetric generalized Taub-NUT solutions}

\author{Florian Beyer and J\"org Hennig}
\address{Department of Mathematics and Statistics,
University of Otago, P.O. Box 56, Dunedin 9054, New Zealand}
\eads{\mailto{fbeyer@maths.otago.ac.nz}, \mailto{jhennig@maths.otago.ac.nz}}

\begin{abstract}
  We study a class of $\Sth$-Gowdy vacuum models with a regular past
  Cauchy horizon which we call smooth Gowdy symmetric generalized
  Taub-NUT solutions.  In particular, we prove existence of such solutions by
  formulating a singular initial value problem with asymptotic data on
  the past Cauchy horizon. We prove that also a future Cauchy horizon
  exists for generic asymptotic data, and derive an explicit expression
  for the metric on the future Cauchy horizon in terms of the
  asymptotic data on the past horizon. This complements earlier
  results about $\So\times \St$-Gowdy models.
\end{abstract}



\section{Introduction}

Studies of cosmological solutions of Einstein's field equations have a
long tradition and led to astonishing results about our own universe.
In particular, observations indicate that there was a big bang in the
distant past, and indeed, the simplest cosmological models, namely the
Friedmann solutions for reasonable matter fields, predict precisely
this behavior. The question arises as to whether such curvature
singularities occur for generic solutions of Einstein's field
equations or if strong symmetry assumptions, e.g.\ those made for the
Friedmann models, are necessary for this. The Hawking-Penrose
singularity theorems \cite{Hawking:1973tb} shed some light on this
question. They predict incompleteness of causal geodesics in a wide
class of situations. However, the information about the geometric reason for the
incompleteness which is provided by these theorems is very limited, and it is
indeed not always caused by a geometric singularity.

Let us restrict all of our investigations to vacuum with a vanishing
cosmological constant and to four spacetime dimensions.  The
corresponding Cauchy problem for Einstein's field equations is
well-posed and leads to the notion of the \keyword{maximal globally hyperbolic
development} (MGHD) of a given Cauchy data set
\cite{Andersson:2004um,Ringstrom:2009cj}.  The example of the Taub solution
\cite{Taub:1951vk}, which we discuss in more detail below, shows that incompleteness of causal geodesics, as
predicted by the singularity theorems, can signal a different kind of
phenomenon; in particular it is possible to extend the MGHD. The extended
solutions are, however, not globally hyperbolic. There exist closed causal
curves and indeed, there are several non-equivalent extensions.  This
unexpected property has caused an ongoing debate in the literature.
Are such pathological phenomena a typical feature of Einstein's theory
of gravity -- in which case it could not be considered as a ``proper''
physical theory -- or, again, do such phenomena only occur under very
strong and special assumptions, for example the high symmetry of the
Taub solutions?  An interesting  hypothesis in
this context is the strong cosmic censorship conjecture whose widely
accepted formulation was given for the first time in \cite{Chrusciel:1991uk},
based on ideas by Eardly and Moncrief \cite{Moncrief:1981bq} and Penrose
\cite{Penrose:1969tw}. More details and references can be found in
\cite{Ringstrom:2009cj,Rendall:540392}. This conjecture states that, for spatially compact or asymptotically flat spacetimes,  the MGHD
of \keyword{generic} Cauchy data is
\keyword{inextendible}\footnote{This conjecture, of course, only makes
  sense if one is able to give a precise and reasonable meaning to the
  terms ``generic'' and ``inextendible''.}. If this were true, it would imply that, in the
generic situation, incompleteness of causal geodesics is indeed caused
by a geometric singularity in some sense, while the pathologies encountered for the Taub spacetimes only occur in special circumstances. At this stage, however, this conjecture has not been confirmed in general situations; see e.g.\
  \cite{Ringstrom:2009cj}.

In his effort to generalize the family of Taub solutions and hence to
show that there is a large (but presumably still non-generic) class of solutions
of the field equations with similar ``undesired'' properties, Moncrief
defines the family of \keyword{generalized Taub-NUT spacetimes} in
\cite{Moncrief:1984js}. He is able to prove an existence result under an
analyticity assumption. The motivation in our paper is
two-fold. First, we want to extend this existence result to the smooth
case\footnote{We use the term ``smooth'' for infinitely differentiable
  objects, as opposed to (real) ``analytic'', which entails the convergence of the Taylor series.} and formulate it as a singular initial value
problem with ``asymptotic data'' on the Cauchy horizon. Then our second motivation is to study the global dynamics of such
solutions. To make this feasible, we restrict to the case with Gowdy
symmetry.  By means of so-called soliton methods, we find that the existence of a past Cauchy horizon generically implies the
existence of a future Cauchy horizon, at least under certain
topological restrictions on the horizons. The reader
should compare these results to the \SoXSt case in \cite{Hennig2010}.

Note that the global existence of Gowdy-symmetric solutions with Cauchy horizons, which is our main result, does not necessarily imply that they are generic (and thus violate strong cosmic censorship). Indeed, cosmic censorship remains a fundamental open problem in GR and studying existence and properties of solutions with acausal extensions is of crucial importance for a deeper understanding of the mathematical structure of the Einstein equations and their solutions. Our considerations are meant to contribute some insights in that direction. 

The paper is organized as follows. In \Sectionref{sec:geomprelim} we
discuss some background material, in particular the symmetry reduction
introduced by Geroch which is later applied to write the metric
and the field equations in a useful way. \Sectionref{sec:GenTaubNut}
is devoted to the definition and discussion of our class of
\keyword{smooth Gowdy symmetric generalized Taub-NUT solutions} based
on Moncrief's earlier mentioned class. The existence and uniqueness
theory and the corresponding singular initial value problem are
considered in \Sectionref{sec:existence}. The basic ingredients for
these investigations are Fuchsian methods which we describe  in
the appendix.  The next main part is \Sectionref{sec:linearproblem}
where we discuss the global-in-time properties of smooth Gowdy
symmetric generalized Taub-NUT solutions. We finish the paper with
conclusions in \Sectionref{sec:discussion}.


\section{Geometric preliminaries}
\label{sec:geomprelim}

\setcounter{footnote}{0}

\subsection{General symmetry reduction by Geroch} 
\label{sec:Geroch}
We briefly present here the symmetry reduction introduced by Geroch in
\cite{Geroch:1971ix}. Let $M=\R\times H$ be an oriented and time-oriented
globally hyperbolic $4$-dimensional spacetime endowed with a
metric\footnote{Depending on the context, we denote any tensor field
  either by a symbol like $T$, or we use the abstract index notation
  ${T^{ab\ldots}}_{cd\ldots}$ where $a$, $b$, $c$, $d$, \ldots denote
  abstract indices.}
$g_{{a}{b}}$ of signature $(-,+,+,+)$, a global time-function $t$ and a Cauchy
surface $H$. We denote the volume form of $g_{{a}{b}}$ by
$\epsilon_{{a}{b}{c}{d}}$ and the hypersurfaces given by $t=t_0$
for any constant $t_0$ by $H_{t_0}$. Each $H_{t_0}$ is homeomorphic to
$H$.

Now, let $\xi$ be a smooth spacelike\footnote{Geroch
  considers the case of a timelike Killing vector field. As a
  consequence of this, some signs in certain expressions are different
  than in Geroch's reference. } Killing
vector field which is tangent to the hypersurfaces $H_t$ and set
\[\lambda:=g(\xi,\xi).
\] 
The twist $1$-form of $\xi$ is 
\[\Omega_{a}:=\epsilon_{{a}{b}{c} {d}}\xi^{b}\nabla^{c}\xi^{d},\]
where $\nabla$ is the covariant derivative compatible with $g$.
The field $\xi$ is hypersurface orthogonal if and only if $\Omega_{a}
\equiv 0$, which will, however, not be assumed in the following. 
We define the ``$3$-metric''
\[h_{{a}{b}}:=g_{{a}{b}}-\frac 1\lambda \xi_{a}\xi_{{b}},
\] 
on $M$, and, by raising indices with the inverse of $g$, we define also
${h^{a}}_{b}$ and $h^{{a}{b}}$ on $M$. The first of these tensors is the projector to
the space orthogonal to $\xi$ in $T_p M$, $p\in M$. From the volume form
$\epsilon_{{a}{b}{c}{d}}$ of $g$, we furthermore introduce 
\[\epsilon_{{a}{b}{c}}:=\frac
1{\sqrt{\lambda}}\epsilon_{{a}{b}{c}{d}}\xi^{d}.
\] 
Let $\alpha_{a}$ be any $1$-form. One defines the derivative
operator $D$ as
\[D_{a} \alpha_{b}:={h^{{a}'}}_{a} \, {h^{{b}'}}_{b} \nabla_{{a}'}
\alpha_{{b}'}.
\]

Note that at this stage we are only interested in local patches of
$M$. Then, the flow generated by $\xi$ induces a map $\pi$ from $M$ to
the space of orbits $S$, i.e.\ $\pi$ maps every $p\in M$ to the
(locally) uniquely determined integral curve of $\xi$ starting at
$p$. The requirement that $\pi$ is a smooth map induces a
differentiable structure on $S$, and hence $S$ can be considered as a
smooth manifold. The relations $\mathcal L_\xi h=0$ and $h(\xi,\cdot)=0$,
are used by Geroch to show that there is a unique metric on $S$ which pulls back to
$h_{{a}{b}}$ along $\pi$. We write this metric on $S$ again
as $h_{{a}{b}}$. The same can be done for ${h^{a}}_{b}$ and $h^{{a}{b}}$,
which are henceforth considered as objects on the quotient space
$S$. The first becomes the identity operator on $T_pS$, while the second is the
inverse of the $3$-metric $h_{{a}{b}}$ on $S$. We can proceed in the
same way with the function $\lambda$, the $1$-form $\Omega_{a}$, the
tensor $\epsilon_{{a}{b}{c}}$ and the covariant derivative operator
$D_{a}$, which can henceforth all be considered as objects on $S$. Then
$\epsilon_{{a}{b}c}$ becomes the volume form of $h_{{a}{b}}$ and
$D_{a}$ the covariant derivative operator compatible with
$h_{{a}{b}}$.  Geroch shows that Einstein's vacuum field equations on
$(M,g)$ imply that the $1$-form $\Omega$ is closed, $\dd\Omega=0$. At this stage we focus on 
local considerations, which allows us to introduce a twist potential
$\omega$ so that $\Omega=\dd\omega$.  The quantities $\lambda$,
$\omega$ and $h_{{a}{b}}$ on $S$ therefore completely characterize the local geometry
of $(M,g)$.

Now Geroch introduces a
conformal rescaling
\[\hat h_{{a}{b}}:=\lambda h_{{a}{b}}.\]
We refer to the associated covariant derivative operator as $\hat D_a$,
Ricci tensor as $\hat S_{{a}{b}}$ etc. He shows that the vacuum field
equations for $(M,g)$ (and certain geometric identities) are
equivalent to the following set of equations
\begin{eqnarray}
  \label{eq:Gerochevollambda}
  \hat D_a\hat D^a\lambda
  &=\frac 1\lambda\left(\hat D^{a}\lambda\hat D_{a}\lambda
    -\hat D^{a}\omega\hat D_{a}\omega\right),\\
  \label{eq:Gerochevolomega}
  \hat D_a\hat D^a\omega
  &=\frac 2\lambda\hat D^{a}\lambda\hat D_{a}\omega,\\
  \label{eq:GerochRicci3}
  \hat S_{{a}{b}}&=\frac 1{2\lambda^2}\left(
    \hat D_{a}\lambda\hat D_{b}\lambda
    +\hat D_{a}\omega\hat D_{b}\omega\right).
\end{eqnarray}
These equations are the Euler-Lagrange equations of the Lagrangian density
\[\mathcal L=\sqrt{-\det\hat h}\left[\hat S
  +\frac 1{2\lambda^2}\left(\hat D^{a}\lambda\hat D_{a}\lambda +\hat
    D^{a}\omega\hat D_{a}\omega\right)\right].
\] 
Hence, the equations can be interpreted as $2+1$-dimensional gravity on
$S$ coupled to a wave map $u:S\rightarrow\mathcal H$ where $\mathcal
H$ is the $2$-dimensional hyperbolic space represented by the
components $(\lambda,\omega)$.


\subsection{Symmetry reduction for spacetimes of spatial \texorpdfstring{$3$}{3}-sphere 
topology}
We
think of \Sth as the submanifold of $\R^4$ determined by
$x_1^2+x_2^2+x_3^2+x_4^2=1$. The \keyword{Euler coordinates} $(\theta,\lambda_1,\lambda_2)$ of $\Sth$ are
\begin{eqnarray*}
    x_1&=\cos\frac\theta 2\cos\lambda_1,  \quad
    x_2&=\cos\frac\theta 2\sin\lambda_1,\\
    x_3&=\sin\frac\theta 2\cos\lambda_2,  \quad
    x_4&=\sin\frac\theta 2\sin\lambda_2,
  \end{eqnarray*}
with $\theta\in (0,\pi)$ and $\lambda_1,\lambda_2\in (0,2\pi)$.
Clearly, these coordinates break down at the points $\theta=0$
and $\pi$ which we refer to as ``poles'' or ``axes'' of \Sth in the following.  We also make use of the coordinates
$(\theta,{\rho_1},{\rho_2})$ (which we also call Euler coordinates) with $\theta$ as above and 
\begin{equation}
  \label{eq:eulerangleparm2}
  \lambda_1=:({\rho_1}+{\rho_2})/2,\quad
  \lambda_2=:({\rho_1}-{\rho_2})/2.
\end{equation}
Note that the coordinate fields $\partial_{\rho_1}$ and
$\partial_{\rho_2}$ are smooth non-vanishing vector fields on \Sth,
while the smooth fields $\partial_{\lambda_1}$ and $\partial_{\lambda_2}$
vanish at certain places. We remark that these fields can be characterized
geometrically (without making reference to coordinates) in terms of
left- and right-invariant vector fields of the standard
action of $SU(2)$ on \Sth, see for example \cite{Beyer:2008gg,Beyer:2009vw}.

Now we specialize the general discussion of \Sectionref{sec:Geroch} to the case
$M=\R\times\Sth$ with $H=\Sth$ and assume that the smooth  spacelike Killing vector field $\xi$ generates a Hopf bundle on every Cauchy surface $H_t$. This means in particular that the quotient manifold $S$ is $\R\times\St$ (in particular, it is a  smooth manifold \textit{globally}). Moreover, the group $U(1)$ acts on $H_t$ effectively, and the integral curves of $\xi$, which are in $H_t$, are closed. 
The quotient map $\pi$, which was introduced for the general case in the previous section, becomes the Hopf map 
$M\rightarrow S$ whose explicit coordinate representation can be given as follows. In every $H_t$ we introduce Euler coordinates, and we assume that $\xi=\partial_{\rho_1}$ on every $H_t$ (this is a gauge condition for the coordinates, which certainly must be justified in our later discussion).
Then the map $\pi$ can be written as
 \begin{eqnarray}\fl
   \quad \pi: \R\times\Sth\rightarrow\R\times\St,\nonumber\\
   \fl
    \qquad\qquad (t,x_1,x_2,x_3,x_4)\mapsto
  (t,y_1,y_2,y_3)\nonumber\\
\fl
  \label{eq:hopf}
 \qquad\quad\qquad=(t,2(x_1 x_3+x_2x_4),2(x_2x_3-x_1x_4),  x_1^2+x_2^2-x_3^2-x_4^2)\\
\fl\qquad\quad\qquad
 =(t,\sin\theta\cos{\rho_2},\sin\theta\sin{\rho_2},\cos\theta).\nonumber
\end{eqnarray}
Here we consider \St as the submanifold of $\R^3$ determined by
$y_1^2+y_2^2+y_3^2=1$. When we introduce standard polar coordinates on
\St, namely
\begin{equation}
  \label{eq:sphericalcoordinates}
  y_1=\sin\vartheta\cos{\phi},\quad
  y_2=\sin\vartheta\sin{\phi},\quad
  y_3=\cos\vartheta,
\end{equation}
then $\pi$ reduces to
\begin{equation}
  \label{eq:hopfspherical}
\pi: (t,\theta,{\rho_1},{\rho_2})\mapsto
(t,\vartheta,\phi)=(t,\theta,{\rho_2}).
\end{equation}
We see explicitly that the
push-forward of $\xi=\partial_{\rho_1}$ to $\R\times\St$ along $\pi$
vanishes (which must be the case of course). Moreover, the push-forward of $\partial_{\rho_2}$ equals the
coordinate vector field $\partial_\phi$ on $\St$.  


Let us now consider the case of an additional smooth spacelike Killing vector field $\eta$ with closed orbits which commutes with $\xi$. We assume that $\xi$ and $\eta$ together generate a global smooth effective action of the
direct product group $U(1)\times U(1)$. It can be shown that all smooth effective actions of $U(1)\times U(1)$ on $\Sth$ are equivalent in the sense that any other smooth
effective action of $U(1)\times
U(1)$ on $\Sth$ can be reduced to a canonical one by applying a
diffeomorphism of \Sth into itself and an automorphism of $U(1)\times
U(1)$ to itself \cite{Chrusciel:1990ti}. On any $H_t$, we can assume that the Euler coordinates above have been chosen so that $\eta=\partial_{{\rho_2}}$, in addition to our previous assumption that $\xi=\partial_{{\rho_1}}$. The assumption that $\xi$ and $\eta$ can be identified with these coordinate vector fields for \textit{all} $t$ is again a restriction on the coordinate gauge which we choose later. According to the definition of Euler coordinates, we see that the two Killing fields generate two families of ``conjugate circles'' in $\Sth$ which yield a foliation of a dense subset of $\Sth$ in terms of $2$-tori; this is related to the Clifford parallelism discussed in \cite{Berger:1987tga,Berger:1987voa}.
Our symmetry action (and hence the foliation in terms of $2$-tori) degenerates,
in the sense that the group orbits become $1$-dimensional, precisely
at the ``poles'' $\theta=0$ and at $\theta=\pi$. When $\theta=0$, the vector
$\partial_{\lambda_2}$ vanishes, while $\partial_{\lambda_1}$ vanishes
at $\theta=\pi$. The vectors $\partial_{{\rho_1}}$ and
$\partial_{{\rho_2}}$ on the other hand never vanish, but both become parallel at
$\theta=0$ and $\theta=\pi$. 

Since $\xi$ and $\eta$ commute, we can apply
Geroch's  symmetry reduction successively for both
fields. However, the result is then not a smooth manifold, but rather
a manifold with boundary. While our discussion in \Sectionref{sec:linearproblem} is not effected by this and is therefore carried out in this fully reduced picture, we would run into serious problems for the PDE analysis in \Sectionref{sec:existence}. In that section we therefore only perform the reduction with respect to $\xi$ and hence
obtain the smooth orbit manifold $S=\R\times\St$; the push-forward of the other Killing vector field $\eta$ along $\pi$, which we denote again by $\eta$, is then a smooth Killing vector field of the $3$-metric $h$ and we have $\eta=\partial_\phi$ (in the
standard polar coordinates on \St given by \Eqref{eq:sphericalcoordinates} associated with the Hopf map by \Eqref{eq:hopf}). 

According to \cite{Geroch:1972bp,Chrusciel:1990ti}, Einstein's vacuum field equations imply that the \keyword{twist quantities}
\begin{equation}
  \label{eq:twistconstants}
  \kappa_1:=\epsilon_{{a}{b}c{d}}\eta^{a} \xi^{b}\nabla^c \xi^{d},\quad
  \kappa_2:=\epsilon_{{a}{b}c{d}}\eta^{a} \xi^{b}\nabla^c
  \eta^{d},
\end{equation}
vanish for spatial \Sth-topology. The geometrical interpretation is that the $2$-space
orthogonal to the $2$-space spanned by $\xi$ and $\eta$ in $T_pM$ ($p\in M$) is
integrable and hence forms a $2$-surface everywhere.  This suggests the
following ansatz for the metric
\begin{equation}
  \label{eq:generalGowdy}
  g=g_{AB}\dd x^A \dd x^B
  +R\left[\ee^L (\dd{\rho_1}+Q \dd{\rho_2})^2+\ee^{-L} \dd{\rho_2}^2\right],
\end{equation}
where $A,B=0,1$ label coordinates $t$ and $\theta$ on the submanifolds orthogonal to
the Killing vector fields; the metric $g_{AB}$ is so far
unspecified. The functions $R$, $L$ and $Q$ only depend on $t$ and
$\theta$, i.e.\ are constant along the Killing vector fields.  Chru\'sciel \cite{Chrusciel:1999dk} shows that under a certain genericity condition, $U(1)\times U(1)$-symmetric vacuum solutions imply the existence of a smooth function\footnote{The
  function $M$ must not be confused with the symbol for the manifold $M$.} $M$, a
constant $R_0>0$, and functions $Q$ and $L$ as above  such
that the spacetime $(0,\pi)\times\Sth$ with $g$ of the form
\Eqref{eq:generalGowdy} and
\begin{equation}
  \label{eq:arealgauge}
  R=R_0 \sin t\sin\theta,\quad (g_{AB})=\ee^M\mathrm{diag}\,(-1,1),
\end{equation}
can be isometrically embedded into a maximally extended globally hyperbolic vacuum spacetime.
 One calls such a
spacetime a \keyword{Gowdy spacetime} and the coordinate condition \Eqref{eq:arealgauge} for $t\in (0,\pi)$ the \keyword{areal gauge}.

By performing the symmetry reduction of a Gowdy spacetime with respect to the Killing field $\xi$ as mentioned above, it follows that $\lambda$ and $\omega$, as objects on $S$, are constant
along $\eta$. 
We can compute from \Eqref{eq:generalGowdy} that
\begin{equation}
  \label{eq:relationLlambda}
  \lambda=R \ee^L,
\end{equation}
and
\begin{equation}
  \fl
  \hspace{5ex}\partial_t\omega=-R\ee^{2L}\sqrt{|\det(g_{CD})|}g^{\theta A}\partial_A Q,\quad
   \partial_\theta\omega=R\ee^{2L}\sqrt{|\det(g_{CD})|}g^{t A}\partial_A Q,
\end{equation}
which simplifies to
\begin{equation}\label{eq:relationPQomegatheta}
 \partial_t\omega = -R_0\ee^{2L}\sin t\sin\theta\,\partial_\theta Q,\quad
 \partial_\theta\omega = -R_0\ee^{2L}\sin t\sin\theta\,\partial_t Q
\end{equation}
for the diagonal metric \eref{eq:arealgauge}.
The $3$-metric is
\begin{equation}
  \label{eq:Gowdyh}
  h=g_{AB}\dd x^A\dd x^B+R \ee^{-L}\dd{\rho_2}^2
  =g_{AB}\dd x^A \dd x^B+\frac{R^2}\lambda \dd{\rho_2}^2.
\end{equation}

\subsection{Smoothness conditions for the metric components}
\label{sec:boundarybehavior}

We have found above that the $3$-metric $h$ is a smooth Lorentzian metric on $\R\times\St$ with Killing field $\eta$. The condition that the twist constants \Eqref{eq:twistconstants} vanish, implies in addition that $\eta$ is
hypersurface orthogonal with respect to $h$.  Of particular importance now is the behavior of
the metric components in standard polar coordinates, with $\pi$ given\footnote{In many of the following expressions, we will therefore replace $\rho_2$ by $\phi$ and $\theta$ by $\vartheta$, and vice versa, in accordance with \Eqref{eq:hopfspherical}.} by \Eqref{eq:hopfspherical}, at the poles of the $2$-sphere at $\vartheta=0,\pi$.
Recall that $\eta=\partial_\phi$ in this representation. 

Any metric $l_{ab}$ on \St, for which $\eta$ is a hypersurface orthogonal Killing vector field, must be of the form
\begin{equation}
  \label{eq:2metricpolcoords}
  l=F(\vartheta)d\vartheta^2+G(\vartheta)d\phi^2,
\end{equation}
for some functions $F$ and $G$ at all points except $\theta=0,\pi$.
In order to find necessary and sufficient conditions on those functions that imply the smoothness of the metric at the poles $\vartheta=0,\pi$, we introduce further \textit{regular} local coordinate patches in a neighborhood of each pole: at the north pole $\vartheta=0$, we introduce coordinates $(y_1,y_2)$ by
\[y_1=\sin\vartheta\cos{\phi},\quad y_2=\sin\vartheta\sin{\phi},
\] 
so that the north pole corresponds to $(y_1,y_2)=(0,0)$,
and at the south pole $\vartheta=\pi$, 
$(\tilde y_1, \tilde y_2)$ by
\[\tilde y_1=\sin\vartheta\cos{\phi},\quad \tilde y_2=\sin\vartheta\sin{\phi},
\] 
so that the south pole corresponds to $(\tilde y_1,\tilde y_2)=(0,0)$. Both of these local patches break down at the equator given by $\vartheta=\pi/2$.
Close to the north pole, every smooth metric can be written as
\begin{equation}\label{eq:LE1}
 l=l_{11} \dd y_1^2+2 l_{12} \dd y_1\dd y_2+l_{22} \dd y_2^2, 
\end{equation}
where $l_{11}$, $l_{12}$ and $l_{22}$ are smooth functions of the coordinates $(y_1,y_2)$. When we transform this metric to polar coordinates and use the conditions  that (i) the resulting metric should be diagonal and (ii) the metric components should be independent of $\phi$, we find that there must be smooth functions $f_1$ and $f_2$  so that
\begin{eqnarray}\label{eq:coeff}
  l_{11}(y_1(\vartheta,{\phi}),y_2(\vartheta,{\phi}))
  &=f_1(\cos\vartheta)-f_2(\cos\vartheta)\sin^2\!\vartheta\cos (2{\phi}),\nonumber\\
  l_{22}(y_1(\vartheta,{\phi}),y_2(\vartheta,{\phi}))
  &=f_1(\cos\vartheta)+f_2(\cos\vartheta)\sin^2\!\vartheta\cos (2{\phi}),\\
  l_{12}(y_1(\vartheta,{\phi}),y_2(\vartheta,{\phi}))
  &=-f_2(\cos\vartheta)\sin^2\!\vartheta\sin(2{\phi}).\nonumber
\end{eqnarray}
By transforming the metric \Eqref{eq:LE1} with coefficients \Eqref{eq:coeff} to polar coordinates $(\vartheta,\phi)$, and comparing the result with \Eqref{eq:2metricpolcoords}, we find that $F$ and $G$ must be smooth functions of $\cos\vartheta$ so that
\begin{equation}
  \label{eq:FGleadingorderpols}
  F(\cos\vartheta)=O(1),\quad G(\cos\vartheta)=\sin^2\!\vartheta\, (F(\cos\vartheta)+O(\sin^2\!\vartheta)),
\end{equation}
at $\vartheta=0$. The same can be carried out for the south pole, and we find that \Eqref{eq:FGleadingorderpols} must also hold at $\vartheta=\pi$.

For the following, we rather parametrize the metric $l$ as  
 \begin{equation*}
l=\ee^{M} \dd\vartheta^2+\sin^2\!\vartheta\, \ee^{2U}\dd{\phi}^2,
\end{equation*} 
for functions $M$ and $U$ instead of the functions $F$ and $G$ in \Eqref{eq:2metricpolcoords}.
According to our above result, the functions  $M$ and $U$ must hence be two smooth functions of $\cos\vartheta$ for which the \keyword{smoothness condition}
\begin{equation}
  \label{eq:boundarycondM}
  \ee^{M}=\ee^{2U}+
  \hat M(\cos\vartheta)\sin^2\!\vartheta,
\end{equation}
holds with some smooth function $\hat M$. 
Using this together with \Eqref{eq:arealgauge}, we can rewrite \Eqref{eq:Gowdyh} for the general metric $h$ on
$S=\R\times\St$ as
\begin{equation}
  \label{eq:ansatzh}
  h=\ee^{M}(-\dd t^2+\dd\vartheta^2)+\sin^2\!\vartheta\, \ee^{2U}\dd\phi^2,
\end{equation}
where
\begin{equation}
  \label{eq:gggaugechoice}
  U=\ln R_0+\ln\sin t-\frac 12\ln\lambda=(\ln R_0-L-\ln\sin\vartheta+\ln\sin t)/2.
\end{equation}
This is a bounded function for every fixed  $t\in (0,\pi)$ under the previous
assumptions because the quantity $\lambda$ is finite and bounded away from
zero (including the poles). Clearly \Eqref{eq:relationLlambda} yields that
\[\ee^L=O(R^{-1})=O(\sin^{-1}\!\vartheta)\]
at the poles, and the smoothness condition \Eqref{eq:boundarycondM}
translates to
\begin{equation}
  \label{eq:boundarycondM2}
  \ee^{M}=\frac{R^2}{\lambda\sin^2\vartheta}
+\hat M(\cos\vartheta)\sin^2\!\vartheta.
\end{equation}
A consequence is that $\ee^M$ is bounded and non-vanishing at the
poles for every fixed $t\in (0,\pi)$.

Let us also derive a smoothness restriction for the function $Q$ in  \Eqref{eq:generalGowdy}.
Our choice of Killing basis
$\{\xi,\eta\}=\{\partial_{{\rho_1}},\partial_{{\rho_2}}\}$,
for which
$\partial_{{\rho_1}}=\partial_{{\rho_2}}$ at $\theta=0$ and
$\partial_{{\rho_1}}=-\partial_{{\rho_2}}$ at $\theta=\pi$ holds, has the consequence that
$g(\xi,\xi)=g(\eta,\eta)=\pm g(\xi,\eta)$ at the poles for every fixed $t\in(0,\pi)$.  Therefore
there must exist a smooth function $\hat Q$ which only depends on $t$
and $\cos\theta$ so that
\begin{equation}
  \label{eq:QBound}
  Q(t,\theta)=\cos\theta+\hat Q(t,\cos\theta)\sin^2\!\theta.
\end{equation}
In particular it follows from
Eq.~\eref{eq:relationPQomegatheta} that for each fixed $t\in (0,\pi)$ 
\[\fl\quad -2=Q(t,\pi)-Q(t,0)=\int_0^\pi Q_\theta\, \dd\theta =-\int_0^\pi R^{-1}
\ee^{-2L}\partial_t\omega\, \dd\theta =-\int_0^\pi R
\lambda^{-2}\partial_t\omega\, \dd\theta.
\] 

\subsection{Reparametrizations of the Gowdy orbits}
\label{sec:reparametrization}
All of our discussions so far are based on the choice
$\{\partial_{{\rho_1}},\partial_{{\rho_2}}\}$ as the Gowdy Killing
basis on $M$. Now we study general reparametrizations of the Gowdy
Killing orbits in $M$, i.e.\ arbitrary bases of the same Gowdy Killing
algebra. 

Let $(\phi_1,\phi_2)\in\R^2$ be coordinates on the Killing orbits so
that a Gowdy invariant metric has the form analogous to
Eq.~\eref{eq:generalGowdy}, i.e.\
\[g=\ee^M(-\dd t^2+\dd\theta^2)
+R\left[ \ee^L (\dd\phi_1+Q \dd\phi_2)^2+\ee^{-L} \dd\phi_2^2\right].\]
We are allowed to reparametrize the orbits by means of constants
$a,b,c,d\in\R$, so that
\[ad-bc\not=0\]
and
\[\phi_1=a\tilde\phi_1+b\tilde\phi_2,\quad
\phi_2=c\tilde\phi_1+d\tilde\phi_2.
\] 
The coordinates $t$ and $\theta$ are not changed. In terms of the new
coordinates, we want to write the metric as
\[g=\ee^M(-\dd t^2+\dd\theta^2)+\tilde R\left[ \ee^{\tilde L} (\dd\tilde\phi_1
+\tilde Q \dd\tilde\phi_2)^2+\ee^{-\tilde L} \dd\tilde\phi_2^2\right].\]
One finds that
\begin{eqnarray}
 \label{eq:trans1}
  \tilde R&=|ad-bc| R,\\
  \ee^{\tilde L}&=\frac{(a+c\, Q)^2 \ee^L+c^2 \ee^{-L}}{|ad-bc|},\\
 \label{eq:trans3}
  \tilde Q&=\frac{(a+c\,Q)(b+d\,Q)\ee^L+c\, d\, \ee^{-L}}
  {(a+c\, Q)^2 \ee^L+c^2 \ee^{-L}}.
\end{eqnarray}

A particularly useful transformation is the \keyword{inversion}, i.e.\ the
interchange of the Killing basis fields. Then we have $a=d=0$,
$b=c=1$, and hence
\begin{equation*}
  \tilde R=R,\quad \ee^{\tilde L}=\ee^L Q^2+\ee^{-L},\quad
  \tilde Q=\frac{\ee^L Q}{\ee^L Q^2+\ee^{-L}}.
\end{equation*}

Another useful reparametrization is the following. Let us consider a metric in the parametrization $(\rho_1,\rho_2)$
of the Killing orbits as given by Eq.~\eref{eq:generalGowdy}, i.e.\ we pick
$\phi_1={\rho_1}$ and $\phi_2={\rho_2}$. Now let
$\tilde\phi_1=\lambda_1$ and $\tilde\phi_2=\lambda_2$ and hence
\[g=\ee^M(-\dd t^2+\dd\theta^2)+\tilde R\left[ \ee^{\tilde L} (\dd\lambda_1 +\tilde Q
\dd\lambda_2)^2+\ee^{-\tilde L} \dd\lambda_2^2\right].
\] 
For this we must choose $a=1$, $b=1$, $c=1$, $d=-1$ from
Eq.~\eref{eq:eulerangleparm2}.  It follows that
\begin{equation}
  \label{eq:transformationQ}
  \tilde R=2R,\quad
  \ee^{\tilde L}=\frac{(1+Q)^2 \ee^L+\ee^{-L}}{2},\quad  
  \tilde Q=\frac{-(1-Q^2)\ee^L+\ee^{-L}}{(1+Q)^2 \ee^L+\ee^{-L}}.
\end{equation}
The inverse of this reparametrization is
\begin{equation*}
  R=\frac{\tilde R}{2},\quad
  \ee^{L}=\frac{(1+\tilde Q)^2 \ee^{\tilde L}+\ee^{-\tilde L}}{2},\quad
  Q=\frac{(1-\tilde Q^2)\ee^{\tilde L}-\ee^{-\tilde L}}
  {(1+\tilde Q)^2 \ee^{\tilde L}+\ee^{-\tilde L}}.
\end{equation*}

From this and the discussion in \Sectionref{sec:boundarybehavior}, we
can easily derive the behavior of the functions $\tilde R$, $\tilde
L$, $\tilde Q$ at the poles for $t\in(0,\pi)$ in areal coordinates,
\begin{eqnarray}
  \label{eq:BoundQlambda}
  \tilde R&=\hat R\sin\theta,\quad  \ee^{\tilde L}=\ee^{\hat L}\cot\frac{\theta}{2},\quad
  \tilde Q=(1-\cos\theta)\hat Q,\\
  \ee^M&=\frac{\hat R}{4}\left[\ee^{\hat L}(1-\cos\theta)
    +\ee^{-\hat L}(1+\cos\theta)\right]+\hat M\,\sin^2\!\theta,\nonumber
\end{eqnarray}
with smooth functions $\hat R$, $\hat L$, $\hat Q$ and $\hat M$ which
only depend on $t$ and on $\cos\theta$.

A particular consequence is the following interesting fact about \keyword{polarized
  Gowdy spacetimes}. A Gowdy spacetime is called polarized if there
exists an everywhere orthogonal basis of Gowdy Killing fields. With
respect to this basis, the function $Q$ must hence vanish
identically. Now, Eq.~\eref{eq:QBound} shows that this can never
happen for the Killing basis
$\{\partial_{{\rho_1}},\partial_{{\rho_2}}\}$, but it is possible for
the basis $\{\partial_{\lambda_1},\partial_{\lambda_2}\}$ according to
Eq.~\eref{eq:BoundQlambda}. Indeed, one can show that $Q$ can only
vanish identically for a smooth Gowdy symmetric metric on \Sth if the
Killing basis is chosen such that one of the two fields is
proportional to $\partial_{\lambda_1}$ and the other to
$\partial_{\lambda_2}$. This fact will be used in our later discussion.

\section{The class of smooth Gowdy symmetric generalized 
  Taub-NUT solutions}
\label{sec:GenTaubNut}

\subsection{The Taub solutions}
\label{sec:Taubsolution}
The \keyword{Taub solutions} were discovered by Taub \cite{Taub:1951vk} as a family
of cosmological solutions of the vacuum field equation with spatial
\Sth-topology. They are a family of spacetimes
\begin{equation*}
\fl\qquad
  g=l^2\left(-\frac{4(1+\tau^2)}{V(\tau)}\dd\tau^2 +(1+\tau^2)(\omega_1\otimes\omega_1+\omega_2\otimes\omega_2)+\frac{V(\tau)}{1+\tau^2}\omega_3\otimes\omega_3\right),
\end{equation*}
with two free parameters $l>0$ and $m\in\R$, and where
\[V(\tau):=-4\tau^2 -8\frac{m}{l} \tau +4.\] 
Here,
\begin{eqnarray*}
  \omega_1 &=& \sin\rho_1 \dd\theta-\cos\rho_1\sin\theta \dd\rho_2,\\
  \omega_2 &=& \cos\rho_1 \dd\theta+\sin\rho_1\sin\theta \dd\rho_2,\\
  \omega_3 &=& \dd\rho_1+\cos\theta \dd\rho_2,
\end{eqnarray*}
are the standard invariant one-forms with respect to the  standard transitive action of \SU on \Sth. In particular it follows that \SU is a $3$-dimensional subgroup of the isometry group -- hence the Taub solutions are spatially homogeneous. The full isometry group is $4$-dimensional where the fourth symmetry is generated by the smooth right-invariant vector field $\partial_{\rho_2}$; the details can be found e.g.\ in \cite{Beyer:2007vp,Beyer:2008gg}. The total symmetry group is the direct product $U(1)\times\SU$ and therefore these spacetimes can be classified as LRS-Bianchi IX \cite{Wainwright:2005wss}. Taking the $U(1)$ subgroup of $\SU$ generated by $\partial_{\rho_1}$, it follows that Taub solutions are particular Gowdy solutions. They are not polarized, and we can bring them to the form \Eqref{eq:generalGowdy}
with \eref{eq:arealgauge}
\begin{equation*}
  g=\ee^M(-\dd t^2+\dd\theta^2)
  +R\left[\ee^{L} (\dd{\rho_1}+{Q} \dd{\rho_2})^2
    +\ee^{-L} \dd{\rho_2}^2\right].
\end{equation*}
For arbitrary parameters $l>0$ and
$m\in\R$, the Taub solutions are then given by
\begin{eqnarray*}
  R&=2l\sqrt{l^2+m^2}\sin t\sin\theta, \quad &\ee^M=l^2+(m+\sqrt{l^2+m^2}\cos t)^2,\\
  \ee^{L}&=\frac{R}{\ee^M\sin^2\!\theta},\quad
  &Q=\cos\theta.
\end{eqnarray*}

The Taub metric  is smooth and globally hyperbolic where $V$ is positive, i.e.\ for
\[\tau\in (\tau_-,\tau_+),\quad \tau_\pm:=-m/l\pm\sqrt{1+m^2/l^2}.\]
It was demonstrated for the first time in \cite{Newman:1963up} that the
solutions can be extended analytically through the apparently
singular times $\tau_\pm$. 
By this we mean the following \cite{Ringstrom:2009cj}. A spacetime $(M,g)$ is called \keyword{extendible} (in which case we say that ``it can be extended'') if there exists a spacetime $(\tilde M,\tilde g)$ of the same dimension and an isometric embedding $\Psi:M\rightarrow\tilde M$ which is not surjective. If $(M,g)$ and $(\tilde M,\tilde g)$ are analytic (smooth) manifolds and $\Psi$ is an analytic (smooth) map, then we say that the extension is analytic (smooth). 

In the case of the Taub spacetimes, we can find $(\tilde M,\tilde g)$ and $\Psi$ by appropriate coordinate transformations. We discuss more of these issues in  \Sectionref{sec:GenTaubNUT}. The extensions are not globally
hyperbolic and the surface corresponding to $\tau=\tau_-$ in the extended spacetime is a smooth null hypersurface with a closed null generator; in particular this implies that there exist closed causal curves. This surface is therefore a past Cauchy horizon. In the same way, there is a future Cauchy horizon at $\tau=\tau_+$. It has turned out that there are several non-equivalent analytic extensions of the Taub solutions. All these extended spacetimes were christened
\keyword{Taub-NUT solutions}.

\subsection{Generalizations of the
  Taub-NUT solutions}
\label{sec:GenTaubNUT}

Motivated by these intriguing properties of the Taub solutions above, Moncrief \cite{Moncrief:1984js} introduces the family of \keyword{generalized Taub-NUT solutions}. The idea is to obtain a family of solutions of Einstein's vacuum equations -- which is \textit{not} restricted to spatial homogeneity -- with similar properties as for the Taub solutions; in particular there should exist smooth compact Cauchy horizons.  It was shown in
\cite{Moncrief:1983ir} that if the spacetime is an analytic solution of the
vacuum field equations and if the Cauchy horizon is ruled by closed
null generators in the sense of an $\So$-bundle, in particular the null
generator coincides with the generators of the bundle, then the
spacetime necessarily has a $1$-dimensional isometry group and the
corresponding Killing field is proportional to the null generators of
the Cauchy horizon on the horizon. The result was generalized to the
case of non-analytic (i.e.\ smooth) solutions in \cite{Friedrich:1999de}. 

Motivated by all this, Moncrief restricts to the class of analytic vacuum solutions $(M,g)$ with $M=(0,\delta]\times \Sth$ for some sufficiently small $\delta>0$ with one spatial Killing vector field with the following properties. The spacetime is globally hyperbolic and the level sets of a global time function $t$ are Cauchy
surfaces homeomorphic to \Sth. The spacetime can be extended analytically through $t=0$ (in the sense described before) so that the points corresponding to $t=0$ in the extended spacetime form an analytic
null hypersurface with $\Sth$-topology. In particular, this null hypersurface is then a compact Cauchy horizon and hence global hyperbolicity breaks down at $t=0$. Additionally, he assumes that the Cauchy horizon is a Hopf bundle whose $U(1)$-generator is a Killing field. 

Let us study the consequences of these assumptions. Introducing Euler coordinates as before, Moncrief shows that the metrics of all such spacetimes can be written as
\begin{equation}
  \label{eq:moncriefmetric}
\fl\qquad g=\ee^{-2\gamma}(-\tilde N^2 \dd t^2+\tilde g_{\alpha\beta}\dd x^\alpha \dd x^\beta) +\sin^2\! t\,
\ee^{2\gamma}[k(\dd{\rho_1}+\cos\theta\, \dd{\rho_2})+\beta_\alpha \dd x^\alpha]^2,
\end{equation} 
for all $t\in(0,\delta]$, where the $U(1)$-Killing field is $\partial_{{\rho_1}}$.  
The functions
$\gamma$ and $\tilde N$ only depend on $t$, $\theta$ and
${\rho_2}$. The index $\alpha$ takes the values $1$ (corresponding to the
coordinate $\theta$) and $3$ (corresponding to the coordinate
$\rho_2$).  The field $\tilde g_{\alpha\beta}$ is a symmetric $2$-tensor field
and $\beta_\alpha$ a $1$-form. The function $\tilde N$ is supposed to be uniformly positive. Moreover, $k>0$ is a constant. Moncrief assumes that all fields $\gamma$, $\tilde N$, $\tilde g_{\alpha\beta}$ and
$\beta_\alpha$ are analytic on $(0,\delta]\times\Sth$ for some small
$\delta>0$.

However, these assumptions are not yet sufficient to guarantee that the spacetime can be extended through $t=0$. 
In order to write down the conditions for this, we here restrict to the case of interest, namely
to the case of Gowdy symmetry; the details for the general $U(1)$-symmetric case can be found in Moncrief's paper. For Gowdy symmetry, the metric coefficients above must be independent of both $\rho_1$ and
${\rho_2}$.  Let us
define
the function $N$ by
\begin{equation}
  \label{eq:ansatzforN}
  \ee^L=\frac{R\ee^{-M}}{\sin^2\!\theta} N^2.
\end{equation}
In order to identify Moncrief's metric with the metric given by \Eqsref{eq:generalGowdy}--\eref{eq:arealgauge} for
$t\in(0,\delta]$ (and $\delta\in (0,\pi)$),
we set
\begin{eqnarray*}
  \fl\qquad
    \tilde g_{\theta\theta}&=\ee^M,\quad \tilde g_{\theta{\rho_1}}=0,\quad 
    \tilde g_{{\rho_1}{\rho_1}}=\frac{\ee^M\sin^2\!\theta}{N^2},\quad
    \beta_\theta&=0,\quad \beta_{\phi}=R_0(Q-\cos\theta),\\
\fl\qquad
    \tilde N&=\frac{R_0 N}{k},\quad \gamma=-\frac{M}{2}+\ln\frac{R_0 N}{k}.
\end{eqnarray*}
Now, in order to find the extensions through $t=0$, let us introduce new coordinates
$(t',\theta',{\rho_1}',{\rho_2}')$ by
\begin{equation}
  \label{eq:extensionthroughCH}
  t=\arcsin\sqrt{t'},\quad \theta=\theta',\quad {\rho_1}={\rho_1}'+\frac
  \kappa {R_0} \ln t',\quad {\rho_2}={\rho_2}'.
\end{equation}
The quantity $\kappa$ is a constant which has, so far, not been
fixed. In these new coordinates, the metric becomes
\begin{eqnarray*}
  \fl\quad
  g=-\left(\frac{\ee^M}{4(1-t')t'}-\frac{\ee^{-M}N^2\kappa^2}{t'}\right)
  {\dd t'}^2
  +\ee^M \dd\theta^2\\ \fl
  \qquad\quad+\ee^{-M}N^2\left[
    2R_0\kappa(\dd{\rho_1}'+Q \dd{\rho_2}')\dd t'
    +R_0^2 t'\left(\dd{\rho_1}'+Q d{\rho_2}'\right)^2\right]
  +\frac{\ee^M\sin^2\!\theta}{N^2}{\dd{\rho_2}'}^2.
\end{eqnarray*}
The metric extends analytically through $t'=0$ if all the functions $M$, $N^2$, $Q$ and $(4\kappa^2 N^2-\ee^{2M})/t'$ -- for some choice of the constant $\kappa$ -- extend analytically through $t'=0$ when expressed in terms of the new coordinates, and if $N^2$ extends as a strictly positive function. Note that it is necessary for this that $M$, $N^2$, $Q$ -- expressed in terms of the original coordinates -- extend as analytic functions to the manifold $[-\delta,\delta]\times\Sth$, which are even in $t$, for some sufficiently small $\delta>0$.

If the spacetime can be extended through $t=0$ in this way, then the field
$\partial_{\rho_1'}$ (which equals $\partial_{\rho_1}$ wherever the latter is defined) is a null\footnote{We have $g_{\rho_1'\rho_1'}=R_0^2 t' e^{-M} N^2$ which vanishes at $t'=0$. } generator of the
surface given by $t'=0$. This surface is therefore an analytic null
hypersurface with \Sth-topology whose null generators are closed, and so is an
analytic past Cauchy horizon.

As an example, the Taub solutions satisfy all the above assumptions when we choose
\[\kappa=\pm \left(l^2 + m (m + \sqrt{l^2 + m^2})\right).\]

As we have mentioned above, Moncrief restricts to the analytic case. One of our main contributions here is a generalization of his results to the non-analytic (i.e.\ smooth) case; this means that the reader must replace every occurrence of ``analytic'' in the previous discussion by ``smooth''. Due to our restriction to Gowdy symmetry here, we call spacetimes with all the above properties \keyword{smooth
  Gowdy symmetric generalized Taub-NUT spacetimes}. The name is supposed to reflect the fact that this family of spacetimes is motivated  by Moncrief's
\keyword{generalized Taub-NUT spacetimes}.

Note that if an analytic spacetime as above solves
Einstein's field equations in vacuum for $t>0$, then the analytically extended spacetimes are
necessarily also solutions of the vacuum field equations. In the
non-analytic smooth case, we do not know in general whether the
extensions are vacuum solutions. We will not address this problem in
this paper. As another interesting side-remark: Chru\'sciel et al.\ note in
\cite{Chrusciel:1999dk} that there are no smooth extensions through a
Cauchy horizon of $\Sth$-topology -- solution of the field equations or
not -- in the \textit{polarized} Gowdy case. We find here that
none of the spacetimes which we consider are
polarized, and hence there is no contradiction.

\subsection{Existence of smooth Gowdy symmetric generalized 
  Taub-NUT solutions}
\label{sec:existence}
\subsubsection{The main existence result.}
In this section we show the existence of a non-trivial family of smooth Gowdy
symmetric generalized Taub-NUT spacetimes, which solve Einstein's
vacuum field equations. A central technique here is the
Fuchsian method introduced in
\cite{Beyer:2010fo,Beyer:2010tb,Beyer:2011ce,Ames:yV5l9m6A,Ames:uh}, which we must reformulate for the particular spatial topology used here. This is done in the appendix (\Sectionref{sec:backgroundFuchsian}).

In the following we call a function on $\St$ \keyword{rotationally symmetric}
 if it does not depend on the azimuthal angle $\phi$ in
standard spherical coordinates \Eqref{eq:sphericalcoordinates}. The
Hopf map allows to lift any such function to a smooth $U(1)\times
U(1)$-invariant function on $\Sth$
.

\begin{theorem}
  \label{th:existence}
  Let $S_{**}$ and $Q_{*}$ be rotationally symmetric functions
  in $C^\infty(\St)$ with the property
  \begin{equation}
    \label{eq:dataconditions}
    S_{**}(0)=S_{**}(\pi),
  \end{equation}
  and $R_0>0$ a constant.  Then there exists a unique smooth
  Gowdy symmetric generalized Taub-NUT solution of the form \Eqsref{eq:generalGowdy}--\eref{eq:arealgauge} for all $t\in(0,\pi)$ which
  has the following uniform expansions at $t=0$:
  \begin{eqnarray*}
    R(t,\theta)\ee^{L(t,\theta)}&=t^2 \ee^{S_{**}(\theta)}+O(t^4),\\
    Q(t,\theta)&=\cos\theta+Q_*(\theta)\sin^2\theta 
    +O(t^2),\\
    M(t,\theta)&=S_{**}(\theta)-2S_{**}(0)+2\ln R_0+O(t^2).
  \end{eqnarray*}
  Corresponding expansions
  hold for all derivatives.
\end{theorem}

Let us make a couple of comments before we proceed with the proof of this theorem in \Sectionref{sec:equn}.
\Theoremref{th:existence} implies that we can prescribe arbitrary smooth \keyword{asymptotic data} functions $S_{**}$ and $Q_{*}$ subject to
  the condition \eref{eq:dataconditions} and find a unique smooth Gowdy symmetric generalized Taub-NUT solution of the vacuum equations so that the \keyword{leading-order} behavior at $t=0$ is determined by these data functions. In this sense, the functions $S_{**}$ and $Q_{*}$ can be considered as ``data on the Cauchy horizon'' at $t=0$. Hence we solve here a \textit{singular initial value problem} with leading-order terms as above; this is discussed in greater depth later. In particular, we find the same number of free functions as in Moncrief's class of generalized Taub-NUT solution (after factoring out gauge transformations in his
  class \cite{Moncrief:1984js}). 

We stress that \Theoremref{th:existence}  implies the existence only of a \textit{past} Cauchy horizon at $t=0$, but says nothing about the properties of the solution at $t=\pi$, where the areal coordinates break down; \Sectionref{sec:linearproblem} is devoted to the question of what happens at $t=\pi$ for the solutions of \Theoremref{th:existence}. 

Particular examples are the Taub solutions (\Sectionref{sec:Taubsolution}), which  correspond to the asymptotic data
  \begin{equation*}
\fl\qquad
    R_0=2l\sqrt{l^2+m^2},\quad
    S_{**}=2\ln R_0-\ln\left(l^2+(m+\sqrt{l^2+m^2})^2\right),\quad
    Q_*=0.
  \end{equation*}
We can therefore interpret \Theoremref{th:existence} as a statement about the ``stability'' of the Cauchy horizon of the Taub solutions at $t=0$ with respect to smooth Gowdy symmetric perturbations.

It is interesting to compare our theorem with earlier results by various authors. For general spacetimes which are ``singular'' in some sense at $t=0$, one expects that the full degree of freedom corresponds to four free data functions. For the case of Gowdy symmetry with spatial $\mathbb{T}^3$-topology \cite{Kichenassamy:1999kg,Rendall:2000ki,Beyer:2010tb}, one can indeed show the well-posedness of a singular initial value problem with the full number of free functions. One obtains a large variety of solutions: on the one hand solutions whose curvature blows up at $t=0$, and, on the other hand, solutions with Cauchy horizons at $t=0$, which can be both compact or non-compact. An earlier attempt to obtain a similarly general result in the case of  spatial $\Sth$-topology (and $\SoXSt$-topology) \cite{Stahl:2002bv} has partly failed: the author successfully constructs a general family of singular solutions of \textit{some} of Einstein's vacuum equations, but the remaining ``constraints'', which require 
certain matching conditions to hold, see below, are ignored. St{\aa}hl's statement ``In what follows we will assume without further comment that the solutions are chosen such that the matching conditions hold.''\ on page 4489 in \cite{Stahl:2002bv} is vacuous: while it is clear how to choose \textit{Cauchy} data so that the matching conditions are satisfied \cite{Chrusciel:1990ti}, this is in general not clear for asymptotic data of the \textit{singular} initial value problem, which we here -- and St{\aa}hl -- consider. Indeed, these matching conditions turn out to be a major difficulty in our proof of \Theoremref{th:existence}, see \Propref{prop:solM}. There are reasons to believe that the matching conditions can never be satisfied for this singular initial value problem in more general situations, in particular if a neighborhood of the axes of the $t=0$-surface is supposed to represent a curvature singularity. This case is, however, not covered in this paper.

It is interesting to note that for the Gowdy case with spatial
  $\mathbb{T}^3$-topology, the asymptotic data have to satisfy an integral constraint. Such a constraint must be imposed
  in order to guarantee that the function $M$ in the metric, as a solution of the
  field equations, is consistent with the periodic topology. Here, it is rather the smoothness condition \Eqref{eq:boundarycondM2} for $M$
  at the ``poles'' of the three-sphere which gives rise to the non-integral constraint
  \Eqref{eq:dataconditions}.

We also point out that that none of the solutions of
  \Theoremref{th:existence} is polarized. Recall
  from \Sectionref{sec:reparametrization} that a smooth Gowdy
  symmetric spacetime is polarized if and only if $Q$ vanishes with respect to the
  $(\lambda_1,\lambda_2)$-parametrization of the symmetry
  orbits and hence if $1-(1-Q^2)\ee^{2L}\equiv 0$ with respect to the
  $(\rho_1,\rho_2)$-parametrization according to \Eqref{eq:transformationQ}. However, for
  our solutions, $Q$ is bounded in a neighborhood of $t=0$ and
  $\ee^{2L}$ is $O(t^4)$; from this we find that $1-(1-Q^2)\ee^{2L}=1+O(t^4)$.

As a last comment, let us point out that \Theoremref{th:existence} can be generalized to asymptotic data with only finitely many derivatives. We, however, do not discuss this here.

\subsubsection{Equations and unknowns.}
\label{sec:equn}
For our proof of \Theoremref{th:existence}, let us
make the following convenient choices. We consider Geroch's reduction with respect to the field $\xi=\partial_{\rho_1}$ and hence with the projection map $\pi$ of the form \Eqref{eq:hopfspherical} as discussed before. In areal coordinates, the $3$-metric $h_{ab}$ on the quotient manifold $(0,\delta]\times\St$ is therefore given by \Eqsref{eq:ansatzh}--\eref{eq:gggaugechoice}. When we define
\begin{equation}
  \label{eq:defS}
  S:=\ln\lambda=L+\ln R,
\end{equation}
which we expect to be a smooth function wherever the spacetime is defined in $(0,\delta]\times\St$,
it becomes
\begin{equation}
  \label{eq:hfinal}
  h=\ee^{M}(-\dd t^2+\dd\vartheta^2)+R_0^2\,\sin^2\!t\,\sin^2\!\vartheta\, \ee^{-S}\dd\phi^2.
\end{equation}
As outlined before, the geometry of the Gowdy spacetimes is completely determined by the quantities $S$, $\omega$ and $h$ on the  quotient manifold $(0,\delta]\times\St$.
Then,
\Eqsref{eq:Gerochevollambda}--\eref{eq:Gerochevolomega}, as equations on the quotient manifold, imply\footnote{Derivatives of functions along coordinate vector fields are written either as e.g.\ $\partial_\vartheta\omega$ or as e.g.\ $\omega_{,\vartheta}$ in all of what follows.}
\begin{eqnarray}\fl
  \label{eq:FuchsS}
  \quad D^2S-t^2\Delta_{\St} S&=&(1-t\cot t) DS-\ee^{-2S}\left((D\omega)^2
  -(t\partial_\vartheta\omega)^2\right),\\ 
  \label{eq:Fuchsomega}
  \fl
  \quad D^2\omega-4D\omega-t^2\Delta_{\St} \omega    
  &=&(1-t\cot t)D\omega+2(DS-2) D\omega
  -2 (t\partial_\vartheta S)(t\partial_\vartheta\omega).    
\end{eqnarray}
 We use the notation $D:=t\partial_t$ and
$D^2=t\partial_t(t\partial_t)$.
Note that we have added a term $-4D\omega$ to both sides of the second equation for
later convenience. The operator $\Delta_{\St}$ is the
Laplace operator of the standard metric on the unit sphere
\[\Delta_{\St}=\partial_\vartheta^2+\cot\vartheta\partial_\vartheta
+\frac 1{\sin^2\!\vartheta}\partial^2_\phi.
\] 
In our case, we restrict to solutions which are independent of the azimuthal angle
$\phi$ and hence the $\partial_\phi^2$-term in this operator does not appear. In
this case, all terms in the equations above have a geometric
coordinate independent meaning with respect to the scalar product of the standard Riemannian metric on $\St$
\[\tilde h=\dd\vartheta^2+\sin^2\vartheta\, \dd\phi^2.\]
\Eqsref{eq:FuchsS}--\eref{eq:Fuchsomega} therefore constitute a coupled semilinear system of geometric wave equations for the scalar quantities $S$ and $\omega$ with respect to the standard
metric on the unit sphere as long as $t\in (0,\pi)$. 

The remaining equations are found from the $2+1$-dimensional Einstein equations coupled to the wave map given by $S$ and $\omega$ \Eqref{eq:GerochRicci3}. On the one hand, the above form of the $3$-metric $h$ implies
\begin{equation}
  \label{eq:waveequationM}
  0=-M_{,tt}+\Delta_{\St} M+\cot t (M_{,t}+2S_{,t})-S_{,t}^2-\ee^{-2S}\omega_{,\vartheta}^2+2=:H,
\end{equation}
and on the other hand, 
\begin{equation}
  \label{eq:constraints}
  0=4R_{,\pm\vartheta}+R (S_{,\pm}^2+\ee^{-2S}\omega_{,\pm}^2)-2R_{,\pm}(S_{,\pm}+M_{,\pm})=:2R_{,\pm}C_{\pm},
\end{equation}
with $\partial_\pm:=\partial_\vartheta\pm\partial_t$. While for the first of these equations all terms have a geometric coordinate independent meaning as before, this is not the case for \Eqsref{eq:constraints}. This is not a major problem, as we will be able to analyze \Eqsref{eq:constraints} completely in the particular coordinate system.

\subsubsection{Steps of the proof of \Theoremref{th:existence}.}
\label{sec:analyseequations}

The strategy for the proof of \Theoremref{th:existence} is to solve only \Eqsref{eq:FuchsS}--\eref{eq:Fuchsomega} in a first step; we refer to the latter as the \keyword{main evolution system} or the \keyword{Gowdy equations}. Given any such solution $(S,\omega)$ of that system with the ``correct'' behavior at $t=0$, the remaining equations, \Eqsref{eq:waveequationM}--\eref{eq:constraints}, form an overdetermined system for the other unknown $M$. We must therefore ask for the conditions for which these can be solved for $M$. As already discussed in \cite{Chrusciel:1990ti,Garfinkle:1999ix,Stahl:2002bv}, one encounters certain ``matching conditions'' which must be satisfied in order to obtain smooth solutions for $M$. The quantities $H$, $C_+$ and $C_-$ introduced above are of particular importance since they measure the violation of \Eqsref{eq:waveequationM}--\eref{eq:constraints}.

The first step is the construction of such solutions of \Eqsref{eq:FuchsS}--\eref{eq:Fuchsomega} in a small time neighborhood of $t=0$, which are compatible with the notion of smooth Gowdy symmetric generalized Taub-NUT solutions given in \Sectionref{sec:GenTaubNUT}. We find the following central result.

\begin{proposition}
  \label{prop:basicexistence}
  Let $\omega_*\in\R$, and $\omega_{**},S_{**}\in C^\infty(\St)$ be
  rotationally symmetric functions. Choose an exponent vector
  $\mu=(\mu_1,\mu_2)$ with $1<\mu_1(x)<2$ and
  $(4+\sqrt{17})/2<\mu_2(x)<4+\mu_1(x)$ for all $x\in\St$. Then there exists a unique solution
  $(S,\omega)$ of the Gowdy equations \eref{eq:FuchsS}--\eref{eq:Fuchsomega} with
  \begin{eqnarray}
    \label{eq:Sleadingorder}
    S(t,\vartheta)&=2\ln t+S_{**}(\vartheta)+w_1(t,\vartheta),\\
    \label{eq:omegaleadingorder}
    \omega(t,\vartheta)&=\omega_*+\omega_{**}(\vartheta)t^4+w_2(t,\vartheta),
  \end{eqnarray}
  where $w:=(w_1,w_2)\in \tilde X_{\delta,\mu,\infty}(\St)$ and $D^2w\in X_{\delta,\mu,\infty}(\St)$ for a
  sufficiently small $\delta>0$. In particular, the functions $w_1,w_2$ are
  rotationally symmetric for each $t\in (0,\delta]$.
\end{proposition}

 Before we prove this proposition let us make the following remarks. The meaning of this proposition can be summarized as follows (the technical details of the spaces $X_{\delta,\mu,\infty}(\St)$ and $\tilde X_{\delta,\mu,\infty}(\St)$ are listed in the appendix). Suppose that $\omega_{**}$ and $S_{**}$ are
smooth rotationally symmetric asymptotic data functions on $\St$, and $\omega_*$ is a constant. Then
there exists a smooth solution of \Eqsref{eq:FuchsS}--\eref{eq:Fuchsomega} of the form \Eqsref{eq:Sleadingorder}--\eref{eq:omegaleadingorder} on the time interval $(0,\delta]$ with $w_1\in \tilde X_{\delta,\mu_1,\infty}(\St)$ and $w_2\in \tilde X_{\delta,\mu_2,\infty}(\St)$.
The fact that the \textit{remainder} $w=(w_1,w_2)$ is in such a space implies that it, together with all of its
spatial derivatives, decays at a rate $t^{\mu_1}$ and $t^{\mu_2}$, respectively, uniformly at each spatial point in the limit $t\searrow 0$. Moreover, the second time derivative $D^2w$, and in the same way all higher time derivatives of the remainder, decay with the same rate at $t=0$. Notice here that $(4+\sqrt{17})/2\approx 4.06$. All this shows that \Eqsref{eq:Sleadingorder}--\eref{eq:omegaleadingorder} can be interpreted as describing the leading-order behavior, which we may prescribe by means of the asymptotic data $\omega_*$, $\omega_{**}$ and $S_{**}$.
We stress that this \textit{singular} initial value problem differs significantly from a standard \textit{Cauchy} initial value
  problem. The equations are singular at $t=0$ (of semilinear Fuchsian wave equation
  type, see the appendix). A consequence is the fourth power of $t$ in the leading-order term of
  $\omega$ and the logarithm in the expansion of $S$ on the one hand,
  and, on the other hand, we do not obtain four free data functions. Notice in particular that $\omega_*$ must be a constant. All this is justified below.  

 As mentioned before, the quantities $\mu_1$ and $\mu_2$ control the decay of the
  remainders $w_1$ and $w_2$ at $t=0$. We 
  argue below that  the bounds for these constants in \Propref{prop:basicexistence} are not yet optimal; in fact we find below that $\mu_1\le 2$ and $\mu_2\le 6$ are upper bounds.
Moreover, the uniqueness result, as it is given by the lower bounds, is not yet optimal. It guarantees uniqueness only among remainders in the spaces $X_{\delta,\mu,\infty}$ with $\mu_1>1$ and $\mu_2>(4+\sqrt{17})/2$. The general class of remainders which is of interest here, however, is given by the larger space $X_{\delta,\mu,\infty}$ with $\mu_1>0$ and $\mu_2>4$. At this stage, we cannot rule out possible further solutions in this larger space. However, by computing expansions of higher order below, we can show that the solutions of \Propref{prop:basicexistence} are indeed unique in the full space.

The form of the leading-order term \Eqsref{eq:Sleadingorder}--\eref{eq:omegaleadingorder} has been chosen to be compatible with the notion of smooth Gowdy symmetric generalized Taub-NUT solutions. For example,  the term $2\ln t$ in \Eqref{eq:Sleadingorder} guarantees that the coefficient of ${\dd\rho_1}^2$ in the metric is $O(t^2)$ at $t=0$ in agreement with \Eqref{eq:moncriefmetric}.

We see in the proof of \Propref{prop:basicexistence} that the same result can be obtained under much less stringent regularity assumptions. Indeed, we only need to require that the asymptotic data has a certain finite number of derivatives. This will, however, not be discussed in any more detail.

\begin{proof}[Proof of \Propref{prop:basicexistence}]
  The proof is an application of \Theoremref{th:Wellposedness1stOrderFiniteDiff} in the appendix. First, we realize that our equations are a system of the form \Eqref{eq:1stordersystem} with $u=(S,\omega)^T$, and $d=2$. The coefficient matrix $A$ is
\[A=\left(\begin{array}{cc}
         0 & 0\\
         0 & -2
       \end{array}\right).
\] 
From this, we compute the energy dissipation matrix \Eqref{eq:energydissipationmatrix} which is positive definite -- and hence \Conditionref{en:cond2} of \Theoremref{th:Wellposedness1stOrderFiniteDiff} is satisfied -- if $\mu_1>1$ and $\mu_2>(4+\sqrt{17})/2$.
The leading-order term is
\begin{equation}
  \label{eq:leadingorderterm}
  u_0=(2\ln t+S_{**},\omega_*+\omega_{**}t^4)^T,
\end{equation}
so that
\[L[u_0]=(-t^2\Delta_{\St}S_{**},-t^6\Delta_{\St}\omega_{**})^T,\]
where the operator $L$ is given by \Eqref{eq:DefLPDE}. The expressions for the operators $F(u_0)$ and $F_{red}(u_0)$, see \Eqsref{eq:defF} and \eref{eq:defFLu}, can then be found from the right-hand sides of \Eqsref{eq:FuchsS}--\eref{eq:Fuchsomega}. Suppose that $q>1$ and $0<\mu_1<2$ and $4<\mu_2<4+\mu_1$ (the lower bound $4$ is sufficient here and we do not need to require the slightly larger value $(4+\sqrt{17})/2$ for this part of the argument). Then one can find easily that \Conditionref{en:cond4N} of \Theoremref{th:Wellposedness1stOrderFiniteDiff} is satisfied. Notice that this would not be true if $\omega_*$ was not a constant due to terms in the equation that are proportional to $\partial_\vartheta\omega_*$ which would be too singular at $t=0$. The remaining \Conditionref{cond:LipschitzF} follows automatically; one can argue in the same way as in the proof of Lemma~3.4 in \cite{Ames:yV5l9m6A}.

In order to prove that the solutions, which we have just found, are rotationally symmetric at every $t$, we can take a $\phi$-derivative of both sides of \Eqsref{eq:FuchsS}--\eref{eq:Fuchsomega}. One obtains a system of evolution equations for the unknowns $\partial_\phi S$ and $\partial_\phi\omega$, which has exactly the same form as the original system. We can hence solve the same singular initial value problem, but now with a vanishing leading-order term. Then, uniqueness implies that $\partial_\phi S(t,x)=0$ and $\partial_\phi\omega(t,x)=0$ for all $(t,x)\in (0,\delta]\times\St$.
\end{proof}

Let us gather further information about the solutions $(S,\omega)$ of \Propref{prop:basicexistence}.
\begin{lemma}
  \label{lem:extension}
  Let $(S,\omega)$ be the functions $(0,\delta]\times\St\rightarrow\R$ determined in \Propref{prop:basicexistence} as solutions of \Eqsref{eq:FuchsS}--\eref{eq:Fuchsomega} from smooth rotationally symmetric asymptotic data functions $\omega_{**}$ and $S_{**}$, and the constant $\omega_*$. Let $\lambda:=\ee^S$ (cf.\ \Eqref{eq:defS}). Then $\lambda$ and $\omega$ are solutions of
\begin{eqnarray}\fl
  \label{eq:Fuchslambda}
  D^2\lambda-t^2\Delta_{\St} \lambda&=&(1-t\cot t) D\lambda-\lambda^{-1} (t\partial_\vartheta\lambda)^2-\lambda^{-1}\left((D\omega)^2
    -(t\partial_\vartheta\omega)^2\right),\\ 
  \label{eq:Fuchsomegalambda}
  \fl
  D^2\omega-4D\omega-t^2\Delta_{\St} \omega    
  &=&(1-t\cot t)D\omega+2\lambda^{-1}(D\lambda-2\lambda) D\omega
  -2\lambda^{-1} (t\partial_\vartheta \lambda)(t\partial_\vartheta\omega),    
\end{eqnarray}
derived from \Eqsref{eq:FuchsS}--\eref{eq:Fuchsomega},
on $(0,\delta]\times\St$. The expansions of $\lambda$ and $\omega$ at $t=0$ are
\begin{eqnarray*}
  \lambda(t,\vartheta)&=t^2 \ee^{S_{**}(\vartheta)}+O(t^4),\\
  \omega(t,\vartheta)&=\omega_*+\omega_{**}(\vartheta)t^4+O(t^6),
\end{eqnarray*}
and all higher terms are proportional to even powers of $t$. Thus, $\lambda$ and $\omega$ can be extended to functions in $C^\infty([-\delta,\delta]\times\St)$ which are even in $t$: $\lambda(t,x)=\lambda(-t,x)$, $\omega(t,x)=\omega(-t,x)$. These extended functions satisfy \Eqsref{eq:Fuchslambda}--\eref{eq:Fuchsomegalambda} on $[-\delta,\delta]\times\St$ in the sense of uniform limits at $t=0$.
\end{lemma}

\begin{proof}
  The main technical tool here is the theory of \keyword{(order n)-leading order terms}, see \cite{Ames:yV5l9m6A} which can be  generalized directly to our case. This is an algorithm to compute leading-order expansions of arbitrarily high order for solutions with a given leading-order term (here the leading-order term is of the form \Eqref{eq:leadingorderterm}). When we use this method to compute expansions of $(S,\omega)$ of any order in $t$ at $t\searrow 0$, we find that only the first term in the expansion of $S$ is a log-term, and all other terms in the expansions of both $S$ and $\omega$ are positive even integer powers of $t$. Hence $(\lambda,\omega)$ can be extended as smooth, even functions as claimed above.  Moreover, it is easy to see that \Eqsref{eq:Fuchslambda}--\eref{eq:Fuchsomegalambda} for $(\lambda,\omega)$  are invariant under the transformation $t\mapsto -t$. This proves that the extended functions $(\lambda,\omega)$ satisfy  the equations on both intervals $t\in [-\delta,0)$ and $t\in (0,\delta]
$. 
Both limits $t\searrow 0$ and $t\nearrow 0$ of the equations, which involve formally singular terms at $t=0$, exist uniformly in space, and as a result, the equations are also satisfied at $t=0$.
\end{proof}

In the following, we  often consider functions, whose definition involves the quantities $S$ and $\omega$, and we attempt to extend those to the time interval $[-\delta,\delta]$. Then we understand that we first replace $S$ by $\ln\lambda$, where $\lambda$ is the extended function in \Lemref{lem:extension}. If  it is possible to find a smooth extension in this sense, then we say that the function is extendible to $[-\delta,\delta]$.

Given a solution of the main evolution equations as above, we now attempt to solve the remaining equations implied by Einstein's field equations thereby determining the remaining unknown $M$. We repeat that while there is a general result for existence of solutions of the main evolution equations also in \cite{Stahl:2002bv}, the problem of solving these remaining equation has not been considered there. We address this problem now and find the following.

\begin{proposition}
  \label{prop:solM}
  Let $R_0>0$ and $\omega_*$ be constants, and $\omega_{**}$ and $S_{**}$ be
  rotationally symmetric functions in $C^\infty(\St)$ satisfying
  \[S_{**}(0)=S_{**}(\pi).\] 
  Suppose that $(S,\omega)$ is the
  corresponding solution of \Eqsref{eq:FuchsS}--\eref{eq:Fuchsomega} according to \Propref{prop:basicexistence} on $(0,\delta]\times\St$, and $(\lambda,\omega)$ the smooth continuation to $[-\delta,\delta]\times\St$ according to \Lemref{lem:extension}.
  Then there is a unique
  smooth function $M\in C^\infty((0,\delta]\times\St)$, which is rotationally symmetric and satisfies the smoothness condition
  \Eqref{eq:boundarycondM2} at each
  $t\in (0,\delta]$, so that the $3$-metric $h$ given by \Eqref{eq:hfinal} is smooth and so that
the $2+1$-Einstein equations \Eqref{eq:GerochRicci3}, which are represented by \Eqsref{eq:waveequationM}--\eref{eq:constraints}, are satisfied
on $(0,\delta]\times\St$.
This function $M$ can be extended as a smooth
  function to $[-\delta,\delta]\times\St$ which is even in $t$ and satisfies the smoothness condition on the extended time interval. The expansion of $M$ at $t=0$ is
  \[M(t,\vartheta)=S_{**}(\vartheta)-2S_{**}(0)+2\ln R_0+O(t^2).\]
\end{proposition} 

We note that the smoothness condition \Eqref{eq:boundarycondM2}, which takes the form $\ee^M = (R_0^2 \sin^2 t)\, \ee^{-S}$ at $\vartheta=0,\pi$, is non-trivial in the limit $t\rightarrow 0$ due to the presence of the singular factor $\ee^{-S}$. However, the limit turns out to be well-defined since the singular behavior of $\ee^{-S}=O(t^{-2})$ is canceled exactly by the factor $\sin^2\! t$. 

The steps of the proof are as follows; the details will be discussed afterwards. We consider \Eqsref{eq:constraints} and first recall that those equations are not in a coordinate independent geometric form, and are, in particular, not well-defined at $\vartheta=0$, $\pi$. However, each equation of the system \Eqsref{eq:waveequationM}--\eref{eq:constraints} is a linear combination of the components of these $2+1$-dimensional Einstein equations \Eqref{eq:GerochRicci3}, which can be understood as the difference of the Einstein tensor of the $3$-metric $h$ and the energy-momentum tensor corresponding to the scalar fields $\lambda$ and $\omega$. Therefore, as long as the $3$-metric $h$ is a smooth metric on the quotient manifold $(0,\delta]\times\St$, and since the functions $\lambda$ and $\omega$ as solutions of the previous proposition are smooth, the $2+1$-dimensional Einstein equations are satisfied on $(0,\delta]\times\St$ if and only if they are satisfied on the dense subset $(0,\delta]\times\tilde\St$ with 
$\tilde\St:=\St\backslash\{\mathrm{north\ pole, south\ pole}\}$, that is, precisely where all terms in \Eqsref{eq:constraints} are well-defined. This insight allows us to treat \Eqsref{eq:waveequationM}--\eref{eq:constraints} as follows: for given smooth solutions $\lambda$ and $\omega$ above, we look for a function $M$ which (i) is smooth on $(0,\delta]\times\St$, (ii) satisfies the smoothness condition \Eqref{eq:boundarycondM2} at the poles $\vartheta=0$ and $\vartheta=\pi$, and (iii) satisfies \Eqsref{eq:constraints} on $(0,\delta]\times\tilde\St$. In a subsequent step we can then show that this implies that \Eqref{eq:waveequationM}, and hence the full set of $2+1$-Einstein equations is satisfied on $(0,\delta]\times\tilde\St$, and therefore on the full domain $(0,\delta]\times\St$. When this is achieved, we show that this function $M$ can be extended as a smooth function to $[-\delta,\delta]\times\St$ which is even in $t$ at $t=0$ and therefore satisfies the smoothness condition \Eqref{eq:boundarycondM2} on 
this extended time interval. In total, this gives a smooth $3$-metric $h$, which, on the one hand, satisfies the $2+1$-dimensional Einstein equations on $(0,\delta]\times\St$, and which, on the other hand, turns out to be compatible with the hypotheses of smooth Gowdy symmetric generalized Taub-NUT solutions.

We use the following technical trick. We consider the map 
\[\fl\qquad\Psi: \Tt\rightarrow\St: (\alpha,\beta)\mapsto (y_1,y_2,y_3)=(\sin\alpha\cos\beta,\sin\alpha\sin\beta,\cos\alpha)\in\St\subset\R^3,\]
where $\alpha$ and $\beta$ are $2\pi$-periodic coordinates on $\Tt$. This allows us to pull-back all smooth rotationally symmetric functions on $\St$ to functions on $\Tt$, which are (a) $2\pi$-periodic with respect to $\alpha$, (b) even in $\alpha$ at $\alpha=0$, and (c) independent of $\beta$. On the other hand, since the above implies that we only need to solve  \Eqsref{eq:constraints} on the spatial domain $\tilde\St$ where $\Psi^{-1}$ is well-defined, we can also push-forward the coordinate vector field $\partial_\vartheta$ to $\Tt$ and extend the result as a global smooth vector field which then coincides with $\partial_\alpha$. In summary, in order to satisfy conditions (i) and (iii) in the previous paragraph for the function $M$, we replace each function in \Eqsref{eq:constraints} by its composition with $\Psi$ (this can be understood as extending the variable $\vartheta$ in each function $2\pi$-periodically) and each $\vartheta$-derivative by an $\alpha$-derivative. Then we look for a solution of 
this version of \Eqsref{eq:constraints} for the unknown $M\circ\Psi$ which is even in $\alpha$ at $\alpha=0$ for every time $t\in (0,\delta]$. The corresponding function $M$, which at first is defined only on $(0,\delta]\times\tilde\St$, then extends as a smooth function to $(0,\delta]\times\St$ and is a solution of the equation \Eqsref{eq:constraints} on $(0,\delta]\times\tilde\St$. 

We follow this approach now; in order to simplify the notation, however, we write $S$, $\omega$, $\lambda$, $M$ and $R$ for the functions defined on $\Tt$ as opposed to $S\circ\Psi$, $\omega\circ\Psi$ etc., and we write $\partial_{\vartheta}$ instead of $\partial_\alpha$ for the vector field on $\Tt$, as long as there is no confusion with the corresponding quantities on $\St$. We then refer to any function $f(t,\theta)$ on $[-\delta,\delta]\times\Tt$, which does not depend on $\beta$, as $t$-even (or $\vartheta$-even) if $f(t,\theta)=f(-t,\theta)$ at every $\theta$ (or $f(t,\theta)=f(t,-\theta)$ at every $t$); similar we define the notion of $t$-odd and $\vartheta$-odd functions.



The most difficult step of the proof of \Propref{prop:solM} is summarized in the following lemma, which we prove first.

\begin{lemma}
  \label{lem:solMmatching}
  Let $R_0>0$ and $\omega_*$ be constants, and $\omega_{**}$ and
  $S_{**}$ be rotationally symmetric functions in
  $C^\infty(\St)$. Suppose that $(S,\omega)$ is the corresponding
  solution of \Eqsref{eq:FuchsS}--\eref{eq:Fuchsomega} according to
  \Propref{prop:basicexistence} on $(0,\delta]\times\St$, and
  $(\lambda,\omega)$ the smooth continuation to
  $[-\delta,\delta]\times\St$ according to
  \Lemref{lem:extension}. Then the two functions
  \begin{equation}
    \label{eq:DefFeFo}
    \Fe:=-\frac{R_{,+}\mu_--R_{,-}\mu_+}{4R_{,+} R_{,-}}-\frac 2t,\quad
    \Fo:=\frac{R_{,+}\mu_-+R_{,-}\mu_+}{4R_{,+} R_{,-} \sin\vartheta},
  \end{equation}
  with 
  \begin{equation}
    \label{eq:defmupm}
    \mu_\pm:=4R_{,\pm\vartheta}+R (S_{,\pm}^2+\ee^{-2S}\omega_{,\pm}^2),
  \end{equation}
  are smooth $\vartheta$-even functions on $[-\delta,\delta]\times\Tt$, where
$\Fe$ is $t$-odd and $\Fo$ is $t$-even. 
We have
\begin{equation}
  \label{eq:limits}
  \Fe(0,\cdot)\equiv 0,\quad 
  \Fo(0,\cdot)\equiv \frac 2{\sin\vartheta}\,\partial_\vartheta S_{**}.
\end{equation}
\end{lemma}

The issue which is addressed in this lemma is the following. Since $R_{,\pm}=R_0\sin(t\pm \vartheta)$, the functions $\Fe$ and $\Fo$ could, a priori, be singular
along the diagonals of the Gowdy square
where one of the functions $R_{,\pm}$ vanishes. Hence, when \Eqsref{eq:constraints} are written as
\begin{equation}
 \label{eq:constraintsSolvedForM}
  M_{,t}=-(S_{,t}-2/t)+\Fe,\quad
  M_{,\vartheta}=-S_{,\vartheta}+\Fo\sin\vartheta,
\end{equation}
it may then follow that there exists no smooth solution for the remaining metric function $M$, even if the functions $S$ and $\omega$ are smooth. Since the functions $\Fe$ and $\Fo$ are smooth, however, as asserted by the previous lemma, we see shortly that a smooth solution for $M$ exists. In this case we say that \Eqsref{eq:constraintsSolvedForM}
satisfy \keyword{matching conditions}. As shown by Chru\'sciel \cite{Chrusciel:1990ti}, there is a generic subset of \textit{Cauchy} data for $S$ and $\omega$ prescribed at any $t\in (0,\pi)$ which guarantee that the matching conditions are satisfied on that Cauchy hypersurface. It is then a consequence of the main evolution equations that the matching conditions are satisfied at every time $t$ where the solution is defined. In our case, however, we do not solve a Cauchy problem for $S$ and $\omega$ and hence this fact does not help. Indeed, it is a rather  surprising result of the previous lemma that \textit{all} choices of asymptotic data for $S$ and $\omega$ are compatible with the matching conditions.

 We also see from  \Eqsref{eq:constraintsSolvedForM} that the term $2/t$ in the
definition of $\Fe$ has been introduced to cancel precisely the most
singular term of $S_t$; we see in particular that $S_t-2/t$ can be extended as a smooth function to $[-\delta,\delta]\times\Tt$ which is $t$-even and $\vartheta$-even.

Another remark about \Lemref{lem:solMmatching} is that the term $2/{\sin\vartheta}\,\partial_\vartheta S_{**}$ in \Eqref{eq:limits} is indeed a smooth function on $\Tt$ because $S_{**}$ is $\vartheta$-even.

\begin{proof}[Proof of \Lemref{lem:solMmatching}]
Let us define the functions
\[f_0:=t (R_{,+}\mu_--R_{,-}\mu_+),\quad 
  f_1:=\frac{R_{,+}\mu_-+R_{,-}\mu_+}{\sin\vartheta},\]
with $\mu_\pm$ given by \Eqref{eq:defmupm}, so that
\begin{equation*}
    \Fe=-\frac 1t\, \frac{f_0}{4R_{,+}R_{,-}}-\frac 2t,\quad
    \Fo=\frac{f_1}{4R_{,+}R_{,-}}.
  \end{equation*}
In a first step, it is straightforward to show that $f_0$ and $f_1$ are smooth $\vartheta$-even functions on $\Tt$ for every $t\in (0,\delta]$ using the definitions of the functions $\mu_\pm$ in \Eqref{eq:defmupm}; notice that this is neither the case for $\mu_+$, $\mu_-$, $R_{,+}$ nor $R_{,-}$ individually. Moreover, by using our knowledge about the behavior of $S$ and $\omega$ as solutions of \Propref{prop:basicexistence} and the corresponding extensions $(\lambda,\omega)$ according to \Lemref{lem:extension} and hence by expressing the quantities above by $\lambda$ and $\omega$ and their derivatives, lengthy computations confirm that $f_0$ and $f_1$ extend as smooth  functions to $[-\delta,\delta]\times\Tt$ which are $t$-even and $\vartheta$-even.

We show now that the smooth functions $f_0$ and $f_1$ vanish precisely where the product $R_{,+} R_{,-}$ vanishes. After some computations, we find that the main evolution equations lead to the following equations for $\mu_+$ and $\mu_-$
\begin{equation}
  \label{eq:evolutionmu}  \fl\quad\partial_+\mu_-=-R_{,-}\left(S_{,+}S_{,-}+\ee^{-2S}\omega_{,+}\omega_{,-}\right),
\quad
\partial_-\mu_+=-R_{,+}\left(S_{,+}S_{,-}+\ee^{-2S}\omega_{,+}\omega_{,-}\right).
\end{equation}
This implies that along the diagonal $\vartheta=t$, where $R_{,-}=0$, we have that $\mu_-=\mu_-^{(0)}=\textrm{constant}$ and hence
\[\left.f_0\right|_{\vartheta=t}=t R_0\sin(2t)\mu_-^{(0)},\quad 
  \left.f_1\right|_{\vartheta=t}=\frac{R_0\sin(2t)\mu_-^{(0)}}{\sin t},\]
for all $t\in (0,\delta]$. Since both are smooth functions, their expansions at $\vartheta=t=0$ along the curve $\vartheta=t$ are
\[\left.f_0\right|_{\vartheta=t}=2R_0 \mu_-^{(0)} t^2+O(t^3),\quad
\left.f_1\right|_{\vartheta=t}=2R_0 \mu_-^{(0)}+O(t).\]
On the other hand, the definitions of $f_0$ and $f_1$ and the uniform  leading-order expansions of $\lambda$ and $\omega$ at $t=0$, 
\[\lambda=t^2 \ee^{S_{**}}+O(t^4),\quad \omega=\omega_*+t^4\omega_{**}+O(t^6),\]
imply
\[\left.f_0\right|_{\vartheta=t}=O(t^3),\quad
\left.f_1\right|_{\vartheta=t}=O(t).\]
We conclude that $\mu_-^{(0)}=0$ and hence that
\[\left.f_0\right|_{\vartheta=t}=0,\quad
\left.f_1\right|_{\vartheta=t}=0,\]
for all $t\in [-\delta,\delta]$. 
The same argument applied at $\vartheta=\pi-t$, where $R_{,+}=0$ and hence $\mu_+=\textrm{constant}$, leads to the similar result
\[\left.f_0\right|_{\vartheta=\pi-t}=0,\quad
\left.f_1\right|_{\vartheta=\pi-t}=0,\]
for all $t\in [-\delta,\delta]$.

Let us consider a small neighborhood of the point $(t,\vartheta)=(0,0)$ where we now introduce coordinates $x=t+\vartheta$ and $y=t-\vartheta$. With our analysis above, we have found that there exist smooth functions $\tilde f_0(x,y)$ and $\tilde f_1(x,y)$ so that
\[f_0(x,y)=x y \tilde f_0(x,y), \quad f_1(x,y)=x y \tilde f_1(x,y).\]
Since $x y=t^2-\vartheta^2$, the functions $\tilde f_0$, $\tilde f_1$ must therefore be $\vartheta$-even and $t$-even. Now we note that the product 
\[R_{,+} R_{,-}=R_0^2(\cos\vartheta-\cos t)(\cos\vartheta+\cos t)=R_0^2\sin x\sin y.\]
Hence, the quotients $f_0/(R_{,+} R_{,-})$ and $f_1/(R_{,+} R_{,-})$, which appear in the definitions of $\Fo$ and $\Fe$, are smooth functions on our small neighborhood of $(0,0)$. The same argument applies to a neighborhood of the point $(t,\vartheta)=(0,\pi)$, and hence $f_0/(R_{,+} R_{,-})$ and $f_1/(R_{,+} R_{,-})$ are smooth functions everywhere on $[-\delta,\delta]\times\Tt$ which are $\vartheta$-even and $t$-even. As a consequence, $\Fo$ is a smooth function on  $[-\delta,\delta]\times\Tt$ which is $t$-odd and $\vartheta$-even, and $\Fe$ a smooth function which is $t$-even and $\vartheta$-even.

It only remains to compute the values of the extended functions $\Fe$ and $\Fo$ at $t=0$, i.e.\ \Eqref{eq:limits}. This can be done directly using the leading-order behavior of $\lambda$ and $\omega$.
\end{proof}

\begin{proof}[Proof of \Propref{prop:solM}]
We have proven in \Lemref{lem:solMmatching} that all terms in
\Eqsref{eq:constraintsSolvedForM} are well-defined when they are considered as functions with spatial domain $\Tt$ as discussed above.
The integrability condition is satisfied, and we can hence conclude that there exists a unique solution $M$ on $[-\delta,\delta]\times \Tt$ which is $\vartheta$-even, as soon as the value $M_0$ of $M$ has been fixed somewhere, say, at $(t,\vartheta)=(0,0)$. This solution can be written as
\begin{eqnarray*}
M(t,\vartheta)&=&M_*-(S(t,\vartheta)-2\ln t)+\int_0^\vartheta\Fo(0,x)\sin x\,\dd x+\int_0^t
\Fe(\tau,\vartheta)\,\dd\tau,\\
&=&M_*-(S(t,\vartheta)-2\ln t)+2S_{**}(\vartheta)-2S_{**}(0)+\int_0^t
\Fe(\tau,\vartheta)\,\dd\tau,
\end{eqnarray*}
for some constant $M_*$. Since $\Fe$ vanishes at $t=0$ and is $t$-odd, the function $M$ must be $t$-even.

Now let us consider the smoothness condition \Eqref{eq:boundarycondM2}, i.e.\ we want to show that 
\[\left.M(t,\vartheta)-2\ln R_0+S(t,\vartheta)-2\ln\sin t\right|_{\vartheta=0,\pi}=0,\]
for all $t\in[-\delta,\delta]$.
Using the expression for $M$ above, we compute
\begin{eqnarray*}
  &M(t,\vartheta)-2\ln R_0+S(t,\vartheta)-2\ln\sin t\\
  &=M_*+2(\ln t-\ln\sin t)+2S_{**}(\vartheta)-2S_{**}(0) -2\ln R_0+\int_0^t\Fe(\tau,\vartheta)\,\dd\tau.
\end{eqnarray*}
On the other hand, we can evaluate $\Fe$ at
$\vartheta=0$ (and below at $\vartheta=\pi$)
using \Eqsref{eq:DefFeFo}--\eref{eq:defmupm} (recall that $R=0$ at $\vartheta=0,\pi$). We find
\begin{eqnarray*}
\fl
\qquad\left.\Fe\right|_{\vartheta=0}&:=&-\left.\frac{R_{,+}R_{,-\vartheta}-R_{,-}R_{,+\vartheta}}{R_{,+}R_{,-}}-\frac 2t\right|_{\vartheta=0}
=-\frac{-\sin t \cos t-\sin t \cos t}{\sin^2 t}-\frac 2t\\
&=&2\left(\frac{\cos t}{\sin t}-\frac 1t\right).
\end{eqnarray*}
The integral of this expression is
\[\int_0^t\Fe(\tau,0)\,\dd\tau=-2(\ln t-\ln\sin t).\]
Hence, 
\begin{equation*}
  M(t,0)-2\ln R_0+S(t,0)-2\ln\sin t
  =M_*-2\ln R_0,
\end{equation*}
which vanishes for all $t$ if and only if
\[M_*=2\ln R_0.\]

Now we perform the same analysis at $\vartheta=\pi$. We find
\begin{equation*}
\left.\Fe\right|_{\vartheta=\pi}
=2\left(\frac{\cos t}{\sin t}-\frac 1t\right).
\end{equation*}
Therefore,
\begin{equation*}\fl
  M(t,\pi)-2\ln R_0+S(t,\pi)-2\ln\sin t
  =M_*+2(S_{**}(\pi)-S_{**}(0))-2\ln R_0.
\end{equation*}
Given the value for $M_*$ above, the smoothness condition at $\vartheta=\pi$ is satisfied if and only if 
$S_{**}(\pi)=S_{**}(0)$.

Now it remains to show that \Eqref{eq:waveequationM} is satisfied on $(0,\delta]\times\tilde\St$.
We define $C_1:=(C_++C_-)/2$ and $C_2:=(C_+-C_-)/2$ from
Eqs.~\eref{eq:constraints}, and $H$ as defined in
Eqs.~\eref{eq:waveequationM}. The system \eref{eq:FuchsS} and
\eref{eq:Fuchsomega} implies the \keyword{subsidiary system}
\[\partial_t C_1-\partial_\vartheta C_2=0,\quad
\partial_t C_2-\partial_\vartheta C_1=H+\cot t\, C_2+\cot\vartheta\, C_1.
\]
Since $C_1$ and $C_2$ vanish identically on $(0,\delta]\times\tilde\St$, so must $H$. 
We conclude that our solutions do solve \Eqref{eq:waveequationM}; indeed,
the full set of Geroch's equations is satisfied on $(0,\delta]\times\St$.
\end{proof}

All functions constructed so far can now be lifted to smooth
functions on $[-\delta,\delta]\times\Sth$ which are invariant along $\partial_{\rho_1}$ and
$\partial_{\rho_2}$. In particular, the quantities $\lambda$, $\omega$ and $h_{ab}$ allow us to determine the metric $g_{ab}$ on $(0,\delta]\times\Sth$. It remains to compute $Q$ on $(0,\delta]\times\Sth$ and to show that it can be extended as a smooth function to $[-\delta,\delta]\times\Sth$, which is even in time.
We must therefore find a smooth function $Q$ which satisfies
\begin{equation}
  \label{eq:detQ}
  \partial_\theta Q=-R \lambda^{-2}\partial_t\omega,\quad
  \partial_t Q=-R \lambda^{-2}\partial_\theta\omega,
\end{equation}
from \Eqref{eq:relationPQomegatheta}.
In the same the spirit as above, we can ignore the fact that, strictly speaking, these equations are not defined at $\theta=0$, $\theta=\pi$. Moreover, $Q$ must also satisfy the smoothness condition \Eqref{eq:QBound}.
Given a solution $(S,\omega)$ of \Propref{prop:basicexistence}, or equivalently the extensions $(\lambda,\omega)$ as above, the integrability condition of \Eqref{eq:detQ}  follows for both time intervals $(0,\delta]$ and $[-\delta,0)$. The expansions of $\lambda$ and $\omega$ can be used to show that the right-hand sides of \Eqsref{eq:detQ} are in fact smooth functions on $[-\delta,\delta]\times\tilde\St$ which can be lifted to a dense subset of $[-\delta,\delta]\times\Sth$. It can be concluded that, once the value of $Q$ is fixed at some point, there is a unique smooth function $Q$ on $[-\delta,\delta]\times\Sth$ which satisfies the equations above and the smoothness condition. The expansion for $Q$ at $t=0$  has no $\log t$-terms and all powers in $t$ are positive even integers. From
\[\partial_\theta Q(0,\theta)
=-4R_0\omega_{**}(\theta)\ee^{-2S_{**}(\theta)}\sin\theta,\] it
follows that
\[Q(0,\theta)=Q_0-4R_0\int_0^\theta \omega_{**}(x)\ee^{-2S_{**}(x)}\sin
x\, \dd x,
\] 
for some constant $Q_0$.  In order to guarantee the smoothness condition
\Eqref{eq:QBound}, namely $Q(t,0)=1$, $Q(t,\pi)=-1$ for all $t>0$, we
must choose $Q_0=1$ and the asymptotic data must satisfy
\begin{equation}
  \label{eq:smoothnessADataQ}
  \int_0^\pi \omega_{**}(x)\ee^{-2S_{**}(x)}\sin
  x\,\dd x=\frac 1{2R_0}.
\end{equation}
Then it follows that $Q(0,0)=1$, $Q(0,\pi)=-1$, and the second of
Eqs.~\eref{eq:detQ} implies that $Q(t,0)=1$, $Q(t,\pi)=-1$ for all
$t\in [-\delta,\delta]$. 
%
We have found the expansion
\[\fl\quad Q(t,\theta)=1-4R_0\int_0^\theta \omega_{**}(x)\ee^{-2S_{**}(x)}\sin x\,
\dd x+\frac 12 R_0\,\partial_\theta\omega_{**}(\theta)\ee^{-2S_{**}(\theta)}\sin\theta\,
t^2 + O(t^4).
\] 
Equivalently to prescribing the data functions $S_{**}$ and $\omega_{**}$ which obey \Eqref{eq:smoothnessADataQ}, we can prescribe free smooth rotationally symmetric data functions $S_{**}$ and $Q_*$, and set
\[w_{**}(\theta)=\ee^{2S_{**}(\theta)}\frac{1-\partial_\theta Q_*(\theta)\sin\theta
  -2Q_*(\theta)\cos\theta}{4R_0}.\]
Then, \Eqref{eq:smoothnessADataQ} follows automatically. The above expansion for $Q$ simplifies to
\[Q(t,\theta)=\cos\theta+Q_*(\theta)\sin^2\theta
+O(t^2),
\]
and the smoothness conditions for $Q$ at $\theta=0,\pi$ become manifest.

Hence, we obtain smooth Gowdy vacuum solutions on the time interval $(0,\delta]$. In order to show that these solutions are indeed smooth Gowdy symmetric
generalized Taub-NUT solutions and hence extendible through $t=0$, we must perform the coordinate transformation \Eqref{eq:extensionthroughCH} near $t'=0$. Since all the metric functions are smooth through $t=0$ and even in $t$ and independent of $\rho_1$ in particular, they can be extended as smooth functions in terms of the new coordinates through $t'=0$. It remains to show that the expression $(4\kappa^2N^2-\ee^{2M})/t'$ is also a smooth function through $t'=0$ for some choice of $\kappa$. 
For this, we compute the uniform limit of the quantity $N$ (defined in
\Eqref{eq:ansatzforN}) at $t=0$,
\[N(0,\theta)=\ee^{2(S_{**}(\theta)-S_{**}(0))}.
\] 
From that it is
easy to determine that the quotient above is extendible if
\[\kappa=\pm \frac{R_0^2}2 \ee^{-S_{**}(0)}.\]

We have thus obtained smooth Gowdy symmetric
generalized Taub-NUT solutions in a (possibly small) time interval
$(0, \delta]$. The global existence theorem of
Chru\'sciel~\cite{Chrusciel:1990ti} (Theorem~6.3 in
\cite{Chrusciel:1990ti}, which makes essential use of results by Christodoulou and Tahvildar-Zadeh \cite{Christodoulou:1993vf}) can now be used to extend the
spacetimes to the time interval $(0,\pi)$ as smooth globally
hyperbolic Gowdy solutions.

\section{The linear problem and global-in-time properties}
\label{sec:linearproblem}

We have seen above that for given smooth asymptotic data at $t=0$
(e.g.\ the values of $S_{**}$ and $Q_{*}$) a smooth
Gowdy-symmetric generalized Taub-NUT solution exists in a vicinity of
$t=0$. Moreover, using Chru\'sciel's theorem, we see that this solution can even be extended smoothly to the whole time interval $(0,\pi)$. However, the surface
$t=\pi$ itself is expected to contain either singularities or Cauchy
horizons. On the other hand, Chru\'sciel's result also allows the case that the $t=\pi$-surface is regular, but just the coordinates break down there.
It is the purpose of the following considerations to find out what happens at $t=\pi$. In particular, we construct explicitly the metric
potentials at this boundary (as well as on the axes $\theta=0,\pi$) in terms of the asymptotic data. For that
purpose, we apply the so-called \emph{soliton methods}, which were used in
\cite{Hennig2010} for the investigation of $\St\times\So$ Gowdy
spacetimes\footnote{The methods described \cite{Hennig2010} have also
  been applied to studying the interior region of axisymmetric and
  stationary black holes with surrounding matter, see
  \cite{Ansorg:2008uc,Ansorg:2009vq,Hennig:2009tj}.}, to the present
$\Sth$-symmetric case. 

In all of what follows we make the same hypotheses as in
  \Theoremref{th:existence}. These assumptions are consistent with
  those listed in \Sectionref{sec:GenTaubNUT}, and hence we consider
  ``smooth Gowdy symmetric generalized Taub-NUT solutions''.

\subsection{Einstein's field equations and the Ernst formulation}

The first important step for the following considerations is the introduction of the complex Ernst formulation of the Einstein equations which will be described in this subsection.

Again we start from the metric
\begin{equation}
  g=\ee^M(-\dd t^2+\dd\theta^2)+R_0\sin t\sin\theta\left[\ee^L (\dd\rho_1+Q \dd\rho_2)^2+\ee^{-L} \dd\rho_2^2\right]
\end{equation}
in the Killing basis $\{\partial_{\rho_1},\partial_{\rho_2}\}$. Here, we express $L$ in terms of a metric potential $u$ via
\begin{equation}\label{eq:Lu}
 \ee^L=\frac{\sin t}{\sin\theta}\,\ee^u.
\end{equation}
In this way, we arrive at
\begin{equation}
  g=\ee^M(-\dd t^2+\dd\theta^2)+R_0\left[\sin^2\!t\,\ee^u (\dd\rho_1+Q \dd\rho_2)^2+\sin^2\!\theta\,\ee^{-u} \dd\rho_2^2\right].
\end{equation}
Note that $u$ is related to the quantity $S$ (defined in Sec.~\ref{sec:equn}) via 
$$u(t,\theta)=S(t,\theta)-\ln(R_0)-2\ln\sin t,$$
i.e.\ the singularity of $S$ at $t=0$ ($S$ behaves as $2\ln t$ for $t\to 0$, see Prop.~\ref{prop:basicexistence}) is removed by subtracting the term $2\ln\sin t$.

Now we reformulate the Einstein equations as equations for $u$, $Q$ and $M$. We obtain two second-order equations for the metric potentials $u$ and $Q$,
\begin{equation}\fl\label{eq:u}
 -\partial_t^2u-\cot t\,\partial_t u+\partial_\theta^2u
 +\cot\theta\,\partial_\theta u
 +\ee^{2u}\frac{\sin^2\! t}{\sin^2\!\theta}\left[(\partial_t Q)^2
 -(\partial_\theta Q)^2\right] +2 = 0,
\end{equation}
\begin{equation}\fl\label{eq:Q}
 -\partial_t^2 Q-3\cot t\,\partial_t Q+\partial_\theta^2 Q
 -\cot\theta\,\partial_\theta Q-2[(\partial_t u)(\partial_t Q)
 -(\partial_\theta u)(\partial_\theta Q)]=0
\end{equation}
and two first-order equations for $M$,
\begin{eqnarray}\fl\label{eq:M1}
 (\cos^2\! t-\cos^2\!\theta)\partial_t M & = &
  \frac{1}{2}\ee^{2u}\frac{\sin^3\! t}{\sin\theta}
 \Big[\cos t\sin\theta[(\partial_t Q)^2+(\partial_\theta Q)^2]
       -2\sin t\cos\theta (\partial_t Q)(\partial_\theta Q)\Big]
 \nonumber\\
 & & +\frac{1}{2}\sin t \sin\theta
 \Big[\cos t\sin\theta[(\partial_t u)^2+(\partial_\theta u)^2]
       -2\sin t\cos\theta (\partial_t u)(\partial_\theta u)\Big]
 \nonumber\\
 & & -(2\cos^2\!t\,\cos^2\!\theta\,-\cos^2\!t-\cos^2\!\theta)
      \,\partial_t u
 \nonumber\\
 & & -2\sin t\cos t\sin\theta\cos\theta(\partial_\theta u+\tan\theta),  
\end{eqnarray}
\begin{eqnarray}\label{eq:M2}\fl
 (\cos^2\! t-\cos^2\!\theta)\partial_\theta M & = &
  -\frac{1}{2}\ee^{2u}\frac{\sin^3\! t}{\sin\theta}
 \Big[\sin t\cos\theta[(\partial_t Q)^2+(\partial_\theta Q)^2]
       -2\cos t\sin\theta (\partial_t Q)(\partial_\theta Q)\Big]
 \nonumber\\
 & & -\frac{1}{2}\sin t \sin\theta
 \Big[\sin t\cos\theta[(\partial_t u)^2+(\partial_\theta u)^2]
       -2\cos t\sin\theta (\partial_t u)(\partial_\theta u)\Big]
 \nonumber\\ 
 & & -2\sin t\cos t\sin\theta\cos\theta(\partial_t u-\tan t)
 \nonumber\\
 & & -(2\cos^2\!t\,\cos^2\!\theta\,-\cos^2\!t-\cos^2\!\theta)
      \,\partial_\theta u.
\end{eqnarray}
Since $M$ does not appear in \eref{eq:u} and \eref{eq:Q} and since we assume the genericity condition of Chru\'sciel, these
equations may be solved as a first step. Afterwards, \eref{eq:M1} and
\eref{eq:M2} can be used to calculate $M$ via a line integral. Note
that the integrability condition $\partial_t\partial_\theta
M=\partial_\theta\partial_t M$ of the system \eref{eq:M1},
\eref{eq:M2} is satisfied as a consequence of \eref{eq:u},
\eref{eq:Q}. Hence, $M$ does not depend on the path of integration.

It turns out that the two Einstein equations \eref{eq:u}, \eref{eq:Q} are equivalent to a single complex equation, namely to the Ernst equation
\begin{equation}\label{eq:EE}
 \Re(\E) \left(-\partial_t^2\E-\cot t\,\partial_t\E
        +\partial_\theta^2\E+\cot\theta\,\partial_\theta\E\right)
 =-(\partial_t\E)^2+(\partial_\theta\E)^2
\end{equation}
for the complex Ernst potential $\E=f+\ii b$. Here, the real part $f$ of $\E$ is defined in terms of the Killing vector $\partial_{\rho_2}$ by
\begin{equation}\label{eq:deff}
 f:=\frac{1}{R_0}g(\partial_{\rho_2},\partial_{\rho_2})
 =Q^2\ee^u\sin^2\! t+\ee^{-u}\sin^2\!\theta
\end{equation}
and the imaginary part $b$ is given by
\begin{equation}\label{eq:defb}
 \partial_t a=\frac{1}{f^2}\sin t\sin\theta\,\partial_\theta b,\quad
 \partial_\theta a=\frac{1}{f^2}\sin t\sin\theta\,\partial_t b
\end{equation}
with
\begin{equation}\label{eq:defa}
 a:= \frac{g(\partial_{\rho_1},\partial_{\rho_2})}
          {g(\partial_{\rho_2},\partial_{\rho_2})}
   = \frac{Q}{f}\ee^u\sin^2 t.
\end{equation}

\begin{figure}\centering
 \includegraphics[scale=0.9]{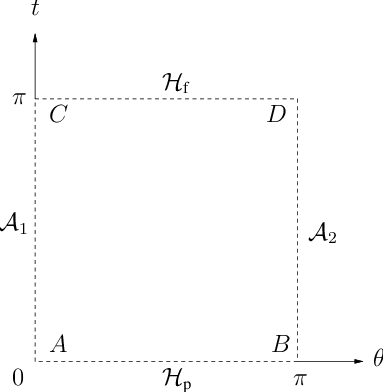}
 \caption{\label{fig:Gowdy}We integrate the LP along the boundaries of the Gowdy square (dashed path) in order to investigate for which asymptotic data the solution can be regularly extended up to the future boundary $\Hf$ ($t=\pi$).}
\end{figure}

Note that for smooth functions $u$ and $Q$ the Ernst potential $\E$ is also smooth: For the real part $f$, smoothness is clear from definition \eref{eq:deff}. In the case of the imaginary part $b$, it can be shown by solving the two equations in \eref{eq:defb} for $\partial_t b$ and $\partial_{\theta} b$ and replacing $a$ and $f$ via \eref{eq:defa} and \eref{eq:deff}. The resulting expressions for $\partial_t b$ and $\partial_{\theta} b$ in terms of $Q$ and $u$ (and their first order derivatives) turn out to be smooth functions, if we use the fact that $Q$ behaves as given in \eref{eq:QBound}. Hence, integration will lead to a smooth function $b$. Therefore, we can conclude from the previous local existence results and Chru\'sciel's global existence theorem that for any given set of asymptotic data (as described in Theorem~\ref{th:existence}) the corresponding Ernst potential $\E$ is a smooth complex function on $(-T,\pi)\times\St$ for some $T>0$.
In the following we investigate under which conditions $\E$ can be extended smoothly to the boundary $t=\pi$ and beyond.
Note that our assumptions imply that $f>0$ holds in the entire Gowdy square with the exception of the points $A$ and $B$ and with the possible exception of the future boundary $\Hf$ (see Fig.~\ref{fig:Gowdy}) which is important since we will divide by $f$ in some of the following formulae.

Once we have obtained an Ernst potential $\E$ as a solution to the Ernst equation \eref{eq:EE}, we can calculate the corresponding metric potentials from it. It turns out that the integrability condition 
$\partial_t\partial_\theta a=\partial_\theta\partial_t a$ of \eref{eq:defb} is satisfied as a consequence of the Ernst equation. Therefore, $a$ may be calculated via line integration from $\E$. The metric potentials $u$ and $Q$ can then be obtained from $a$ and $f$. With \eref{eq:deff} and \eref{eq:defa} we find
\begin{equation}\label{eq:uQ}
 \ee^u=\frac{f a^2}{\sin^2\! t}+\frac{\sin^2\!\theta}{f},\quad
 Q=\frac{f^2 a}{f^2 a^2 + \sin^2\! t\sin^2\!\theta}.
\end{equation}
Finally, $M$ may be calculated using \eref{eq:M1} and \eref{eq:M2}, as mentioned earlier.

As an example, we give the Ernst potential for the Taub solution:
\begin{eqnarray}
 f & = & \frac{2l}{X}\sin^2\!t\,\cos^2\!\theta
         +\frac{X}{2l}\sin^2\!\theta,\\
 b & = & \frac{1}{X}\left[\cos t(\cos^2\! t-3)\sqrt{m^2+l^2}-2m\right]
        \cos^2\!\theta+\cos t
\end{eqnarray}
with $X:=(1+\cos^2\!t)\sqrt{m^2+l^2}+2m\cos t$. (Here we have set an arbitrary additive integration constant in $b$ to zero.)

\subsection{The linear problem}\label{sec:LP}

Interestingly, the Ernst equation \eref{eq:EE} belongs to a remarkable class of nonlinear partial differential equations for which an associated linear problem (LP) exists which is equivalent to the nonlinear equation via its integrability condition. For applications of this LP in the context of \emph{axisymmetric and stationary} spacetimes we refer the reader to, e.g., \cite{Neugebauer:2003uw,Neugebauer2011}. In the Gowdy setting, we use the LP in the form \cite{Neugebauer:1979vc,Neugebauer:1980wa}, which reads in our 
coordinates\footnote{The formal relation between our coordinates (describing Gowdy spacetimes with two spacelike Killing vectors) and the Weyl-Lewis-Papapetrou coordinates $(\rho,\zeta,\varphi,\tilde t\,)$ as used by Neugebauer (describing axi\-symmetric and stationary spacetimes with one spacelike and one timelike Killing field) is given by $\rho=\ii R_0\sin t\sin\theta$, 
$\zeta=R_0\cos t\cos\theta$, $\varphi=\rho_1$, $\tilde t=\rho_2$.} as

\begin{equation}\label{eq:LP}
\eqalign{
 \partial_x\bPhi & = \left[\left(\begin{array}{cc}
                   B_x & 0\\ 0 & A_x\end{array}\right)
                   +\lambda\left(\begin{array}{cc}
                   0 & B_x\\ A_x & 0\end{array}\right)\right]\bPhi,\\
 \partial_y\bPhi & = \left[\left(\begin{array}{cc}
                   B_y & 0\\ 0 & A_y\end{array}\right)
                   +\frac{1}{\lambda}\left(\begin{array}{cc}
                   0 & B_y\\ A_y & 0\end{array}\right)\right]\bPhi,
 }
\end{equation}
where the \emph{pseudopotential}
$\bPhi=\bPhi(x,y,K)$ is a $2\times 2$ matrix depending on the coordinates
\begin{equation}
 x = \cos(t-\theta),\qquad y = \cos(t+\theta)
\end{equation}
as well as on the \emph{spectral parameter} $K\in\C$. The remaining ingredients of the LP are the function
$\lambda$,
\begin{equation}\label{eq:lambda}
 \lambda(x,y,K) := \sqrt{\frac{K-y}{K-x}},
\end{equation}
and the matrix elements $A_x$,
$A_y$, $B_x$ and $B_y$, defined in terms of the Ernst potential as
\begin{equation}
 \label{eq:AB}
 A_i = \frac{\partial_i\E}{2f}, \qquad
 B_i = \frac{\partial_i\bar\E}{2f},\qquad i=x,y.
\end{equation}

Due to the two possible signs of the square root in \eref{eq:lambda}, $\lambda:\C\to\C$, $K\mapsto\lambda$
describes, for fixed values $x$, $y$, a mapping from a two-sheeted Riemann surface (K-plane) onto the complex $\lambda$-plane. The two $K$-sheets are connected at the branch points
\begin{equation}
 K_1 = x\quad (\lambda=\infty),\qquad K_2=y\quad (\lambda=0).
\end{equation}
In general, the pseudopotential $\bPhi$ will take on different values on the two $K$-sheets. Only at the branch points it has to be unique, since both Riemannian sheets coincide there. We will see below that this observation plays an important role for the calculation of the Ernst potential from the solution of the LP.

As already mentioned, the integrability condition $\partial_x\partial_y\bPhi=\partial_y\partial_x\bPhi$ of \eref{eq:LP} is
equivalent to the Ernst equation \eref{eq:EE}. Hence, the Ernst equation
is a consequence of the LP and, on the other hand, for a given 
potential $\E$ as a solution to the Ernst equation,
the matrix $\bPhi$ does not depend on the path of integration.

Finally, we note that for any solution $\bPhi$ to the LP \eref{eq:LP},  
the product $\bPhi{\bf C}(K)$, where ${\bf C}(K)$ is an arbitrary $2\times 2$ matrix, is also a
solution (corresponding to the same Ernst potential). As shown by Neugebauer \cite{Neugebauer:2003uw}, it is always possible to choose ${\bf C}(K)$ in such a way that the transformed pseudopotential takes the form
\begin{equation}\label{eq:struct}
\eqalign{
 \bPhi^>(x,y,K) & =\left(\begin{array}{cc}
       \psi_1^>(x,y,K) & \psi_1^<(x,y,K)\\
       \psi_2^>(x,y,K) & -\psi_2^<(x,y,K)
       \end{array}\right),\\
 \bPhi^<(x,y,K) & =\left(\begin{array}{cc}
       \psi_1^<(x,y,K) & \psi_1^>(x,y,K)\\
       \psi_2^<(x,y,K) & -\psi_2^>(x,y,K)
       \end{array}\right)\\
     &=\left(\begin{array}{cc} 1 & 0\\ 0 & -1\end{array}\right)
      \bPhi^>(x,y,K)
      \left(\begin{array}{cc} 0 & 1\\ 1 & 0\end{array}\right),
 }
\end{equation}
where the superscripts ``$>$'' or ``$<$'' indicate whether the functions are
evaluated on the ``upper'' $(\lambda=1$ for $K=\infty$) or ``lower''
($\lambda=-1$ for $K=\infty$) $K$-sheet. Hence, $\bPhi$ is completely determined by the values of two functions $\psi_1$ and $\psi_2$ on both $K$-sheets. In all of what follows we assume that we have already achieved this form for $\bPhi$.

\subsection{Solution of the linear problem} 

\subsubsection{Coordinate transformation.}

In the following we intend to integrate the LP along the boundaries of the Gowdy square. 
For that purpose,
it turns out to be useful to study the
situation not only in the coordinate system $\Sigma$, corresponding to the Killing basis $\{\partial_{\rho_1},\partial_{\rho_2}\}$, but also in a coordinate frame $\tilde\Sigma$,
\begin{equation}\label{eq:rc}
 \tilde\Sigma:\quad\tilde t=t,\quad \tilde\theta=\theta,\quad
 \tilde\rho_1=\rho_1+q\rho_2,\quad \tilde\rho_2=\rho_2
\end{equation}
with $q=\textrm{constant}$. 
According to \eref{eq:trans1}-\eref{eq:trans3} and \eref{eq:Lu}, the transformed metric potentials are 
\begin{equation}
 \tilde R_0=R_0,\quad \tilde u=u,\quad \tilde Q=Q-q,
\end{equation}
i.e.\ only $Q$ is changed by subtracting a constant.
In particular, we will choose the two systems $\tilde\Sigma$ with $q=1$ or $q=-1$, in which $\tilde Q|_{\Al}=0$ or $\tilde Q|_{\Ar}=0$ holds, respectively.

Since the coordinate transformation \eref{eq:rc} is merely a change of the Killing basis, the Ernst equation \eref{eq:EE} retains its form in $\tilde\Sigma$. This implies the existence of an LP \eref{eq:LP}
for a pseudopotential $\tilde{\bPhi}$ in this frame. As shown by Neugebauer \cite{Neugebauer:2000ut,Neugebauer:2003uw}, the
matrices $\tilde{\bPhi}$ and $\bPhi$ are connected by the transformation \begin{equation}\label{eq:Phis}
 \tilde{\bPhi}
    =\left[\left(\begin{array}{cc} c_- & 0\\ 0 & c_+\end{array}\right)
       +\ii\frac{q}{f}(K-x)\left(\begin{array}{cc}
       1 & \lambda\\-\lambda & -1\end{array}\right)\right]\bPhi
\end{equation}
with
\begin{equation}
 c_\pm:=1-q\left(a\pm\frac{\ii}{f}\sin t\sin\theta\right),
\end{equation}
where all quantities on the right hand side of Eq.~\eref{eq:Phis} belong to the original frame $\Sigma$. As we will see, this transformation becomes particularly simple at the boundaries of the Gowdy square for our choices $q=\pm1$.

\subsubsection{The LP on \texorpdfstring{$\Hp$}{Hp}, \texorpdfstring{$\Al$}{A1} and \texorpdfstring{$\Ar$}{A2}.}
\label{sec:LPboundaries}

From our previous discussions, namely from the local investigation of the singular initial value problem for the Einstein equations with Fuchsian methods (see Sec.~\ref{sec:existence}) and from Chru\'sciel's global existence theorem, we know that for any smooth set of asymptotic data on $\Hp$  a corresponding smooth Gowdy symmetric generalized Taub-NUT solution exists. Moreover, this solution is smooth both on the axes of symmetry $\Al$, $\Ar$ and in the interior of the Gowdy square. The goal of this subsection is to find explicit expressions for the values of the solution on $\Al$ and $\Ar$, which are determined by the data on $\Hp$. Afterwards we will study the behavior as $t\to\pi$ and investigate whether a continuation of the solution to $\Hf$ is possible.



Along the entire integration path, we have $x=y$ and therefore $\lambda=\pm 1$, cf.\ \eref{eq:lambda}. However, it suffices to study the case $\lambda=1$ alone, since the solution on the Riemannian sheet with $\lambda=-1$ can easily be obtained from the solution with $\lambda=1$ using \eref{eq:struct}.

For $x=y$ and $\lambda=1$, the LP \eref{eq:LP} reduces to the
ODE 
\begin{equation}
  \label{eq:LPBound}
 \partial_x\bPhi=\frac{1}{2f}
   \left(\begin{array}{cc}\partial_x\bar\E & \partial_x\bar\E \\
                          \partial_x\E     & \partial_x\E
         \end{array}\right)\bPhi
\end{equation}
with the general solution\footnote{By plugging the solution \eref{eq:gensol} into \eref{eq:LPBound}, we see that the matrix on the right hand side of \eref{eq:LPBound} is proportional to $f$, i.e.\ the factor $1/f$ is canceled and the solution extends smoothly over points with $f=0$. However, it follows from \eref{eq:deff} that $f$ does not vanish on $\Hp$, $\Al$, $\Ar$ with the exception of the corners $A$, $B$ of the Gowdy square (as well as $C$ and $D$, provided $Q^2\ee^u$ is bounded for $t\to\pi$).}
\begin{equation}\label{eq:gensol}
 \bPhi={\bf E}{\bf C}(K),\qquad
      {\bf E}:=\left(\begin{array}{cc}\bar\E & 1\\ \E & -1\end{array}\right)
\end{equation}
in terms of the Ernst potential on the boundary,
where the $2\times 2$ matrix ${\bf C}$ is a $K$-dependent ``integration
 constant''. The solutions on all parts of the integration path have
 the form \eref{eq:gensol}, but with different integration constants:
 \begin{eqnarray}
  \label{eq:L1}
 t=0:\quad & \bPhi={\bf E}{\bf C},\quad
        {\bf C}=\left(\begin{array}{cc}
        C_1 & C_3\\ C_2 & C_4\end{array}\right),\\
 \label{eq:L3}        
 \theta=0: & \bPhi={\bf E}{\bf D},\quad
        {\bf D}=\left(\begin{array}{cc}
        D_1 & D_3\\ D_2 & D_4\end{array}\right),\\
 \label{eq:L4}
 \theta=\pi: & \bPhi={\bf E}\tilde{\bf D},\quad
        \tilde{\bf D}=\left(\begin{array}{cc}
        \tilde D_1 & \tilde D_3\\ \tilde D_2 & \tilde D_4\end{array}\right).    
\end{eqnarray}
 A further simplification can be achieved by normalizing $\bPhi$ at $t=0$
 via
\begin{equation}\label{eq:norm}
 t=0:\quad \psi_1^<=\psi_2^<=\psi(K), 
\end{equation}
 where $\psi(K)$ is an arbitrary gauge
 function (which will later be specified in such a way that the LP has a regular solution, see Eq.~\eref{eq:gauge} below). This is possible since the form \eref{eq:struct} of $\bPhi$ is
invariant under the transformation \cite{Neugebauer2011}
 \begin{equation}
  \bPhi\to\bPhi\cdot\left(\begin{array}{cc}\alpha(K) & \beta(K)\\
  \beta(K) & \alpha(K)\end{array}\right).
 \end{equation}
 The two degrees of freedom $\alpha$, $\beta$ can be used to achieve
 the two conditions in \eref{eq:norm}. As a consequence, we obtain
 \begin{equation}
 C_3=0,\quad C_4=\psi
 \end{equation}
in this gauge.

From \eref{eq:L1}-\eref{eq:L4} we can now calculate the solution of the LP in the frame $\tilde\Sigma$ (cf.\ \eref{eq:rc}) using the transformation formula \eref{eq:Phis}. It follows from \eref{eq:defa} that $a$ takes on the boundary values\footnote{Note that $a$ is automatically discontinuous at the points $A$, $B$, $C$, $D$ as a consequence of the definition \eref{eq:defa}. In contrast, $a$ is a smooth function in the remaining part of the Gowdy square.}
\begin{equation}\label{eq:avals}
 \Hp: \quad a=0,\quad
 \Al:\quad a=\frac{1}{Q}=1,\quad
 \Ar:\quad a=\frac{1}{Q}=-1.
\end{equation}

Plugging this into \eref{eq:Phis}, we obtain for $\lambda=1$
\begin{eqnarray}
 t=0: & \tilde{\bPhi}=\left(\begin{array}{cc}
                           \bar\E \pm 2\ii(K-x) & 1\\
                           \E \mp 2\ii(K-x)     & -1
                          \end{array}\right){\mathbf C}
        \quad\textrm{in}\ \tilde\Sigma\ \textrm{with}\ q=\pm 1,\\
 \theta=0: & \tilde{\bPhi} = +2\ii (K-x)
        \left(\begin{array}{cc} D_1 & D_3\\ -D_1 & -D_3\end{array}\right)
     \quad\textrm{in}\ \tilde\Sigma\ \textrm{with}\ q=1,\\
 \theta=\pi: & \tilde{\bPhi} = -2\ii (K-x)
        \left(\begin{array}{cc} \tilde D_1 & \tilde D_3\\ -\tilde D_1 & -\tilde D_3\end{array}\right)
     \quad\textrm{in}\ \tilde\Sigma\ \textrm{with}\ q=-1.
\end{eqnarray}

%
 As we will see below, $C_1(K)$ and $C_2(K)$ are determined completely by  the data at $t=0$. Now we intend to express the components of the matrices ${\bf D}$ and $\tilde{\bf D}$ in terms of $C_1$, $C_2$. For that purpose, we use that $\bPhi$ has to be continuous at the corners $A$ and $B$ of the Gowdy square (see Fig.~\ref{fig:Gowdy}). This condition leads to an algebraic system of $4$ equations which, however, is not sufficient to calculate the $8$ unknowns $D_1,\dots,D_4$, $\tilde{D_1},\dots,\tilde{D_4}$. This is the reason for introducing the coordinate frame $\tilde{\Sigma}$. From the requirement that also $\tilde{\bPhi}$ (in $\tilde{\Sigma}$ with $q=1$) is continuous at $A$ and $\tilde{\bPhi}$ (in $\tilde{\Sigma}$ with $q=-1$) is continuous at $B$ we find another $4$ algebraic equations\footnote{The reason why just the usage of a different coordinate system can lead to independent algebraic equations is the following. The boundary values of the quantity $a$ enter the transformation law \eref{eq:Phis}. 
Therefore, the additional algebraic equations found in $\tilde{\Sigma}$ ensure that $a$ indeed takes on the boundary values \eref{eq:avals} (and, as a consequence, also $Q$ takes on the correct boundary values $Q=1$ on $\Al$ and $Q=-1$ on $\Ar$). From \eref{eq:defb} alone it would only follow that $a$ is constant on the boundaries without specification of these constants.}. In this way we obtain an algebraic system of $8$ equations for the $8$ unknowns with the following solution for the matrices ${\bf D}(K)$ and ${\bf\tilde D}(K)$ in terms of ${\bf C}(K)$:
\begin{eqnarray*}
 D_1 & = C_1 - \frac{b_AC_1+\ii C_2}{2(K-1)},\quad
     & D_2 = C_2 - \frac{\ii b_A(b_AC_1+\ii C_2)}{2(K-1)},\\
 D_3 & = -\frac{\ii\psi}{2(K-1)},\quad
     & D_4 = \psi\left(1+\frac{b_A}{2(K-1)}\right),
\end{eqnarray*}
\begin{eqnarray*}
 \tilde D_1 & = C_1 + \frac{b_B C_1+\ii C_2}{2(K+1)},\quad
  &\tilde D_2 =  C_2 + \frac{\ii b_B(b_B C_1+\ii C_2)}{2(K+1)},\\
 \tilde D_3 & = \frac{\ii\psi}{2(K+1)},\quad
  & \tilde D_4 = \psi\left(1-\frac{b_B}{2(K+1)}\right).
\end{eqnarray*}

In the following subsection we utilize these results to determine the Ernst potential and the metric potentials on $\Al$ and $\Ar$ in terms of the data on $\Hp$.


\subsubsection{Ernst potential on \texorpdfstring{$\Al$}{Al} and \texorpdfstring{$\Ar$}{Al}.\label{Sec:EP}}
From the solution of the LP obtained in the previous subsection we may, as a first step, calculate the Ernst potential on $\Al$ and $\Ar$ in terms
of the initial potential on $\Hp$. To this end, we start by expressing $C_1$ and $C_2$ in terms of $\E$ on $\Hp$. 

As mentioned in Sec.~\ref{sec:LP}, the mapping $K\mapsto\lambda$ in
\eref{eq:lambda} defines a two-sheeted Riemannian $K$-surface. At the
branch points $K_1$ and $K_2$, where both sheets are connected, any function of $K$ has to be unique, i.e.\ the values on the upper and lower
sheet have to be the same. For $t$-$\theta$-values on the boundaries of the
Gowdy square, we have confluent branch points, i.e.\ $K_1=K_2=x$. The uniqueness of $\bPhi$ at $K=K_1=K_2$ leads to the conditions (see \eref{eq:struct})
\begin{equation}\label{eq:gau}
 \Hp,\Al,\Ar:\quad
 \psi_1^>=\psi_1^<\quad\textrm{and}\quad
 \psi_2^>=\psi_2^<\quad
 \textrm{for}\quad K=x.
\end{equation}
In particular, on $\Hp$ we obtain the two equations
\begin{equation}
 \bar\E\p C_1+C_2=\psi,\quad
  \E\p C_1-C_2=\psi
\end{equation}
with the solution
\begin{equation}\fl
 C_1(x)=\frac{2\psi(x)}{\E\p(x)+\bar\E\p(x)}\equiv
        \frac{\psi(x)}{f\p(x)},\quad
 C_2(x)=\frac{\E\p(x)-\bar\E\p(x)}
             {\E\p(x)+\bar\E\p(x)}\psi
       \equiv\frac{\ii b\p(x)}{f\p(x)}\psi(x),
\end{equation}
where
\begin{equation}
 \E\p(x)=f\p(x)+\ii b\p(x) =\E(t=0,\theta=\arccos x).
\end{equation}

Now we can suggest a possible choice for the gauge function $\psi$, which was introduced in \eref{eq:norm}. If we set
\begin{equation}\label{eq:gauge}
 \psi(K)=(K^2-1)^2,
\end{equation}
then the solution $\bPhi$ (as well as $\tilde{\bPhi}$) is regular, because $\psi$ compensates for the poles that the matrices  ${\bf D}$ and  ${\bf\tilde D}$ would otherwise have at $K=\pm 1$. Note that $C_1$ and $C_2$ are also regular because $f\p(x)=\ee^{-u}(1-x^2)$, cf.\ \eref{eq:deff}. But of course, as we will see below, the Ernst potential on $\Al$, $\Ar$ and $\Hf$ is independent of this gauge choice.

With these expressions for $C_1$ and $C_2$ together with the solution of the LP from the previous section, we can also evaluate
the condition \eref{eq:gau} on $\Al$ and $\Ar$ to obtain explicit formulae for the Ernst potential there. The result that we find independently of the particular choice for the gauge function $\psi$ is
\begin{equation}\fl\label{eq:EA1}
 \Al: \quad \E_1(x):=\E(t=\arccos x, \theta=0)
                       = \frac{\ii[b_A-2(1-x)]\E\p(x)+b_A^2}
                        {\E\p(x)-\ii[b_A+2(1-x)]},
\end{equation}
\begin{equation}\fl\label{eq:EA2}
 \Ar:  \quad \E_2(x):=\E(t=\arccos(-x),\theta=\pi)
                       =  \frac{\ii[b_B-2(1+x)]\E\p(x)+b_B^2}
                         {\E\p(x)-\ii[b_B+2(1+x)]}.
\end{equation}

From the latter equations we can conclude that $\E_1$ and $\E_2$ are smooth functions of $x$. To see this, recall that we assume smooth data at $t=0$. For smooth initial functions $u(0,\theta)$ and $Q(0,\theta)$, the initial Ernst potential $\E\p$ will also be smooth, cf.\ Eqs. \eref{eq:fp}, \eref{eq:bp} below. As a consequence, the numerators and denominators of the fractions in \eref{eq:EA1}, \eref{eq:EA2} are also smooth and an irregularity in the Ernst potentials could only occur if the denominators became zero for some $x\in[-1,1]$. However, it follows from \eref{eq:fp}, \eref{eq:bp} below together with Eq.~\eref{eq:QBound} that the only zeros are at $x=1$ or at $x=-1$. Moreover, these equations show that the numerators have zeros of at least the same multiplicity at these $x$-values. Hence, the zeros in the numerators and denominators cancel each other out and the fractions are smooth functions of $x$ for all $x\in[-1,1]$. The only exceptional cases occur for asymptotic data with $b_B=b_A+4$ or $b_B=b_
A-
4$. In the first case, $\E_1$ diverges at point C and in the second case $\E_2$ diverges at D, cf.\ Fig.~\ref{fig:Gowdy}.

\subsubsection{Metric potentials on \texorpdfstring{$\Al$}{Al} and \texorpdfstring{$\Ar$}{Ar}.}

In the previous subsection we have provided explicit formulae for the Ernst potential on the axes $\Al$ and $\Ar$ in terms of the initial potential on $\Hp$. Now we will see how the metric potentials $u$, $Q$ and $M$ can be obtained from the Ernst potential on these boundaries.

We assume that asymptotic data $u(0,\theta)$ and $Q(0,\theta)$ (or, equivalently, $S_{**}(\theta)$ and $Q_{*}(\theta)$) and a constant $R_0>0$ are given. From these data, we may calculate the initial Ernst potential $\E\p=f\p+\ii b\p$. The real part can be obtained from \eref{eq:deff},
\begin{equation}\label{eq:fp}
 f\p(\theta)=\ee^{-u(0,\theta)}\sin^2\!\theta,
\end{equation}
and the imaginary part can be calculated by integrating the first equation in \eref{eq:defb} with respect to $\theta$, using \eref{eq:defa}. We obtain
\begin{equation}\label{eq:bp}
 b\p(\theta)=b_A+2\int_0^\theta Q(0,\theta')\sin\theta'\,\dd\theta',
\end{equation}
where $b_A=b(0,0)$ is an arbitrary integration constant.

From $\E\p$ we may calculate $\E_1$ and $\E_2$ via \eref{eq:EA1}, \eref{eq:EA2}. Afterwards, we can use these results to determine the potentials $u$, $Q$ and $M$ on $\Al$ and $\Ar$. Using again \eref{eq:deff}-\eref{eq:defa} together with \eref{eq:boundarycondM}, \eref{eq:QBound} and \eref{eq:Lu}, we find
\begin{eqnarray}
 \label{eq:pot1}
 \Al: & \quad\ee^{u(t,0)}=\frac{f_1(t)}{\sin^2\!t},\quad
	     \ee^{M(t,0)}=\frac{R_0\sin^2\!t}{f_1(t)},\quad
             Q(t,0)=1,\\
 \label{eq:pot2}
 \Ar: & \quad\ee^{u(t,\pi)}=\frac{f_2(t)}{\sin^2\!t},\quad
	     \ee^{M(t,0)}=\frac{R_0\sin^2\!t}{f_2(t)},\quad
             Q(t,\pi)=-1.
\end{eqnarray}

\subsection{Situation on \texorpdfstring{$\Hf$}{Hf}}
\label{sec:irrID}

So far we have seen that we can prescribe arbitrary smooth data at $t=0$ and will always find smooth potentials for $t<\pi$ as solutions to the field equations. In particular, we have derived explicit formulae for the Ernst potential and the metric potentials on the axes $\Al$ and $\Ar$. It remains to study under which conditions the solution can even be extended smoothly to the future boundary $\Hf$.
In order to answer this question, we tentatively solve the LP on $\Hf$ and investigate whether this solution can be attached continuously to the solutions on $\Al$ and $\Ar$.

The LP on $\Hf$ reduces to the same ODE as on the other boundaries of the Gowdy square, namely to Eq.~\eref{eq:LPBound}. We write the solution in $\Sigma$ as
\begin{equation}\label{eq:L2}
 t=\pi:\quad \bPhi={\bf E}\tilde{\bf C},\quad
        \tilde {\bf C}=\left(\begin{array}{cc}
        \tilde C_1 & \tilde C_3\\ \tilde C_2 & \tilde
        C_4\end{array}\right).
\end{equation}
In order to obtain the solution in the coordinate frame $\tilde\Sigma$ too, we need to calculate the quantity $a$ on $\Hf$ so we can apply the transformation formula \eref{eq:Phis}. It follows from \eref{eq:defa} that, if the metric potentials $u$ and $Q$ remain bounded for $t\to\pi$, then $a=0$  holds  on $\Hf$ (provided $f$ does not vanish on $\Hf$ with exception of the boundary points $C$, $D$). However, it is not yet clear how $u$ and $Q$ behave as $t\to\pi$. Therefore,  so far we can only say that $a$ is constant on $\Hf$, cf.\ \eref{eq:defb},
\begin{equation}
 t=\pi:\quad a=a_0=\textrm{constant}.
\end{equation}
Using \eref{eq:Phis}, we find therefore in $\tilde\Sigma$
\begin{equation}\label{eq:L2a}\fl
 t=\pi:\quad
 \tilde{\bPhi}=\left(\begin{array}{cc}
                      (1\mp a_0)\bar\E \pm 2\ii (K-x) & 1\mp a_0\\
                      (1\mp a_0)\E \mp 2\ii (K-x) & -(1\mp a_0)\ 
                     \end{array}\right)\tilde{\bf C}
 \quad\textrm{in}\ \tilde\Sigma\ \textrm{with}\ q=\pm 1,
\end{equation}

Now we can investigate whether $\bPhi$ in \eref{eq:L2} and $\tilde{\bPhi}$ in \eref{eq:L2a} can be attached continuously to the corresponding solutions on $\Al$ and $\Ar$, i.e.\ whether $\bPhi$ and $\tilde{\bPhi}$ are continuous at the points $C$ and $D$, see Fig.~\ref{fig:Gowdy}. This question is equivalent to the solvability of an algebraic system of $8$ equations, which involves the matrix components $\tilde C_1,\dots,\tilde C_4$ as well as the constant $a_0$ and the values $b_C$ and $b_D$ of the imaginary part of the Ernst potential at $C$ and $D$.  It turns out that this system can be solved if and only if the initial parameters $b_A$ and $b_B$ satisfy
\begin{equation}
  \label{eq:exceptionalcase}
  b_B\neq b_A+4,\quad\textrm{and}\quad b_B\neq b_A-4.
\end{equation}
(It was already discussed at the end of Sec.~\ref{Sec:EP} that the Ernst potential diverges at $C$ or $D$ if one of these conditions is violated.)
The algebraic equations then fix the functions $\tilde C_1(K),\dots,\tilde C_4(K)$ in dependence of our initial data as well as the values of $b_C$, $b_D$ and $a_0$ in terms of the initial quantities $b_A$ and $b_B$. In particular, we obtain
\begin{eqnarray}
 b_C & = & \frac{4b_B+b_A(b_A-b_B)}{b_A-b_B+4},\\
 b_D & = & \frac{-4b_A+b_B(b_A-b_B)}{b_A-b_B-4},\\
 \label{eq:a0}
 a_0 & = & \frac{8(b_B-b_A)}{16+(b_B-b_A)^2}.
\end{eqnarray}
With these results we can calculate the Ernst potential on $\Hf$. With the same considerations as in Sec.~\ref{Sec:EP} we obtain 
\begin{equation}\label{eq:Ef}
 \E\f:=\E(t=\pi,\theta=\arccos(-x))=\frac{a_1(x)\E\p(x)+a_2(x)}{b_1(x)\E\p(x)+b_2(x)},
\end{equation}
where
\begin{eqnarray}
  a_1 & = & -\ii\left[\left((b_A-b_B)^2+16\right)x^2
             -2(b_A-b_B)(b_A+b_B-4)x\right.\nonumber\\
      &   & \qquad+\left.(b_A-b_B)^2+8(b_A+b_B-2)\right],\\
  a_2 & = & 4(b_A-b_B)(b_Ab_B-2b_A-2b_B)x-8(b_A^2+b_B^2),\\
  b_1 & = & 4\left[(b_A-b_B)x-4\right],\\
  b_2 & = & -\ii\left[\left((b_A-b_B)^2+16\right)x^2
             +2(b_A-b_B)(b_A+b_B-4)x\right.\nonumber\\
      &   & \qquad+\left.(b_A-b_B)^2-8(b_A+b_B+2)\right].
\end{eqnarray}
Similarly to the discussion in Sec.~\ref{Sec:EP} we find that $\E\f$ is a smooth function on the entire boundary $\Hf$ (with our assumption \eref{eq:exceptionalcase}).

 As already mentioned, the auxiliary quantity $a$ would satisfy the boundary condition $a=0$ on $\Hf$ if the metric potentials $u$ and $Q$ were bounded for $t\to\pi$. From \eref{eq:a0} we can read off that this is only the case if the initial parameters $b_A$ and $b_B$ satisfy the condition
 \begin{equation}\label{eq:res}
  b_A=b_B,
 \end{equation}
which can also be expressed in terms of the metric potential $Q$ as 
 \begin{equation}\label{eq:res1}
  \int_0^\pi Q(0,\theta)\sin\theta\,\dd\theta=0,
 \end{equation}
 cf.\ \eref{eq:bp}.
In the following subsections, we study separately the cases $b_A=b_B$ and $b_A\neq b_B$.

\subsubsection{Initial data with \texorpdfstring{$b_A=b_B$}{bA equal to bB}.}
Such data lead to a solution of the field equations with $a=0$ on $\Hf$. As a consequence, the metric potentials $u$ and $Q$ are regular at $\Hf$.

The formula \eref{eq:Ef} for $\E\f$ simplifies in this case to
\begin{equation}\label{eq:EHf}
 \Hf:  \quad \E\f(x)
   = \frac{\ii( b_A-1+x^2)\E\p(x)+ b_A^2}
          {\E\p(x)-\ii(b_A+1-x^2)}
\end{equation}
and in terms of this Ernst potential, the metric potentials are given by
\begin{equation}\fl\label{eq:potf}
 \Hf:\quad \ee^{u(\pi,\theta)}=\frac{\sin^2\!\theta}{f\f(\theta)},\quad
	   \ee^{M(\pi,\theta)}=R_0\ee^{2u_A}\frac{\sin^2\!\theta}{f\f(\theta)} ,\quad
           Q(\pi,\theta)=-\frac{\partial_\theta b\f(\theta)}{2\sin\theta},
\end{equation}
where $f\f=\Re\E\f$, $b\f=\Im\E\f$.
Here we have used that $M-u$ is constant on $\Hf$ as a discussion of Eq.~\eref{eq:M2} in the limit $t\to\pi$ reveals.

It follows from these results that $\Hf$ is a regular Cauchy horizon, generated by the Killing vector $\partial_{\rho_1}$ (just like the past horizon $\Hp$). To see this, we can use a modification of the transformation \eref{eq:extensionthroughCH} to regular coordinates in a vicinity of this boundary,
\begin{equation}
  \pi-t=\arcsin\sqrt{t'},\quad \theta=\theta',\quad {\rho_1}={\rho_1}'+\frac
  \kappa {R_0} \ln t',\quad {\rho_2}={\rho_2}'.
\end{equation}
As a consequence of \eref{eq:potf}, the constant $\kappa$ can always be chosen such that the metric is regular in terms of $t'$, $\theta'$, $\rho_1'$, $\rho_2'$. Moreover, $g(\partial_{\rho_1},\partial_{\rho_1})=R_0\ee^{u}\sin^2\!t$ tends to zero for $t\to\pi$, i.e.\ $\Hf$ is indeed a regular null hypersurface and therefore a Cauchy horizon.

\subsubsection{Initial data with \texorpdfstring{$b_A\neq b_B$}{bA not equal to bB}.} Now we study the case $b_A\neq b_B$ and assume that in addition $b_B\neq b_A\pm4$ holds.
In this case, the auxiliary quantity $a$ tends to $a_0\neq0$ as given in \eref{eq:a0} as $t\to\pi$. As a consequence of \eref{eq:defa}, we see that at least one of the metric potentials $Q$ and $u$ cannot be bounded in this limit. Indeed, we can read off from \eref{eq:uQ} that $\ee^u$ diverges as $1/\sin^2\!t$ as $t\to\pi$ for all $\theta\in(0,\pi)$. 

However, it turns out that this divergence is only a peculiarity of our special choice of metric potentials. A better quantity for discussing regularity is the Ernst potential $\E$ which is defined invariantly in terms of the Killing vectors. And indeed, the Ernst potential also remains regular in the entire Gowdy square for  $b_A\neq b_B$. Moreover, from $\E$ one can calculate the Kretschmann scalar $R_{abcd}R^{abcd}$ on $\Hf$ and find that it remains bounded --- with the exception of the earlier discussed special cases $b_B=b_A\pm 4$ which we have excluded here. (For $b_B=b_A+4$, the Kretschmann scalar on $\Al$ behaves as $1/(\pi-t)^{12}$ for $t\to\pi$, and it has the same behavior on $\Ar$ for $b_B=b_A-4$, i.e.\ there occur scalar curvature singularities at the points $C$ or $D$.)

In order to obtain the metric potentials on $\Hf$ in terms of the Ernst potential and the constant $a_0$, we replace the potential $u$ by a potential $v$ in a neighborhood of $\Hf$ via
\begin{equation}
 \ee^{u(t,\theta)}=\frac{\ee^{v(t,\theta)}}{\sin^2\! t}.
\end{equation}
Then we find
\begin{equation}
 \Hf:\quad Q=\frac{1}{a_0},\quad \ee^{v(\pi,\theta)}=a_0^2 f\f(\theta),\quad
     \ee^{M(\pi,\theta)}=c\,\frac{\sin^2\!\theta}{f\f},
\end{equation}
where the integration constant $c$ in the expression for $M$ can be determined from the requirement of a continuous transition to the axes.

In the present case $b_A\neq b_B$, it turns out that $\Hf$ is a regular Cauchy horizon, generated by the linear combination $\partial_{\rho_1}-a_0\partial_{\rho_2}$ of the two Killing vectors. Regular coordinates can be introduced via
\begin{equation}
 \pi-t=\arcsin\sqrt{t'},\quad \rho_1=\rho_1'+\frac{\kappa_1}{R_0}\ln t'\quad
 \rho_2=\rho_2'+\frac{\kappa_2}{R_0}\ln t',
\end{equation}
where $\kappa_1$ and $\kappa_2$ are two constants that can always be chosen such that the resulting metric potentials are regular.

Finally, we may look again at the singular cases $b_B=b_A\pm 4$. As mentioned earlier, the corresponding Ernst potential and the Kretschmann scalar on $\Al$ (for $b_B=b_A+4$) or $\Ar$ (for $b_B=b_A-4$) diverge in the limit $t\to\pi$. Since we therefore cannot find a solution of the LP on $\Hf$ that is continuously connected to the axes, it is not possible to construct the Ernst potential on $\Hf$ in these two singular cases directly. However, in order to study the situation on $\Hf$ in these cases too, we can consider a sequence of solutions with $b_B\neq b_A\pm 4$ that approaches a solution with $b_B=b_A\pm 4$. Then, for each element of the sequence, the LP can be solved along all four boundaries of the Gowdy square and the corresponding expression for the Ernst potential $\E\f$ on $\Hf$, constructed from this solution, is valid. It turns out that the limit $t\to\pi$ of $\E\f$ remains regular for $0<\theta<\pi$, whereas $\E\f$ diverges as expected at $C$ or $D$. Hence we can conclude that only the boundary 
points $C$ or $D$ of $\Hf$ become singular and 
the interior of $\Hf$ is still  a regular null hypersurface.

\section{Discussion}
\label{sec:discussion}

In this paper we have studied the class of \emph{smooth Gowdy symmetric generalized Taub-NUT solutions} as interesting examples of Gowdy spacetimes with spatial $\Sth$ topology.  This class is characterized by a special behavior of the metric potentials in a vicinity of the initial surface $\Hp$ ($t=0$), see Fig.~\ref{fig:Gowdy}, which, in particular, implies that $\Hp$ is a smooth (past) Cauchy horizon. Utilizing Fuchsian methods, we were able to show that for smooth asymptotic data, describing the spacetime at $\Hp$, there  always exists a unique smooth Gowdy symmetric generalized Taub-NUT spacetime as a solution to the Einstein equations for $t\in(0,\pi)$. In a second step, we have investigated the behavior of these solutions on the symmetry axes $\Al$ ($\theta=0$) and $\Ar$ ($\theta=\pi$). Using the complex Ernst formulation of the field equations and its reformulation in terms of an equivalent linear problem, we have constructed explicit formulae for the metric potentials on $\Al$ and $\Ar$ in terms of 
the data on $\Hp$. Afterwards, it was possible to extend the solution to the future boundary $\Hf$ ($t=\pi$) of the Gowdy square and to find explicit expressions for the metric potentials there, too.
It followed from these expressions, that we have to distinguish between four types of asymptotic data, which  are characterized by the values $b_A$ and $b_B$ of the imaginary part $b$ of the Ernst potential at the points $A$ ($t=\theta=0$) and $B$ ($t=0$, $\theta=\pi$) and which lead to solutions with a completely different behavior on $\Hf$:

\begin{enumerate}
 \item $b_B=b_A+4$:\\
 In this case a scalar curvature singularity occurs at the point $C$ ($t=\pi$, $\theta=0$).

 \item $b_B=b_A-4$:\\
 Here, a scalar curvature singularity occurs at the point $D$ ($t=\theta=\pi$).

 \item $b_B=b_A$:\\
 The spacetime is regular in the entire Gowdy square. In particular, the Ernst potential $\E$ and the metric potentials $u$, $Q$ and $M$ are smooth. Moreover, $\Hf$ is a smooth Cauchy horizon, generated by the Killing vector $\partial_{\rho_1}$. 

 \item $b_B\neq b_A$ and $b_B\neq b_A\pm4$:\\
 The spacetime is regular in the entire Gowdy square and the Ernst potential $\E$  is smooth, but the metric potential $u$ is not well adapted to describing this case and blows up at $\Hf$. However, there is no physical singularity at $\Hf$. Instead, $\Hf$ is a smooth Cauchy horizon, generated by the null vector $\partial_{\rho_1}-a_0\partial_{\rho_2}$.
\end{enumerate}
This shows that --- with the exception of the two singular cases (i) and
(ii) --- smooth Gowdy symmetric generalized Taub-NUT solutions (with a
past Cauchy horizon at $t=0$) always develop a second Cauchy horizon
at $t=\pi$. This future Cauchy horizon, in the same way as the one in
the past, is homeomorphic to \Sth and its null generator has closed
integral curves. Hence our results can in particular be understood as
a partial resolution of a problem which remained open in
\cite{Chrusciel:1990ti}. Namely, at least in our class of spacetimes, the
Gowdy square is isometric to the closure of the MGHD of corresponding Cauchy data.

It is interesting to compare these results with the situation of
spatial $\St\times\So$ topology as investigated in
\cite{Hennig2010}. For this case, it was shown that Gowdy
spacetimes with a regular past Cauchy horizon $\Hp$ develop a regular
future horizon $\Hf$ if and only if a particular quantity $J$, which
can be read-off from the asymptotic data\footnote{In terms of the
  Ernst potential at $\Hp$, $J$ is defined as
  $J=-\frac{1}{8Q\p^2}(b_A-b_B-4Q\p)$, where $Q\p$ denotes the
  constant value of the metric potential $Q$ on $\Hp$ in the
  $\St\times\So$ case.}, does not vanish. In the limit $J\to 0$, $\Hf$
transforms into a scalar curvature singularity. Hence, the behavior
is similar to the $\Sth$ case: with the exception of non-generic singular cases,
spacetimes with a past Cauchy horizon generically develop a future
Cauchy horizon. However, the nature of the singular cases is slightly
different: In the $\St\times\So$ case, the curvature blows up along
the entire future boundary $\Hf$, whereas we find only singularities
at the isolated points $C$ or $D$ on $\Hf$ for $\Sth$ topology.

Do our assumptions rule out important cases? This is not clear and
probably difficult to answer. For example our assumptions do not allow
solutions with past Cauchy horizons ruled by non-closed generators. There
is no reason why such solutions should not exist. Also solutions with  non-compact or
incomplete Cauchy horizons at $t=0$ are excluded by our assumptions so far. Indeed such solutions have been constructed
in the polarized case by the techniques of Moncrief and Isenberg in
\cite{Isenberg:1990gn}. Since the geometry of \Sth-Gowdy spacetimes away from the axes is locally the same as the geometry of $\mathbb T^3$-Gowdy spacetimes, it seems possible to  construct $\Sth$-Gowdy solutions which are smooth concatenations of solutions with pieces of Cauchy horizons covering a neighborhood of the axes of the $t=0$-surface (by means of the same Fuchsian method which we have employed in this paper; in particular the matching conditions would be satisfied) and pieces of solutions which are singular away from the axes at $t=0$ (by means of the Fuchsian method for $\mathbb T^3$-Gowdy solutions). The question of how such solutions evolve globally in time is of course completely open.

Can our results be generalized to situations with less symmetry and
eventually maybe even to generic solutions with Cauchy horizons? We
have employed two, in principle, independent techniques for the two
main steps of our discussion: the Fuchsian method for the basic
existence proof and the soliton method for the study of the global
properties of the solutions. As far as the Fuchsian method and the
underlying singular initial value problem and hence the existence and
uniqueness of solutions with prescribed ``data'' on a Cauchy horizon
is concerned, we can say the following. In general, it cannot be
expected that such an ``initial value problem'' for equations of
hyperbolic type is well-posed. It seems at least necessary that the
generator of the horizon, being a null hypersurface, is a Killing
field. This is the case here and also in the more general
$U(1)$-symmetric case discussed by Moncrief
\cite{Moncrief:1984js}. 
The results in \cite{Moncrief:1983ir,Friedrich:1999de} suggest that there must
be a $U(1)$-symmetry in a neighborhood of a Cauchy horizon in general
vacuum spacetimes, at least if the horizon is compact and the
generator has closed integral curves. So, there is hope that a
similar singular initial value problem can be formulated under quite
general assumptions and that the existence and uniqueness proof
based on Fuchsian methods goes through. 

As far as the Ernst formulation and the associated linear problem are concerned there is probably little hope of a  generalization to situations with fewer symmetries. The introduction of the complex Ernst potential relies essentially on the existence of two Killing vectors, cf.\ \eref{eq:deff}, \eref{eq:defa}, and the linear problem makes use of the special structure of the Ernst equation. However, it should be quite straightforward to apply the methods to Gowdy symmetric spacetimes with additional \emph{electromagnetic fields}. In that case one has to study the coupled system of the Einstein-Maxwell equations, for which, remarkably, a complex Ernst formulation and an associated linear problem exist as well. The corresponding calculations would follow closely the investigation of axisymmetric and stationary black hole spacetimes with electromagnetic fields as presented in \cite{Hennig:2009tj}.

\ack
We would like to thank Helmut Friedrich for valuable discussions and Gerrard Liddell for reading the manuscript
carefully. F.B.\ likes to thank the Albert Einstein Institute in
Potsdam, where part of the work was done, for the invitation and the
hospitality.

\appendix

\renewcommand{\thesection}{\Alph{section}}

\section{Semilinear Fuchsian wave equations on Riemannian manifolds}
\label{sec:backgroundFuchsian}

One of the main tasks of \Sectionref{sec:analyseequations} is to analyze a singular initial value problem of a coupled system of nonlinear wave equations \eref{eq:FuchsS}--\eref{eq:Fuchsomega}. The idea is to use the Fuchsian method which is introduced in \cite{Ames:yV5l9m6A} for general quasilinear symmetric hyperbolic Fuchsian equations and which was based on the Fuchsian theory for semilinear second-order hyperbolic Fuchsian equations in \cite{Beyer:2010fo,Beyer:2010tb}. However, these earlier discussions are restricted to the case where  the spatial topology is a $1$-dimensional circle; the straightforward generalization to the case of an $n$-torus has been considered in \cite{Ames:uh}. Here we discuss the case of more general smooth compact Riemannian manifolds of which the 2-sphere $\St$ in our applications is a particular example. In order to make the presentation as short as possible we restrict to that particular class of equations which is of interest in \Sectionref{sec:analyseequations}: 
semilinear wave equations. In summary, the main difference to the case of spatial $n$-torus topology will be the definition of the function spaces and the way, the energy estimates are obtained. Despite the fact that we deal with second-order equations here, while  \cite{Ames:yV5l9m6A,Ames:uh} focuses on first-order equations, almost all of the other details of the proofs stay the same.

An alternative approach is the one by St\aa hl \cite{Stahl:2002bv}, who has modified the Fuchsian theory in \cite{Kichenassamy:1999kg, Rendall:2000ki} to treat the spatial manifold $\St$. There are several reasons why we do not follow his approach here. One of them is that we do know how to go from semilinear equations to quasilinear equations in our theoretical framework above. It is therefore likely that our discussion here can also be generalized to the quasilinear case. This will be of interest in future work when we study solutions of Einstein's field equations under more general conditions and more general topologies than before. On the other hand, our technique above gives rise to a numerical approximation scheme for the singular initial value problem with practical error estimates. In future work, we plan to compute Gowdy symmetric generalized Taub NUT solutions numerically using our framework.

\paragraph{Notation and function spaces.}
We use the definition of Sobolev spaces on Riemannian manifolds given in Appendix~I of \cite{ChoquetBruhat:2008te}, which we quickly summarize as follows; further details can be found in \cite{Aubin:1982tw}. Let $(H,h_{ab })$ be a smooth orientable $n$-dimensional Riemannian manifold. As before, the indices $a,b $ etc.\ are abstract tensor indices with respect to $H$.
Let $f:H\rightarrow \R^d$ be a $d$-vector-valued function for some integer $d\ge 1$. For any integer $s\ge 0$, we can define $\nabla_{a_1}\cdots\nabla_{a_s} f$ by considering the usual covariant derivative $\nabla$ compatible with $h_{ab}$ which acts on each of the $d$ components of $f$ individually. The result $\nabla_{a_1}\cdots\nabla_{a_s} f$  is then understood as a $d$-vector-valued $(0,s)$-tensor field. In the following, when we write expressions where two such objects are multiplied, e.g.\ $\nabla_{a_1}\cdots\nabla_{a_s} f\, \nabla_{a_1'}\cdots\nabla_{a_s'} f$, we, first, evaluate each of the two factors componentwise and then take the Euclidean scalar product of the resulting $d$-vector-valued quantities, i.e.\
\[\nabla_{a_1}\cdots\nabla_{a_s} f\, \nabla_{a_1'}\cdots\nabla_{a_s'} f
:=\sum_{i=1}^d\nabla_{a_1}\cdots\nabla_{a_s} f^i\, \nabla_{a_1'}\cdots\nabla_{a_s'} f^i,\]
the result being a scalar-valued $(0,2s)$-tensor field.
Our notation in the following will hence mimic the notation for scalar-valued quantities as much as possible, but it should always be kept in mind that many of the following quantities will actually be $d$-vector-valued.

Now pick a non-negative integer $q$. Let $f$ and $g$ be $d$-vector-valued functions such that the $d$-vector-valued tensor fields $\nabla_{a_1}\cdots\nabla_{a_s} f$, $\nabla_{a_1}\cdots\nabla_{a_s} g$ exist in the distributional sense and are square-integrable for all $s=0,\ldots,q$, in the sense that
\[\scalarpr{f}{g}_{H^q(H)}:=\left(\int_H\sum_{s=0}^q h^{a_1a_1'}\cdots h^{a_sa_s'} \nabla_{a_1}\cdots\nabla_{a_s} f\, \nabla_{a_1'}\cdots\nabla_{a_s'} g\,\epsilon\right)^{1/2},\]
is well-defined and finite. Here $\epsilon=\epsilon_{a_1\ldots a_{n}}$ is a volume form corresponding to $h_{ab }$. The summand $s=0$ is understood to involve no covariant derivatives and hence no contractions with the contravariant metric. 
 We also set
\[\|f\|_{H^q(H)}:=\left(\scalarpr{f}{f}_{H^q(H)}\right)^{1/2},\]
whenever this is finite.
Then, the \keyword{Sobolev space} $H^q(H)$ of functions $f$ with $\|f\|_{H^q(H)}<\infty$ equipped with the norm $\|\cdot\|_{H^q(H)}$ is a Banach space. In the case $q=0$, we write $L^2(H)$ instead of $H^0(H)$. We note that this definition of Sobolev spaces is purely geometric, i.e.\ it does not depend on the choice of coordinates. If $(H,h)$ is complete, then $H^q(H)$ equals the completion of the space of smooth functions with compact support with the norm $\|\cdot\|_{H^q(H)}$.
We also point the reader to the following properties \cite{ChoquetBruhat:2008te} which we will use frequently below.
 
\begin{lemma}
\label{lem:sobolevRiemannian}
Suppose that $(H,h_{ab })$ is a complete orientable smooth Riemannian manifold of dimension $n$ and that $q$ is an integer larger than $n/2$. Then:
\begin{enumerate}[label=\textit{(\roman{*})}, ref=(\roman{*})]   
\item $H^q(H)$ is continuously imbedded into $\bar C^0(H)$, i.e.\ the Banach space of continuous and bounded functions on $H$ with the supremum norm, and there exists a constant $C>0$ so that \[\sup_{x\in H} |f|\le C \|f\|_{H^q(H)},\]
for all $f\in H^q(H)$.
\item For all $f,g\in H^{q}(H)$, the pointwise product $f\cdot g$ is in $H^{q}(H)$ and 
\[\|f\cdot g\|_{H^{q}(H)}\le C \|f\|_{H^{q}(H)} \|g\|_{H^{q}(H)},\]
for a constant $C>0$. Hence $H^{q}(H)$ is a Banach algebra.
\end{enumerate}
\end{lemma}

Now, in accordance with \cite{Ames:yV5l9m6A,Ames:uh}, we define spaces $X_{\delta,\mu,q}$ of time-dependent $d$-vector-valued functions on $H$ as follows. We do this parametrically, specifying i) a non-negative constant $\delta$, ii) a non-negative integer $q$, and iii) a smooth $d$-vector-valued function $\mu: H\rightarrow \R^d$ with the corresponding diagonal matrix-valued function  
\begin{equation}
  \label{eq:defR}
 \RR{\mu}(t,x):=\mathrm{diag}\, \bigl(t^{-\mu_1(x)},\ldots,t^{-\mu_d(x)}\bigr), 
\end{equation}
and then defining the norm
\begin{equation*}
  \|w\|_{\delta,\mu,q}:=\sup_{t\in (0,\delta]}
    \|\RR{\mu}(t,\cdot) w(t,\cdot)
    \|_{H^q(H)},
\end{equation*}
for appropriate $d$-vector-valued time-dependent functions $w(t,x)$. Based on the norm $\|..\|_{\delta, \mu,q}$, we define the Banach space $X_{\delta, \mu,q}(H)$ -- also simply written as $X_{\delta, \mu,q}$ -- as the completion of the set of all $d$-vector-valued functions $w\in C^\infty((0,\delta]\times H)$ for which this norm is finite.  We use $B_{\delta,\mu,q,r} \subset X_{\delta,\mu,q}$ to denote the closed ball of radius $r>0$ (measured using the norm  $\|\cdot \|_{\delta,\mu,q}$)
and center $0$. To handle the class of functions which are infinitely differentiable, we define the space 
\begin{equation*}
X_{\delta,\mu,\infty}:=\bigcap_{q=0}^\infty X_{\delta,\mu,q}.
\end{equation*}
In comparing a pair of function spaces $X_{\delta,\mu,q}$ and $X_{\delta, \nu,q}$ for functions with the same dimension $d$, we find it useful to write $\nu>\mu$ to denote
the condition that for each index $i=1,\ldots,d$ and for all $x\in H$, the components of $\nu$ and $\mu$ satisfy the inequality
$\nu_{i}(x)>\mu_i(x)$. Note that if $\nu>\mu$, then $X_{\delta,\nu,q}
\subset X_{\delta,\mu,q}$. 
It is often useful to consider functions $f$ in any of the spaces $X_{\delta,\mu,q}$ as maps $f: (0,\delta]\rightarrow H^q(H)$. In this context, we often use both notations $f(t)$ and $f(t,\cdot)$ to denote the corresponding element in $H^q(H)$ at any $t\in (0,\delta]$.

We also point the reader to the appendix in \cite{Ames:yV5l9m6A}, where we list further properties of the above function spaces. All those results there easily generalize to the case of $n$-dimensional Riemannian manifolds. Only at places, where part (ii) of \Lemref{lem:sobolevRiemannian} above is used, i.e.\ in Lemmas~B.1--B.4 in the appendix of \cite{Ames:yV5l9m6A}, we must replace the lower bound ``$\ge 1$'' for the order of differentiability by ``$>n/2$''.

In our context of \textit{second}-order wave equations, we will see that the following spaces are also convenient. Let $u$ be a time-dependent $d$-vector-valued function on $(0,\delta]\times H$ such that $u$, $Du$, $t\nabla_a u$ are defined and are elements of $X_{\delta,\mu,q}$ for some exponent vector $\mu$ and some integer $q\ge 0$. Then we say that $u\in \tilde X_{\delta,\mu,q}$ and we set
\[\|u\|_{\delta,\mu,q}^\sim:=\|u\|_{\delta,\mu,q}+\|Du\|_{\delta,\mu,q}+\|t\nabla _a u\|_{\delta,\mu,q}.\]
Again, we find that $\tilde X_{\delta,\mu,q}$, also written as  $\tilde X_{\delta,\mu,q}(H)$, equipped with the norm $\|\cdot\|_{\delta,\mu,q}^\sim$, is a Banach space. We use the  notation analogous to that before for closed bounded subsets $\tilde B_{\delta,\mu,q,r}$, and for the spaces $\tilde X_{\delta,\mu,\infty}$.

\paragraph{Semilinear Fuchsian wave equations on $H$ and singular initial value problems.}
From now on, we assume that $(H,h_{ab })$ is compact in addition to the assumptions before. It is a consequence of the Hopf Rinow Theorem, see Theorem~21 and Corollary~23 in \cite{ONeill:1983ua}, that compact Riemannian manifolds are complete.
We restrict to equations of the form
\begin{equation}
  \label{eq:1stordersystem}
  D^2u+2A Du-t^2\Delta_{h} u=f(t,x,u,Du,t\nabla_a u),
\end{equation}
in the following --
the system \Eqsref{eq:FuchsS}--\eref{eq:Fuchsomega} is a particular example -- where $u: (0,\delta]\times H\rightarrow\R^d$ is the time-dependent $d$-vector-valued unknown, $A$ is a $d\times d$-diagonal matrix -- which we assume to be a constant here for simplicity (and which acts by matrix multiplication on the vector-valued function $Du$). The Laplace operator  is
$\Delta_{h}:=h^{ab }\nabla_a\nabla_b$, 
and $f$ is the source term which depends as a $q$-times continuously differentiable vector-valued function on $t$, $x$, $u$, $Du$ and the $1$-form $t\nabla_a u$; the integer $q$ will be fixed later. The latter means that, with respect to any local coordinate patch, $f$ depends as a $q$-times continuously differentiable function in particular on the $n$ coordinate components of $t\nabla_a u$. 
Such a system of equation is referred to as a \keyword{semilinear Fuchsian wave equation on $H$}, and we write
\begin{equation}
  \label{eq:DefLPDE}
  L[u]:= D^2u+2A Du-t^2\Delta_{h} u,
\end{equation}
for its \keyword{principal part}.

Clearly, these equations are singular at $t=0$ and we wish to discuss the properties of solutions at this singular time.
Given the parameters $\delta,\mu,q$ as in the above discussion of the function space $X_{\delta,\mu,q}$, and a chosen function $u_0:(0,\delta]\times H\rightarrow\R^d$ (with so far unspecified regularity), the \keyword{singular initial value problem} consists of seeking a solution $u=u_0+w$ to \Eqref{eq:1stordersystem} whose \keyword{remainder} function $w$ belongs to
 $X_{\delta,\mu,q}$. This problem is said to be \keyword{well-posed} if, for each choice of $u_0$ in some given function space (discussed below), there exists a unique remainder solution $w\in X_{\delta,\mu,q}$. As soon as a leading-order term $u_0$ has been fixed, we use the operator notation
\begin{equation}
\label{eq:defF}
  F(u_0)[w](t,x):=f\left(t,x,u_0+w,D(u_0+w),t\nabla_a (u_0+w)\right).
\end{equation}
The idea is therefore to prescribe the leading-order behavior of a solution at $t=0$ by fixing the \keyword{leading-order term} $u_0$ and then to constructing solutions which approach $u_0$ at some specified rate $\mu$, i.e.\ whose remainder $w$ is in $X_{\delta,\mu,q}$.

Our aim is to prove the following central theorem. 

\begin{theorem}[Well--posedness for semilinear Fuchsian wave equations on $\St$]
  \label{th:Wellposedness1stOrderFiniteDiff}
  Let $H=\St$ and $h=\dd\vartheta^2+\sin^2\!\vartheta \,\dd\phi^2$.
  Choose a leading-order term $u_0$, a constant $\delta>0$, an exponent vector $\mu$, and an integer $q>2$. Then there exists a unique solution $u$ to \Eqref{eq:1stordersystem} whose remainder $w:=u-u_0$ belongs to $\tilde X_{\bar\delta, \mu, q}$ with $D^2 w \in X_{\bar \delta, \mu, q-1}$ for some $\bar \delta \in (0, \delta]$, provided the following structural conditions are satisfied:
  \begin{enumerate}[label=\textit{(\roman{*})}, ref=(\roman{*})] 
  \item \label{en:cond2}%
   The \keyword{energy dissipation matrix}
   \begin{equation}
     \label{eq:energydissipationmatrix}
     M_0(x):= \left(\begin{array}{ccc}
       M(x) & -1/2 & 0\\
         -1/2 & M(x)+2A & 0\\
         0 & 0 & M(x)-1
       \end{array}\right),
   \end{equation}
   with $M(x)= \diag(\mu_1(x),\ldots,\mu_d(x))$,
   is positive definite for all $x\in H$.
  \item \label{en:cond4N}%
    The map
    \begin{equation}
      \label{eq:defFLu}
      F _{red} (u_0): w\mapsto F(u_0)[w]-L[u_0]
    \end{equation}
    with $F(u_0)$ given by \Eqref{eq:defF}
    is well-defined, and maps $w\in \tilde X_{\delta,\mu,q}$ to
    $X_{\delta,\nu,q}$ for some exponent vector $\nu>\mu$.
  \item \label{cond:LipschitzF}%
    For each $s>0$ and for all $\delta'\in(0,\delta]$, there exists a constant $C>0$ with
    \begin{equation*}
        \|F _{red} (u_0)[w]-F _{red} (u_0)[\tilde w]\|_{\delta',\nu,q}        
        \le C \|w-\tilde w\|_{\delta',\mu,q},
    \end{equation*}    
    for all $w, \tilde w\in \tilde B_{\delta',\mu,q,s}$. The constant $C$ may
    depend on $s$, but, in particular, not on $\delta'$.  
  \end{enumerate}
  If all of the previous conditions are satisfied for all values of
  $q>2$, then there exists a unique solution $u$ of
  \Eqref{eq:1stordersystem} with $u-u_0\in \tilde X_{\bar \delta,\mu,\infty}$.
\end{theorem}

The result actually holds for general smooth oriented connected compact Riemannian $n$-dimensional manifolds; the condition $q>2$ above must then be replaced by $q>n/2+1$.

The reader may want to compare this to Theorem~2.4 in \cite{Ames:yV5l9m6A} and its $n$-dimensional version in \cite{Ames:uh}. The hypotheses simplify here thanks to our restriction to \textit{semilinear equations with smooth coefficients}. A further difference is that we consider second-order equations here. Apart from the necessity of introducing the spaces $\tilde X_{\delta,\mu,q}$, this, however, does not lead to any more complications than for the theory for first-order equations.

We remark that in its applications, this theorem often allows one to find an open set of values for the exponent vector $\mu$ for which the singular initial value problem is well-posed. A lower bound for this set can originate in
 \Conditionref{en:cond2}, while an upper bound is usually determined by \Conditionref{en:cond4N}. Although those bounds on the set of allowed values for $\mu$ may not be sharp, both bounds provide useful information on the problem. The upper bound for $\mu$ specifies  the smallest regularity space and, hence, the most precise description of the behavior of $w$ (in the limit $t\searrow 0$), while the lower bound for $\mu$ determines the largest space in which the solution $u$ is guaranteed to be unique. We note  that this \textit{uniqueness} property must be interpreted with care: under the conditions of our theorem, there is exactly one solution $w$ in the space $X_{\delta,\mu,q}$, although we do not exclude the possibility that another solution may exist in a larger space, for example, in $X_{\delta,\tilde\mu,q}$ with $\tilde\mu<\mu$. 

We also remark that results analogous to those stated in Theorem \ref{th:Wellposedness1stOrderFiniteDiff}
for $D^2(u-u_0)$ 
 can be derived for higher time derivatives if the regularity given by $q$ is sufficiently large. 

\paragraph{Outline of the proof of \Theoremref{th:Wellposedness1stOrderFiniteDiff}.}
We follow the strategy outlined in detail in \cite{Ames:yV5l9m6A,Ames:uh}; the proof hence consists of the following steps. First, we consider linear equations and derive energy estimates for the Cauchy problem with data at any $t>0$. Secondly, we use these energy estimates to prove convergence of a certain sequence of approximate solutions with respect to the spaces defined before, which implies the existence of solutions of the singular initial value problem, and then to improve the regularity of the solutions. Uniqueness is obtained by employing the energy estimates one more time. In a third main step, we use these results for linear equations to set up a fixed point iteration where the nonlinear equations are linearized. The resulting limit is then the unique solution of the singular initial value problem for general nonlinear equations of the form \Eqref{eq:1stordersystem}. 
We discuss all this in more detail now.

\subparagraph{Step 1. Energy estimates for the Cauchy problem of linear tensorial Fuchsian wave equations.} 
We start by considering particular linear Fuchsian wave equations derived from \Eqref{eq:1stordersystem}. We see below that it is useful to consider linear equations not only for time-dependent $d$-vector-valued \textit{scalar} quantities, as in \Eqref{eq:1stordersystem}, but equations for time-dependent $d$-vector-valued $(0,s)$-tensor fields $u_{a_1\ldots a_s}$ for some $s\ge 0$; the scalar case is recovered for $s=0$. The class of equations, which is relevant for this step of the proof, is
\begin{equation}
  \label{eq:lineartensorialsystem}
  D^2u_{a_1\ldots a_s}+2A Du_{a_1\ldots a_s}
-t^2\Delta_{h} u_{a_1\ldots a_s}=f_{a_1\ldots a_s}+t^2 G(u)_{a_1\ldots a_s},
\end{equation}
where $f_{a_1\ldots a_s}$  is some time-dependent vector-valued
$(0,s)$-tensor field. The map $G$ is linear and built from contractions of
$u_{a_1\ldots a_s}$ with other smooth scalar-valued time-independent
tensor fields; in our particular case those are the Riemann tensor of the metric $h_{ab}$ and its covariant derivatives, as we see below. For any $u_{a_1\ldots a_s}$, it follows that $G(u)_{a_1\ldots a_s}$ is a time-dependent $d$-vector-valued $(0,s)$-tensor field with the same regularity as $u_{a_1\ldots a_s}$.
Under these assumptions, we refer to \Eqref{eq:lineartensorialsystem} as a \keyword{linear tensorial Fuchsian wave equation}. 

Before we turn our attention to \textit{singular} initial value problems, we focus on the \textit{Cauchy problem} of \Eqref{eq:lineartensorialsystem}. If $f_{a_1\ldots a_s}$ is smooth,  then it is a consequence of $G$ mapping smooth to smooth and Theorem~12.17 in \cite{Ringstrom:2009cj} that the Cauchy problem is well-posed: if we prescribe smooth fields $u^{*}_{a_1\ldots a_s}$ and $u^{**}_{a_1\ldots a_s}$ at some $t_0\in (0,\delta)$, then there exists a unique smooth solution $u_{a_1\ldots a_s}$ on $[t_0,\delta]\times H$ of \Eqref{eq:lineartensorialsystem} such that $u_{a_1\ldots a_s}(t_0)=u^{*}_{a_1\ldots a_s}$ and $\partial_t u_{a_1\ldots a_s}(t_0)=u^{**}_{a_1\ldots a_s}$. 

Next we derive energy estimates for such smooth solutions of the Cauchy problem of \Eqref{eq:lineartensorialsystem}.
We fix a non-negative integer $s$ and assume that $u_{a_1\ldots a_s}$ is a time-dependent smooth vector-valued tensor field. Let us choose two positive constants $\kappa$, $\gamma$ and define, for a given exponent vector $\mu$, the following \keyword{energy}
\begin{eqnarray*}
\fl
\quad E_{\kappa,\gamma,\mu}[u_{a_1\ldots a_s}]
  :=\frac 12 \ee^{-\kappa t^\gamma}\Bigl(\scalarpr{\RR{\mu}u_{a_1\ldots a_s}}{\RR{\mu}u_{a_1\ldots a_s}}_{L^2(H)}\\
\fl
\qquad+\scalarpr{\RR{\mu}Du_{a_1\ldots a_s}}{\RR{\mu}Du_{a_1\ldots a_s}}_{L^2(H)}
+\scalarpr{\RR{\mu}t\nabla_b u_{a_1\ldots a_s}}{\RR{\mu}t\nabla_b u_{a_1\ldots a_s}}_{L^2(H)}\Bigr);
\end{eqnarray*}
recall the definition of $\RR{\mu}$ in \Eqref{eq:defR}.

\begin{lemma}[Basic energy estimates for the Cauchy initial value problem]
  \label{lem:basicenergyestimate}
  Suppose that $(H,h_{ab})$ is a smooth connected orientable compact Riemannian manifold.
  Pick a constant $\delta>0$ and an exponent vector $\mu$.
  Suppose that \Eqref{eq:lineartensorialsystem} is a linear tensorial Fuchsian wave equation for smooth $f_{a_1\ldots a_s}$ so that the {energy dissipation matrix} given by \Eqref{eq:energydissipationmatrix}
is positive definite for all $x\in H$. Then there exist positive constants $\kappa$, $\gamma$, and $C$ such that for any smooth initial data $u^{*}_{a_1\ldots a_s}$ and $u^{**}_{a_1\ldots a_s}$ at $t=t_0\in (0,\delta)$, the solution of the Cauchy problem $u_{a_1\ldots a_s}$ for this system and this initial data satisfies the energy estimate
\begin{equation}
  \fl
  \label{EnergyEstimate}
  \quad\sqrt{E_{\mu,\kappa,\gamma}[u_{a_1\ldots a_s}](t)}
  \le \sqrt{E_{\mu,\kappa,\gamma}[u_{a_1\ldots a_s}](t_0)}
  +C\int_{t_0}^t s^{-1}\|\RR{\mu}(s) f_{a_1\ldots a_s}(s)\|_{L^2(T^1)}\,\dd s,
\end{equation}
for all $t\in [t_0,\delta]$. The constants $C$, $\kappa$, and $\gamma$ may be chosen  independently of $f_{a_1\ldots a_s}$. If one replaces the data $u^{*}_{a_1\ldots a_s}$ and $u^{**}_{a_1\ldots a_s}$ at $t=t_0\in (0,\delta)$ by other smooth data specified at any time $t_1 \in (0, t_0]$, then the energy estimate holds for the \emph{same} constants $C$, $\kappa$, $\gamma$.
\end{lemma}

Before proving this lemma, we make a few remarks.
First, this lemma does \textit{not} imply that the energy estimate
\Eqref{EnergyEstimate} holds for $t<t_0$, in particular not for the limit $t\searrow 0$. 
Finally, we point out that the map $G$ does not appear explicitly in the energy estimate; the constants, however, will in general depend on $G$.

Note that we can clearly find constants $C_1$ and $C_2$, which do not depend on $t$, $\mu$ and $u_{a_1\ldots a_s}$, so that at every $t\in (0,\delta]$
\begin{eqnarray}
\fl
C_2\Bigl(\|\RR{\mu}(t)u_{a_1\ldots a_s}(t)\|_{L^2(H)}
+\|\RR{\mu}(t)Du_{a_1\ldots a_s}(t)\|_{L^2(H)}
+\|\RR{\mu}(t)t\nabla_bu_{a_1\ldots a_s}(t)\|_{L^2(H)}\Bigr)\nonumber\\
\label{eq:equivalencenormenergy}
\fl\qquad
\le \sqrt{E_{\kappa,\gamma,\mu}[u_{a_1\ldots a_s}](t)}\\
\fl\qquad
\le C_1\Bigl(\|\RR{\mu}(t)u_{a_1\ldots a_s}(t)\|_{L^2(H)}
+\|\RR{\mu}(t)Du_{a_1\ldots a_s}(t)\|_{L^2(H)}\nonumber\\
\fl\qquad\qquad\qquad
+\|\RR{\mu}(t)t\nabla_bu_{a_1\ldots a_s}(t)\|_{L^2(H)}\Bigr).\nonumber
\end{eqnarray}
Hence, we can reformulate \Eqref{EnergyEstimate} purely in terms of spatial $L^2$-norms.

\begin{proof}
Under our assumptions, we are allowed to differentiate under the
integral and thereby obtain (writing $E$ instead of $ E_{\kappa,\gamma,\mu}[u_{a_1\ldots a_s}]$)
\begin{eqnarray*}
  DE&=&-\kappa \gamma t^\gamma E\\
&&+\e^{-\kappa t^\gamma}\Bigl(\scalarpr{(D\RR{\mu})u_{a_1\ldots a_s}+\RR{\mu}Du_{a_1\ldots a_s}}{\RR{\mu}u_{a_1\ldots a_s}}_{L^2(H)}\\
&&\quad +\scalarpr{(D\RR{\mu})Du_{a_1\ldots a_s}+\RR{\mu}D^2u_{a_1\ldots a_s}}{\RR{\mu}Du_{a_1\ldots a_s}}_{L^2(H)}\\
&&\quad +\scalarpr{D(\RR{\mu}t)\nabla_b u_{a_1\ldots a_s}+\RR{\mu}t\nabla_b Du_{a_1\ldots a_s}}{\RR{\mu}t\nabla_b u_{a_1\ldots a_s}}_{L^2(H)}\Bigr).
\end{eqnarray*}
Since $u_{a_1\ldots a_s}$ is a solution of our equation, we have
\begin{eqnarray*}
\fl
\quad\scalarpr{\RR{\mu}D^2u_{a_1\ldots a_s}}{\RR{\mu}Du_{a_1\ldots a_s}}_{L^2(H)}\\
\fl
\qquad=\scalarpr{\RR{\mu}\Bigl(-2A Du_{a_1\ldots a_s}
+t^2\Delta_{h} u_{a_1\ldots a_s}+f_{a_1\ldots a_s}+t^2 G(u)_{a_1\ldots a_s}\Bigr)}{\RR{\mu}Du_{a_1\ldots a_s}}_{L^2(H)}.
\end{eqnarray*}
Before we proceed, let us now consider the following expression:
\begin{eqnarray*}
\fl
\quad  t^2 (\RR{\mu}\Delta_{h} u_{a_1\ldots a_s})(\RR{\mu} Du^{a_1\ldots a_s})
+t^2(\RR{\mu}\nabla_{b }Du_{a_1\ldots a_s})(\RR{\mu}\nabla^b  u^{a_1\ldots a_s})\\
\fl
\qquad  =t^2 (\RR{\mu}\nabla_b \nabla^b u_{a_1\ldots a_s})(\RR{\mu} Du^{a_1\ldots a_s})
+t^2(\RR{\mu}\nabla_{b }Du_{a_1\ldots a_s})(\RR{\mu}\nabla^b  u^{a_1\ldots a_s})\\
\fl
\qquad  =\nabla^b \left(t^2 (\RR{\mu} \nabla_b u_{a_1\ldots a_s})(\RR{\mu} Du^{a_1\ldots a_s})\right)
-2 t^2((\nabla_b\RR{\mu})Du_{a_1\ldots a_s})(\RR{\mu}\nabla^b  u^{a_1\ldots a_s}).
\end{eqnarray*}
When this is multiplied with the volume form and integrated over $H$, the first term drops out due to Stokes theorem (see for example \cite{Aubin:1982tw}). We have therefore found
\begin{eqnarray*}
\fl
\quad\scalarpr{\RR{\mu}(t^2\Delta_{h} u_{a_1\ldots a_s})}{\RR{\mu}Du_{a_1\ldots a_s}}_{L^2(H)}
+\scalarpr{\RR{\mu}t\nabla_b Du_{a_1\ldots a_s}}{\RR{\mu}t\nabla_b u_{a_1\ldots a_s}}_{L^2(H)}\\
\fl
\qquad=-2 \scalarpr{t(\nabla_b\RR{\mu})Du_{a_1\ldots a_s}}{\RR{\mu}t\nabla_b  u_{a_1\ldots a_s}}_{L^2(H)}.
\end{eqnarray*}
We get
\begin{equation*}
  DE=\e^{-\kappa t^\gamma}\scalarpr{\RR{\mu}f_{a_1\ldots a_s}}{\RR{\mu}Du_{a_1\ldots a_s}}_{L^2(H)}-\e^{-\kappa t^\gamma} K(U,U),
\end{equation*}
where $K$ is a bilinear form on the space of smooth time-dependent vector-valued tensor fields generated by
\[U:=\left(\RR{\mu}u_{a_1\ldots a_s},\RR{\mu}Du_{a_1\ldots a_s},\RR{\mu}t\nabla_b  u_{a_1\ldots a_s}\right)^T
=:(U_1,U_2,U_3).\]
The expression of $K$ is
\begin{eqnarray*}
\fl
\qquad  K(U,V)&=&
\frac 12\kappa \gamma t^\gamma
\left(\scalarpr{ U_1}{V_1}_{L^2(H)}+\scalarpr{ U_2}{V_2}_{L^2(H)}+\scalarpr{ U_3}{V_3}_{L^2(H)}\right)
-t^2 \tilde G(U,V)\\
&&+2 t\scalarpr{\nabla_b\RR{\mu}\RR{\mu}^{-1}U_2}{U_3}_{L^2(H)}\\
&&-\scalarpr{(D\RR{\mu})\RR{\mu}^{-1}U_1}{V_1}_{L^2(H)}
-\scalarpr{U_2}{V_1}_{L^2(H)}\\
&&-\scalarpr{(D\RR{\mu})\RR{\mu}^{-1}V_2}{V_2}_{L^2(H)}\\
&&-\scalarpr{D(\RR{\mu-1})\RR{\mu-1}^{-1}U_3}{V_3}_{L^2(H)}
+\scalarpr{2A U_2}{V_2}_{L^2(H)},
\end{eqnarray*}
where $\tilde G$ is a bilinear form defined by the map $G$.
%
%
%
The bilinear form $K$ is positive definite uniformly on $(0,\delta]$ if $\kappa$ is sufficiently large and $\gamma$ is sufficiently small and if the $3d\times 3d$ energy dissipation matrix given by \Eqref{eq:energydissipationmatrix}
is positive definite at each spatial point $x\in H$; note that
$(D\RR{\mu})\RR{\mu}^{-1}=-\diag(\mu_1,\ldots,\mu_d)$,
as follows directly from the definition of $\RR{\mu}$.
If this is the case, then
\begin{eqnarray*}
  DE&\le&\e^{-\kappa t^\gamma}
  \scalarpr{\RR{\mu}f_{a_1\ldots a_s}}{\RR{\mu}Du_{a_1\ldots a_s}}_{L^2(H)}\\
  &\le& \ee^{-\kappa t^\gamma}\|\RR{\mu}f_{a_1\ldots a_s}\|_{L^2(H)}\,\|\RR{\mu}Du_{a_1\ldots a_s}\|_{L^2(H)}\\
  &\le& C \ee^{-\kappa t^\gamma}\|\RR{\mu}f_{a_1\ldots a_s}\|_{L^2(H)}\,\sqrt{E},
\end{eqnarray*}
for some constant $C>0$.
Using the Gr\"onwall inequality, this can be integrated as usual, and we obtain \Eqref{EnergyEstimate}.
\end{proof}

Now we want to obtain similar estimates for higher derivatives of the unknown.
First, we show that the form of \Eqref{eq:lineartensorialsystem} is retained, when we take a covariant derivative of both sides and then interpret the result as an equation for $\nabla_b u_{a_1\ldots a_s}$:
\begin{eqnarray*}
  &D^2&\nabla_b  u_{a_1\ldots a_s}+2A D \nabla_b  u_{a_1\ldots a_s}
-t^2h^{{c}{d}}\nabla_b \nabla_{{c}}{\nabla_{d}} u_{a_1\ldots a_s}\\
&=&\nabla_b  f_{a_1\ldots a_s}
+t^2 G_1(u)_{b  a_1\ldots a_s}+t^2 G_2(\nabla u)_{b  a_1\ldots a_s}.
\end{eqnarray*}
Here $G_1$ and $G_2$ are maps obtained from $G$ by covariant differentiation.
Using the definition of the Riemann and Ricci tensors of $h_{ab }$, we compute
\begin{eqnarray*}
  &h^{{c}{d}}&\nabla_b \nabla_{{c}}{\nabla_{d}} u_{a_1\ldots a_s}\\
  &=&h^{{c}{d}}\nabla_{{c}}\nabla_b {\nabla_{d}} u_{a_1\ldots a_s}
  -h^{{c}{d}}{R_{b {c}{d}}}^{{d}'}{\nabla_{{d}'}} u_{a_1\ldots a_s}
  -\sum_{l=1}^s h^{{c}{d}}{R_{b {c} a_l}}^{a_{l'}}{\nabla_{{d}}} u_{a_1\ldots a_{l'}\ldots a_s}\\
  &=&h^{{c}{d}}\nabla_{{c}}{\nabla_{d}}\nabla_b  u_{a_1\ldots a_s}
  -\sum_{l=1}^s h^{{c}{d}}\nabla_{{c}}({R_{b {d} a_l}}^{a_{l'}} u_{a_1\ldots a_{l'}\ldots a_s})\\
  &&+{R_{b }}^{{d}'}{\nabla_{{d}'}} u_{a_1\ldots a_s}
  -\sum_{l=1}^s h^{{c}{d}}{R_{b {c} a_l}}^{a_{l'}}{\nabla_{{d}}} u_{a_1\ldots a_{l'}\ldots a_s}\\
  &=&h^{{c}{d}}\nabla_{{c}}{\nabla_{d}}\nabla_b  u_{a_1\ldots a_s}
  -\sum_{l=1}^s h^{{c}{d}}\nabla_{{c}}{R_{b {d} a_l}}^{a_{l'}} u_{a_1\ldots a_{l'}\ldots a_s}\\
  &&-\sum_{l=1}^s h^{{c}{d}}{R_{b {d} a_l}}^{a_{l'}} \nabla_{{c}}u_{a_1\ldots a_{l'}\ldots a_s}
  +{R_{b }}^{{d}'}{\nabla_{{d}'}} u_{a_1\ldots a_s}\\
  &&-\sum_{l=1}^s h^{{c}{d}}{R_{b {c} a_l}}^{a_{l'}}{\nabla_{{d}}} u_{a_1\ldots a_{l'}\ldots a_s}\\
  &=&h^{{c}{d}}\nabla_{{c}}{\nabla_{d}}\nabla_b  u_{a_1\ldots a_s}
  +{R_{b }}^{{d}'}{\nabla_{{d}'}} u_{a_1\ldots a_s}\\
  &&-\sum_{l=1}^s\Bigl( \nabla^{{d}}{R_{b {d} a_l}}^{a_{l'}} u_{a_1\ldots a_{l'}\ldots a_s}
  + 2{R_{b {d} a_l}}^{a_{l'}} \nabla^{{d}}u_{a_1\ldots a_{l'}\ldots a_s}\Bigr).
\end{eqnarray*}
Hence, for the time-dependent $d$-vector-valued $(0,s+1)$-tensor field $\hat u_{a_0a_1\ldots a_s}:=\nabla_{a_0} u_{a_1\ldots a_s}$, we have found the system
\begin{equation}
\fl
\label{eq:lineartensorialsystemder}
  \qquad D^2\hat u_{a_0a_1\ldots a_s}+2A D\hat u_{a_0a_1\ldots a_s}
-t^2\Delta_{h} \hat u_{a_0a_1\ldots a_s}=\hat f_{a_0a_1\ldots a_s}
+t^2 \hat G(\hat u)_{a_0 a_1\ldots a_s},
\end{equation}
with
\begin{equation}
  \label{eq:fhat}
\fl
  \qquad\hat f_{a_0a_1\ldots a_s}
  :=\nabla_{a_0} f_{a_1\ldots a_s}+t^2 G_1(u)_{b  a_1\ldots a_s}-t^2\sum_{l=1}^s \nabla^{{d}}{R_{b {d} a_l}}^{a_{l'}} u_{a_1\ldots a_{l'}\ldots a_s},
\end{equation}
and
\begin{equation*}
\fl
 \qquad\hat G(\hat u)_{a_0 a_1\ldots a_s}:=G_2(\hat u)_{a_0  a_1\ldots a_s}+{R_{a_0}}^{{d}'}\hat u_{{d}' a_1\ldots a_s}
 -2\sum_{l=1}^s{R_{a_0{d} a_l}}^{a_{l'}} {{\hat u}^{{d}}}_{\,\,\,\,a_1\ldots a_{l'}\ldots a_s}.
\end{equation*}
Hence, when $u_{a_1\ldots a_s}$ is considered as a fixed known field and \Eqref{eq:lineartensorialsystemder} is understood as an equation for $\hat u_{a_0a_1\ldots a_s}$, then it is of the same form as \Eqref{eq:lineartensorialsystem} with $s$ replaced by $s+1$.
In fact, we can apply \Lemref{lem:basicenergyestimate} to \Eqref{eq:lineartensorialsystemder} 
\begin{eqnarray*}  
  \sqrt{E_{\mu,\kappa,\gamma}[\nabla_{a_0} u_{a_1\ldots a_s}](t)}
  -\sqrt{E_{\mu,\kappa,\gamma}[\nabla_{a_0}u_{a_1\ldots a_s}](t_0)}\\
  \le C\int_{t_0}^t s^{-1}\|\RR{\mu}(s) \hat f_{a_0a_1\ldots a_s}(s)\|_{L^2(H)}\,\dd s,\\
  \le
 C\int_{t_0}^t s^{-1}\Bigl(\|\RR{\mu}(s) \nabla_{a_0} f_{a_1\ldots a_s}(s)\|_{L^2(H)}
+s^2\|\RR{\mu}(s) u_{a_1\ldots a_s}(s)\|_{L^2(H)}\Bigr)\dd s,
\end{eqnarray*}
where we have used \Eqref{eq:fhat} and the constant $C$ has been adapted in the last step in order to incorporate contributions from the map $G_1$ and from the covariant derivatives of the Riemann tensor. 
This can be understood as an estimate for the first spatial derivatives of the unknown. When we add this inequality to the estimate \Eqref{EnergyEstimate} for the undifferentiated unknown, we obtain an estimate for the $H^1(H)$-norm of the unknown by a relation which is similar to the equivalence estimate \Eqref{eq:equivalencenormenergy}. However, we must be cautious when $\mu$ depends on space: then the spatial derivatives of the function $\RR{\mu}$ introduce additional $\ln t$-factors which must be controlled by introducing an appropriate constant $\epsilon>0$ as shown below. We have hence proved the following lemma for the case $q=1$; the case $q>1$ be can obtained iteratively.

\begin{lemma}[Energy estimates of higher order]
\label{lem:higherorderenergyestimate}
Pick an integer $q\ge 1$ and any sufficiently small constant $\epsilon>0$.
Under the otherwise same conditions as \Lemref{lem:basicenergyestimate}, there exist positive constants $C$, $\rho$ such that for any smooth initial data $u^{*}_{a_1\ldots a_s}$ and $u^{**}_{a_1\ldots a_s}$ at $t=t_0\in (0,\delta)$, the solution $u_{a_1\ldots a_s}$ of the Cauchy problem for this system and this initial data satisfies the estimate
\begin{eqnarray*}
\fl\quad
  \|\RR{\mu-\epsilon}u_{a_1\ldots a_s}(t)\|_{H^q(H)}
+\|\RR{\mu-\epsilon}Du_{a_1\ldots a_s}(t)\|_{H^q(H)}+\|\RR{\mu-\epsilon}t\nabla_bu_{a_1\ldots a_s}(t)\|_{H^q(H)}\\
\fl\qquad
\le C\Bigl(\|\RR{\mu}u_{a_1\ldots a_s}(t_0)\|_{H^q(H)}
+\|\RR{\mu}Du_{a_1\ldots a_s}(t_0)\|_{H^q(H)}+\|\RR{\mu}t\nabla_bu_{a_1\ldots a_s}(t_0)\|_{H^q(H)}\\
\fl\qquad\qquad
+\int_{t_0}^t s^{-1}\left(\|\RR{\mu}(s) f_{a_1\ldots a_s}(s)\|_{H^q(H)}
+s^\rho\|\RR{\mu}(s) u_{a_1\ldots a_s}(s)\|_{H^{q-1}(H)}\right)\dd s\Bigr),
\end{eqnarray*}
  for all $t\in [t_0,\delta]$. The constants $C$, $\rho$ may be chosen  independently of $f_{a_1\ldots a_s}$. If one replaces the data $u^{*}_{a_1\ldots a_s}$ and $u^{**}_{a_1 \ldots a_s}$ at $t=t_0\in (0,\delta)$ by other smooth data specified at any time $t_1 \in (0, t_0]$, then the energy estimate holds for the \emph{same} constants $C$, $\rho$.
\end{lemma}

From now on, we restrict to the scalar case $s=0$; it is, however, clear from these energy estimates that the tensorial case works in exactly the same way.

\subparagraph{Step 2. Existence and uniqueness of the singular initial value problem for linear Fuchsian wave equations.}
We now use the same arguments as in \cite{Ames:yV5l9m6A} to show existence of solutions of the \textit{singular} initial value problem for scalar equations of the form \Eqref{eq:lineartensorialsystem} with leading-order term $u_0=0$. We set out to use
solutions of the Cauchy problem via
an approximation scheme which works as follows: We first choose a
monotonically decreasing sequence of times $t_n \in (0,\delta]$ which
converges to zero. Then for each $n$, we construct a function $v_n:
(0,\delta] \times H \rightarrow \R^n$ which vanishes on $(0,t_n] $,
and which is equal on $(t_n, \delta]$ to the solution of the Cauchy
problem with zero initial data at $t_n$. One readily verifies for smooth source-term functions $f$ that for
every choice of $\mu$, one has $v_n \in C^1((0,\delta]\times H)\cap
\tilde X_{\delta,\mu,q}$ for all integers $q\ge 0$. The central result of this section is that if
certain hypotheses hold, then the sequence $(v_n)$ -- whose elements
we label \keyword{approximate solutions} -- converges to a solution of
the singular initial value problem for the linear system with
vanishing leading term.

\begin{proposition}[Existence of solutions of the
   singular initial value problem]
  \label{prop:linearfirstexistence}
   Suppose that $(H,h_{ab})$ is a smooth connected orientable compact Riemannian manifold of dimension $n$.
  Pick a constant $\delta>0$, an exponent vector $\mu$ and an integer $q\ge 1$.
  Suppose that \Eqref{eq:lineartensorialsystem} is a linear Fuchsian wave equation with $s=0$ and $f$ is an element of $X_{\delta,\nu,q}$ for some exponent vector $\nu>\mu$ so that the {energy dissipation matrix} given by \Eqref{eq:energydissipationmatrix}
is positive definite for all $x\in H$. Then, there exists a solution $u=w:(0,{\delta}] \times H\rightarrow \R^d$ to the singular initial value problem with vanishing leading term $u_0=0$  which is an element of $\tilde X_{{\delta}, \mu, q}$. If $q>n/2+1$, then the solution is unique in the space $\tilde X_{{\delta}, \mu, q}$.
The solution operator
$\mathbb H:X_{\delta,\nu,q}\rightarrow\tilde X_{\delta,\mu,q}$, $f\mapsto w$ satisfies
  \begin{equation}
    \label{eq:continuityHPDENew}
    \|\mathbb H[f]\|^{\sim}_{\delta,\mu,q} \le \delta^\rho C\|f\|_{\delta,\nu,q},
  \end{equation}
  for all $f\in X _{\delta,\nu,q}$ for constants $C>0$ and $\rho>0$. 
\end{proposition}

The reader should be aware of the fact that we do not require the source-term $f$ to be smooth here, while this was a condition in \Lemref{lem:basicenergyestimate} and \Lemref{lem:higherorderenergyestimate}. In particular, we do not restrict to smooth solutions here; in fact the regularity of the solutions is determined by the integer $q$. Only if the conditions of \Propref{prop:linearfirstexistence} hold for \textit{all} integers $q\ge 1$, is the remainder of the solution $w$ in $\tilde X_{\delta,\mu,\infty}$ and hence $w$ has infinitely many spatial derivatives. In general one finds that the constants in \Eqref{eq:continuityHPDENew} may depend on $q$. In particular, $C$ may be unbounded and $\rho$ may go to zero in the limit $q\rightarrow\infty$.
Concerning time derivatives, we note that the remainder $w$ is a solution of the equations and hence $w$ is at least twice differentiable in $t$ under the hypothesis of the above proposition with $D^2 w\in X_{\delta,\mu,q-1}$. If $q$ is sufficiently large, then corresponding statements can be derived for higher time derivatives of $w$. In the case $q\rightarrow\infty$, it follows that $w$ has infinitely many time derivatives and all time derivatives are in $X_{\delta,\mu,\infty}$.

\begin{proof}[Proof of \Propref{prop:linearfirstexistence}]
The proof follows the arguments in \cite{Ames:yV5l9m6A} where all the details can be found. 
We only list the steps here and give short discussions of each.
In a first step we we consider the case $q=0$ and assume that the source term $f$ is in $C^\infty((0,\delta]\times H)\cap X_{\delta,\nu,0}$. The sequence of approximate solutions $(v_n)$ defined above is in the Banach space $\tilde X_{\delta,\mu,0}$ as mentioned above. Using the energy estimates of \Lemref{lem:basicenergyestimate}, our hypothesis is sufficient to guarantee that this is a Cauchy sequence; this can be shown in exactly the same way as in the proof of Proposition~2.10 in \cite{Ames:yV5l9m6A}. Next we could show, as done in that aforementioned proposition in \cite{Ames:yV5l9m6A}, that the corresponding limit $w$ in $\tilde X_{\delta,\mu,0}$ is a solution of the equation in a weak sense; however, we will not consider weak solutions here.

Next, following the argument of Proposition~2.12 in \cite{Ames:yV5l9m6A}, we consider the case $q=1$, i.e.\ that $(v_n)$ is a sequence in $\tilde X_{\delta,\mu,1}$ when we assume that $f\in C^\infty((0,\delta]\times H)\cap X_{\delta,\nu,1}$. By combining the energy estimates in \Lemref{lem:higherorderenergyestimate} for $q=1$ and by estimating the $q=0$-terms by means of the estimates which were obtained in the previous step, we find that $(v_n)$ is a Cauchy sequence in $\tilde X_{\delta,\mu,1}$. If $f\in C^\infty((0,\delta]\times H)\cap X_{\delta,\nu,q}$ for any integer $q\ge 1$ as given in the hypothesis, then we prove that $(v_n)$ is a Cauchy sequence in $\tilde X_{\delta,\mu,q}$ by successively using \Lemref{lem:higherorderenergyestimate} and by estimating the terms with lower derivatives by the estimates obtained from the previous step. We call the limit $w$.

We have hence gathered the following information about $w$: since $w\in \tilde X_{\delta,\mu,q}$, (i) the function $w$ is in $X_{\delta,\mu,q}$, (ii) the function $w$ is differentiable in time (see more details about this in the appendix of \cite{Ames:yV5l9m6A}) with $Dw\in X_{\delta,\mu,q}$, and (iii) $t \nabla_a w\in X_{\delta,\mu,{q}}$. In order to check whether $u=w$ is a solution of \Eqref{eq:lineartensorialsystem} with vanishing leading-order term $u_0$, we must first guarantee that all terms in \Eqref{eq:lineartensorialsystem} are well-defined at each $t\in (0,\delta]$ when evaluated on $w$. Since we suppose that $q\ge 1$, it remains to show that $w$ is in fact twice time differentiable. We do this in the same way as in the proof of Proposition~2.12 in \cite{Ames:yV5l9m6A}: we solve \Eqref{eq:lineartensorialsystem} algebraically for $D^2 w$ and replace $u=w$ on the right-hand side by $v_n$. Then we define
\[\hat v_n:=-2A Dv_n
+t^2\Delta_{h} v_n+f+t^2 G\cdot v_n.\]
Let us fix a compact subinterval $[\delta_0,\delta]$. By choosing $n$ sufficiently large, we can assume without loss of generality that $\hat v_n(t,x)=D^2 v_n(t,x)$ for all $(t,x)\in [\delta_0,\delta]\times H$, and we find that $\hat v_n\in X_{\delta,\mu,q-1}$ for all $n$. In fact $(\hat v_n)$ is a Cauchy sequence in  $X_{\delta,\mu,q-1}$ and therefore has a limit $\hat v$. Using standard arguments of uniform convergence, we can then show that $w$ is twice time differentiable and that $D^2 w=\hat v\in X_{\delta,\mu,q-1}$. It is therefore clear that $w$ is really a solution (in the strong sense) of \Eqref{eq:lineartensorialsystem}.

The same arguments as in Proposition~2.11 in \cite{Ames:yV5l9m6A} lead to the conclusion that the thus constructed solution of the singular initial value problem does not depend on the choice of sequence $(t_n)$ which we have used above to define the sequence $(v_n)$. We can thus define a solution operator $\mathbb H$ which maps the source-term $f$ to the thus constructed solution $w$ (but we do not claim uniqueness of the solution yet). It is, at this stage, a map $C^\infty((0,\delta]\times H)\cap X_{\delta,\nu,q}\rightarrow \tilde X_{\delta,\mu,q}$. In the same way as in the proof of Propositions~2.11 and 2.12 in \cite{Ames:yV5l9m6A}, we derive the continuity estimate \Eqref{eq:continuityHPDENew} for $\mathbb H$. It is a standard result that a linear operator with an estimate of the form \Eqref{eq:continuityHPDENew} can be uniquely extended from the dense subspace $C^\infty((0,\delta]\times H)\cap X_{\delta,\nu,q}$ to the full space $X_{\delta,\nu,q}$. It follows easily that this extended operator, which 
we 
refer to with the same name $\mathbb H$, maps any source-term $f\in X_{\delta,\nu,q}$  to  the remainder of solution $w\in \tilde X_{\delta,\mu,q}$.

The last step is to show uniqueness of the solution in $\tilde X_{\delta,\mu,q}$. Suppose that there are two solutions $w_1,w_2\in\tilde X_{\delta,\mu,q}$ of the same equation. Hence, their difference $\xi:=w_1-w_2\in \tilde X_{\delta,\mu,q}$ satisfies the same equation with vanishing source-term function $f$. We notice that the same computation as in the proof of \Lemref{lem:basicenergyestimate} is valid if the solution is not necessarily smooth, but, say, if $w$, $Dw$, $\nabla_a w$, $D^2 w$, $\nabla_a Dw$ and $\Delta_h w$ are all continuous on $(0,\delta]\times H$. According to \Lemref{lem:sobolevRiemannian}, this is the case if $q-1>n/2$. Now, we can consider  the function $\xi$ as a solution of the Cauchy problem of the equation with vanishing source term corresponding to data $\xi^*:=\xi(t_0,\cdot)$, $\xi^{**}:=\partial_t \xi(t_0,\cdot)$.  Then \Lemref{lem:basicenergyestimate} implies that
\begin{eqnarray*}
\fl\qquad
  \|\RR{\mu}\xi(t,\cdot)\|_{L^2(H)}+\|\RR{\mu} D\xi(t,\cdot)\|_{L^2(H)}+\|\RR{\mu} t\nabla_a\xi(t,\cdot)\|_{L^2(H)}\\
\fl\qquad\quad
  \le C \left(\|\RR{\mu}\xi(t_0,\cdot)\|_{L^2(H)}+\|\RR{\mu} D\xi(t_0,\cdot)\|_{L^2(H)}+\|\RR{\mu} t_0\nabla_a\xi(t_0,\cdot)\|_{L^2(H)}\right).
\end{eqnarray*}
In the same way as in the proof of Proposition~2.14 in \cite{Ames:yV5l9m6A} we take the limit $t_0$ of such an estimate and find that $\xi$ must vanish. It follows that $w_1=w_2$.
\end{proof}

\subparagraph{Step 3. Existence and uniqueness for nonlinear equations.}

We now have the essential tools needed to prove \Theoremref{th:Wellposedness1stOrderFiniteDiff} for general semilinear Fuchsian wave equations of the form \Eqref{eq:1stordersystem}. Again, we follow the approach of \cite{Ames:yV5l9m6A} and we refer to that work for the details.

\begin{proof}[Proof of \Theoremref{th:Wellposedness1stOrderFiniteDiff}] 
We restrict to $H=\St$ and $h=\dd\vartheta^2+\sin^2\!\theta\,\dd\phi^2$ here, but we remark that the result holds under more general assumptions.

The idea is to construct the following iteration scheme. We start with some seed function $u_{[0]}=u_0+w_{[0]}$ where $u_0$ is the fixed leading-order term (which does not need to vanish here), and where $w_{[0]}$ is some element in $\tilde X_{\delta,\mu,q}$. Without loss of generality, we can assume that $w_{[0]}=0$. We can linearize the equation around $u_{[0]}$, i.e.\ consider the equation
\[L[w_{[1]}]=F_{red}(u_0)[w_{[0]}]=F(u_0)[w_{[0]}]-L[u_0],\]
for an unknown $w_{[1]}$. Because of the hypothesis of
\Theoremref{th:Wellposedness1stOrderFiniteDiff}, the source-term $F(u_0)[w_{[0]}]-L[u_0]$ satisfies the hypothesis of \Propref{prop:linearfirstexistence}, and hence there is a unique $w_{[1]}$ in $\tilde X_{\delta,\mu,q}$ which solves this equation. In a next step, we solve the linear equation
\[L[w_{[2]}]=F_{red}(u_0)[w_{[1]}],\]
and obtain a unique solution $w_{[2]}$ in $\tilde X_{\delta,\mu,q}$ etc. This yields a sequence $(w_{[i]})$ in $\tilde X_{\delta,\mu,q}$ and we wish to show that this sequence converges to the unique solution of the nonlinear equation. The operator $\mathbb G=\mathbb H\circ F_{red}(u_0)$ maps $w_{[i]}$ to $w_{[i+1]}$, and solutions of the nonlinear equations are precisely the fixed points of this operator.  As in the proof of Theorem~2.4 in \cite{Ames:yV5l9m6A} we find that the Lipschitz requirement on $F_{red}$ in the hypothesis of \Theoremref{th:Wellposedness1stOrderFiniteDiff} and the estimate \Eqref{eq:continuityHPDENew} for a sufficiently small value of $\delta$ is enough to show that the sequence is bounded and $\mathbb G$ is a contraction. The result follows therefore from the Banach fixed point theorem.

Note that in the semilinear case, which we restrict to here, we do not lose a derivative by this argument; this is an issue which occurs for general quasilinear equations as discussed in \cite{Ames:yV5l9m6A}, and which needs to be fixed by a duality argument.

Now we consider the ``smooth case'' $q=\infty$. Hence, we assume that the hypothesis of \Theoremref{th:Wellposedness1stOrderFiniteDiff} holds for all integers $q>2$. In particular, there exists a sequence of constants $(C_q)$ in \Conditionref{cond:LipschitzF} of \Theoremref{th:Wellposedness1stOrderFiniteDiff} for each value of $q$, which may be unbounded in the limit $q\rightarrow\infty$. While it allows us to choose a sufficiently small $\delta_q$ to make the above argument work for every finite value of $q$, it may turn out that the sequence $(\delta_q)$ of these numbers has to approach $0$ in the limit $q\rightarrow\infty$. Hence there would be no smooth solution since the time interval of existence would be empty. However, this problem can be circumvented by means of standard continuation arguments for wave equations (or more general hyperbolic equations). In the literature one finds these arguments mainly for nonlinear wave equations on the spatial domain $\R^n$; see for example Lemma~9.14 in 
\cite{Ringstrom:2009cj}. 
The point is that for this class of equations, the energy of all higher derivatives of the solutions is bounded by  the time integral of a norm of the solution (a function called $m$ in \cite{Ringstrom:2009cj}). This function can be bounded by the $H^p$-norm of the solution for a sufficiently large $p$ (dependent on the dimension; the particular value is not important in the smooth case here).  In particular, if the $H^p$-norm is bounded, then the solution can be extended as an $H^q$-solution for every integer $q>p$ which is compatible with the coefficients of the equations and the data. This argument can easily be adapted to our compact spatial manifold $H=\St$: first, cover the manifold $H$ by a finite number of open coordinate patches $U\subset H$, and hence use the coordinate maps $\Phi$ to transfer the solution and the equation to the corresponding open patches $V$ in $\R^n$ (with $n=2$ in order case). If the $H^p(H)$-norm is finite at a certain time, then the $H^p(V)$-norm is finite for any of 
those open subsets $V\subset\R^n$. According to the above theorem, the solution can be extended for a sufficiently small time interval inside the domain of dependence of each of these local domains as a $H^q(\tilde V)$-solution where $\tilde V\subset V$ is sufficiently small. A standard ``patching'' argument then implies that we obtain a $H^q(H)$ solution on a sufficiently small extension of the time interval.  Such patching arguments have been used often in the literature in general relativity, see for example \cite{Friedrich:1986eo,Friedrich:1991nn,Ringstrom:2008kx}. Now, in our case, we know that there is a solution with remainder in $X_{\delta_p,\mu,p}$ for an arbitrary sufficiently large $p$. So far, we know that it is also in $X_{\delta_q,\mu,q}$ for all $q>p$ where, possibly, $\delta_q<\delta_p$. Thanks to this extension argument, we can, however, always choose $\delta_q=\delta_p$. After we have applied this argument to all $q>p$, we find that there exists a solution with remainder $w\in X_{\delta_p,\mu,\infty}$.
\end{proof}

\section*{References}


\end{document}